\documentclass[fleqn,10pt]{wlscirep}

\usepackage{float}
\usepackage{framed}

\usepackage{bm}
\usepackage{physics}
\usepackage{amsfonts}
\usepackage[utf8]{inputenc}
\usepackage[T1]{fontenc}
\usepackage{mathpazo} % get Palatino math  
\usepackage{amsmath}

\usepackage[symbol*]{footmisc}

\newcommand{\JJ}{ {\bf J} }%current density
\newcommand{\TT}{ {\bf T} }%torque
\newcommand{\MM}{ {\bf M} }%magnetization
\newcommand{\MD}{ {\bf m} }%normalized magnetization
\newcommand{\blue}{}

\title{Magnetism, symmetry and spin transport in van der Waals layered systems}

\author[1,*]{Hidekazu Kurebayashi}
\author[2,**]{Jose H. Garcia}
\author[1]{Safe Khan}
\author[3,4]{Jairo Sinova}
\author[2,5]{Stephan Roche}
\affil[1]{London Centre for Nanotechnology, UCL,  17-19 Gordon Street, London, United Kingdom}
\affil[2]{Catalan Institute of Nanoscience and Nanotechnology (ICN2), CSIC and BIST, Campus UAB, Bellaterra, 08193 Barcelona, Spain}
\affil[3]{Institut f\"ur Physik,  Johannes  Gutenberg  Universit\"at  Mainz,  D-55099 Mainz, Germany}
\affil[4]{Institute of Physics Academy of Sciences of the Czech  Republic, Cukrovarnickaa 10, 16200 Praha 6, Czech Republic}
\affil[5]{ICREA--Instituci\'o Catalana de Recerca i Estudis Avan\c{c}ats, 08010 Barcelona, Spain}

\affil[*]{e-mail: h.kurebayashi@ucl.ac.uk}
\affil[**]{e-mail:josehugo.garcia@icn2.cat}

\begin{abstract}
The discovery of an ever increasing family of atomic layered magnetic materials, together with the already established vast catalogue of strong spin-orbit coupling (SOC) and topological systems, calls for some guiding principles to tailor and optimize novel spin transport and optical properties at their interfaces. Here we focus on the latest developments in both fields that have brought them closer together and make them ripe for future fruitful synergy. After outlining fundamentals on van der Waals (vdW) magnetism and SOC effects, we discuss how their coexistence, manipulation and competition could ultimately establish new ways to engineer robust spin textures and drive the generation and dynamics of spin current and magnetization switching in 2D materials-based vdW heterostructures.  Grounding our analysis on existing experimental results and theoretical considerations, we draw a prospective analysis about how intertwined magnetism and spin-orbit torque (SOT) phenomena combine at interfaces with well-defined symmetries, and how this dictates the nature and figures-of-merit of SOT and angular momentum transfer. This will serve as a guiding role in designing future non-volatile memory devices that utilize the unique properties of 2D materials with the spin degree of freedom.
\end{abstract}

\begin{document}

\flushbottom
\maketitle

\thispagestyle{empty}

%\noindent \textbf{Key points:} Please suggest $\sim 5$ key points, which should be single-sentence bullet points that %summarize the article and remind readers of the take-home messages. An example of key points can be found at %\url{https://www.nature.com/articles/s42254-018-0001-7#Abs3}

\noindent \textbf{Key points:}
\begin{itemize}
    \item Fabrication of 2D vdW magnetic systems offers unprecedented opportunities for controlling magnetism and spin transport phenomena down to the monolayer limit. 
    \item Many vdW magnetic systems possess a low-symmetry crystalline structure, providing an array of exotic spin-orbit Hamiltonians, together with added richness arising from interface phenomena driven by layer-to-layer registry.
    \item Understanding the intertwined contribution of spin-spin interaction and interfacial symmetries is crucial for reaching the upper limit of SOT.  
    \item This exciting research field of spin transport at the frontier of layered spin-orbit coupling and magnetism will lead to discoveries of new materials, novel transport effects, topological phenomena and unconventional electron correlation physics.
\end{itemize}

%\noindent \textbf{Website summary:} Please suggest an $\sim 40$ word summary for the website.Please begin with a %general sentence setting the background, then outline the topics discussed in the article. You can find example %summaries at \url{https://www.nature.com/natrevphys/reviews}

\noindent \textbf{Website summary:} We present a foresight on how the spintronic properties of the emerging class of layered materials combining magnetism and strong spin-orbit coupling can be tailored by proper optimization of chemical interactions and structural material symmetries. This perspective draws a route to achieve best performing material design for reaching the upper limit of spin-orbit torque efficiency in switching magnetization.

%\section*{Section headings max 38 characters for Reviews and 32 for Perspectives, with no punctuation beside commas}
%\textit{Nature Reviews Physics} welcomes submissions in \LaTeX. For  peer-review purposes please just submit a .pdf %file via our manuscript tracking system \url{https://mts-natrevphys.nature.com/}. Following peer-review and subsequent %revisions we will ask for the .tex file and we will edit it using Overleaf. You will be able to follow the track %changes and answer the editor's queries in Overleaf.

\section*{Overall context of magnetism and spin-orbit coupling materials} 

\begin{figure}[ht!]
\centering
\includegraphics[width=15cm]{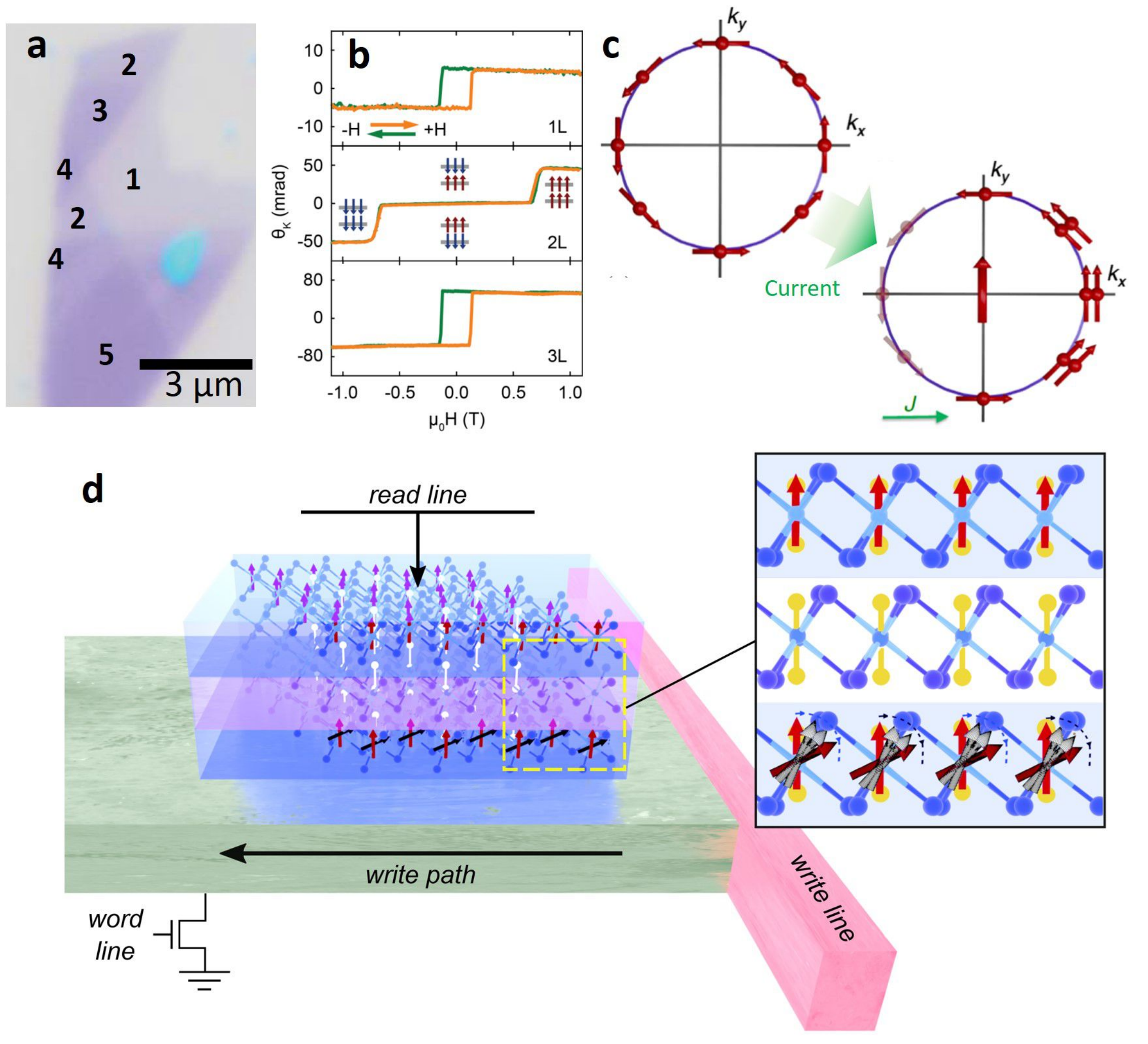}
\caption{\textbf{a} Microscope image of an exfoliated CrI$_3$ flake. The numbers in the image represent expected monolayer numbers calculated by the image contrast\cite{HuangNature2017}. \textbf{b} Magneto-optical Kerr rotation angle measurements applied to the different monolayer number regions while sweeping magnetic field. The monolayer and trilayer regions exhibit a ferromagnetic response, whereas the bilayer region shows an antiferromagnetic behaviour.  \textbf{c} Rashba spin texture in 2D momentum (reciprocal) space k$_x$ and k$_y$. For the sake of simplicity, we only show one spin-subband with electrons at each Fermi surface (Left panel). When we apply an electric current along the $x$ direction, electrons are redistributed and as a result, they become spin-polarized due to the spin texture. This is a source of SOT from the Edelstein effect. \textbf{d} Schematic of a single-cell of an SOT memory device with vdW magnetic materials. An information bit is stored by the relative orientation of two vdW magnets, which can be read by e.g. magnetoresistance through a vertical current through this stack. An external current is sourced by the write line/path electrode which is a low-resistive material such as copper. This current causes magnetization switching due to SOTs in the bottom vdW magnetic layer as highlighted in the zoom-in figure. The current-supplying bottom electrode could be another vdW material that acts as an additional spin source for boosting the switching efficiency. Figure 1a-b are adapted and modified from Ref. \cite{HuangNature2017}. }
\label{fig1}
\end{figure}

The realm of two-dimensional (2D) materials has considerably extended recently with the demonstration of magnetic orders in vdW material systems \cite{GongNature2017,HuangNature2017,Burch2018a,fei2018two,Li2019b,Gong2019b}. Weak mechanical bonding of vdW gaps in these materials offers unprecedented control of individual layer separations/fabrications by exfoliation, from which we are able to create truly monolayer magnets\cite{GongNature2017,HuangNature2017} (See Fig. 1a\&b as an example) and their heterostructures. This capability in sample fabrication is not readily within reach by other established thin-film growth techniques such as molecular beam epitaxy and sputtering, where the formation of grown layers significantly depends on surface energetics and kinetics together with lattice-matching between crystal structures of seed and grown layers. In contrast, exfoliation of vdW materials allows to stack different vdW monolayers without any thermodynamic constraints, enabling to access an unlimited number of stack combinations to explore fascinating, novel electronic structures\cite{GeimNature2013,LiuNRMater2016,AndreiNRMater2021}. This will provide tremendously exciting opportunities to generate new magnetism, spin-transport and electron-correlation physics by using 2D building blocks and currently intensive and extensive research activities on vdW magnetic systems are taking place worldwide. This class of new layered vdW materials is particularly vast, and includes transition metal dichalcogenides \cite{Freitas2015,Freitas2013} (TMDs), such as CrSe$_2$, VSe$_2$, CrTe$_2$, transition metal trihalides \cite{Zhang2015j,McGuire2014,McGuire2017} (CrI$_3$, CrCl$_3$) and transition metal phosphorous trichalcogenides \cite{Joy_PRB1992} (NiPS$_3$, MnPS$_3$, etc.). 2D magnetism has been measured at low temperatures in Cr$_2$Si$_2$Te$_6$ \cite{Lin2016}, CrI$_3$ \cite{HuangNature2017} and Cr$_2$Ge$_2$Te$_6$ \cite{GongNature2017} mono- and bilayers, or in VSe$_2$ \cite{Bonilla2018}, Fe$_3$GeTe$_2$ \cite{fei2018two,Deng2018} and up to room temperature (RT) in single layer MnSe$_2$\cite{OHara2018,OHara2018b}. In all those materials, the existence of stable magnetic properties for few layered systems is believed to be strongly related to their own magnetic anisotropies as well as the interaction with the substrate that prevents charge density wave instabilities \cite{Walker1983,Eaglesham1986,Feng2018,Yang2014b,Coelho2019}. 

In spintronic applications, magnetic moments are informational bits such as those in hard-disc drives and we process the bits by reversing the moment’s direction from one to the other between their equilibrium states\cite{HIROHATAJMMM2020,GrollierNElec2020,LinNElec2019}. Since electronic devices operate either with current- or voltage-regulated inputs, efficient control of magnetic moments by an electric means is therefore the primary interest for developing future commercially-viable functional spintronic devices such as spin random access memory\cite{KAWAHARA2012613,BHATTI2017530}. The spin-orbit interaction fits to this perfectly because it microscopically couples the electron motion (i.e. current) and the spin degree of freedom, hence making it possible to manipulate spins by electric currents (Fig. 1c). The electrically-generated spin polarization can exert spin torques on local magnetic moments via the exchange interaction. This phenomenon in solid states is termed as spin-orbit torque (SOT) which has been discovered and studied mainly in magnetic multilayers with 3D crystals \cite{Chernyshov2009,Miron2011b,Liu2012} and new SOT memory architectures\cite{Cubukcu2018} have been proposed and examined in the past several years. Microscopically, SOT originates from two leading effects, the spin-Hall effect \cite{SinovaRMP2015} and Edelstein effect\cite{Edelstein1990}. The efficiency and mechanisms of SOT are determined by the electronic states of materials as well as interfaces, in particular inversion symmetry breaking in combination with other crystalline symmetries\cite{Manchon2019}. In this regard, 2D vdW materials and heterostructures hold great potential to produce unique SOT phenomena from their exotic electronic structures which can be in principle modified by interface engineering. Furthermore, the efficiency of magnetization switching by SOT is proportional to magnetic volume\cite{Manchon2019} and so the smaller the better, ideal to the 2D limit since the volume is minimised along one dimension. In this extreme monolayer regime, it is important to stabilize magnetic orders against thermal fluctuation. This calls for detailed understanding of magnetism at that limit in terms of their driving mechanisms and the strength of them, which can be designed and controlled by material choice and external stimuli. When we have full understanding of both magnetism and SOTs in novel vdW heterostructures, we can design highly-efficient spintronic memory cells based on vdW materials such as one shown in Fig. 1d. In this cell, an information bit is stored by the relative orientation of two vdW magnetic layers separated by another vdW non-magnetic system in the middle. Tunnelling magneto-resistance effect would be best in terms of outputting a large voltage difference as a read-out. Writing actions for this cell is carried out by a horizontal current which will produce SOT effects to exert magnetic switching of the bottom layer. The efficiency of switching, in terms of electric power, is a key figure of merit since the size of a transistor to drive a current for each cell is determined by this efficiency (a large current implies bigger transistor size) and the cell density is currently limited by the transistor size. This directly means that maximising the SOT efficiency leads not only to low-power consumption of memory cells but also to the high density of data processing. In this sense, detailed understanding of physical origins and mechanisms of SOTs in any novel materials is of direct relevance of future ultracompact and low-power spin memory devices. It is important to note that non-magnetic high-spin-orbit vdW materials, such as WTe$_2$, could be inserted as an additional spin source for SOT-driven magnetization switching. In this regard, a whole variety of vdW heterostructures should be extensively studied for the development of novel vdW spin memory devices.   

This Review particularly focuses on two fundamental aspects of 2D vdW properties for future spintronic applications. The first is on the exchange interaction as a source of magnetic order by which 2D vdW magnetic systems display different exchange interaction mechanisms depending on their crystalline structures and transport properties from insulators to metals. Understanding of these underlying mechanisms is useful to visualize different magnetic orders at play in a variety of materials and provides a good guide for designing monolayer magnets that are magnetically stable above RT. After this, we also review recent development on electric and other control of equilibrium magnetic states, which is the current state-of-the-art in the relevant field. We then focus on the potential and opportunities of SOC effects arising within 2D vdW magnetic materials as well as in cases when they are interfaced to an additional material such as substrates. Their low-symmetry point groups exhibit a colorful picture of SOT effects and by applying high-throughput point group analysis, we generate and summarize the expected symmetry of SOTs in all the class currently available in the literature. This will aid future SOT experiments using listed materials as well as those newly discovered but possessing the same crystalline symmetry as listed. We further catergorize field-like and damping-like torques existing in these materials as well as more exotic anisotropy-like torques arising from their low-symmetry properties. These two fundamental aspects have not been the focus of prior review articles on 2D vdW magnetic systems \cite{Burch_Nature2018,Gibertini_NatNano2019,Gong_eaav4450,Mak_NatRevPhys2019,Huang_NatMater2020,Huang_Nanoscale2020,Wei_2DMater2020}, which extensively cover the recent development of 2D vdW magnetic systems as well as spintronics with non-magnetic vdW materials\cite{Han_APLMater2016,Avsar_RevModPhys2020,Husain_APRev2020,Ahn_npj2020}.

\section*{Different exchange interactions to stabilize magnetic orders in van der Waals systems} \label{sec:exchange}

We here briefly summarize that the magnetic order driven by the exchange interaction of electron spins has different mechanisms, depending on the conductivity regime of materials. This is very relevant to vdW magnetic systems since their conductivity spans across a wide range, from insulating to metallic regimes, possessing a very colorful nature of magnetism due to their conductivity as shown in Fig.\ref{fig2}. This is one of the unique potentials of vdW magnetic systems as research topics for current/future materials science for nano-magnetism and spintronic applications. The origin of long-range magnetic orders in condensed matter arises from the microscopic exchange interaction between individual spins. The exchange interaction between an array of $n$ spins is usually captured by $H_\text{ex}=-\sum_{i,j}^{n}J_{ij}\mathbf{S}_i \cdot \mathbf{S}_j$ where the dot product of $\mathbf{S}_i$ and $\mathbf{S}_j$ represents the relative orientation between the two spins. This Hamiltonian describes how much the spin system gains/loses their total energy by ordering their spin orientations as a whole. 

\subsection*{The super-exchange coupling in insulators}
The exchange interaction model is better suited to describe spins that are well-isolated, such as in the case of magnetic insulators and low-conducting semiconductors, where isolated spins interact with each other via non-magnetic ions (ligands) such as oxygen in their crystal lattices. Such indirect exchange coupling is often called the super-exchange mechanism \cite{Anderson_PR1950} and a schematic to highlight this is shown in Fig.\ref{fig2}. This can be explained by the virtual excitation of electrons in a magnetic cation and subsequent transmission to the neighbouring cation through the non-magnetic ligand within a timescale of Heisenberg uncertainty. This virtual process effectively lowers the total energy of the system when a long-range magnetic order is present. Magnetic ions for the super-exchange interaction possess the same ionic charge states. 

Many of vdW materials exhibit semiconductor/insulator properties and this extends to those possessing long-range magnetic ordering. CrX$_3$ (X= I, Br and Cl) exhibits good insulating properties with a typical bandgap energy of 1.2-3.1 eV \cite{McGuire2017,McGuire2014,HuangNature2017,Lado_2017}. In a single layer of CrX$_3$, each Cr atom is surrounded by X atoms in an octahedral configuration and the Cr atoms form a honeycomb lattice. This crystal field splits the Cr $d$ states into $e_g$ and $t_{2g}$ manifolds where three electrons in their Cr$^{3+}$ state occupy the fully-polarized $t_{2g}$ states with $S$ = 3/2 in the valence bands, following the first Hund's rule. These Cr$^{3+}$ moments are coupled through the super-exchange mechanism via an X atom with an approximately 90 degree bonding angle \cite{KANAMORI1959,Goodenough_PR1955}. It is noted that due to the presence of finite antiferromagnetic coupling between Cr-Cr direct exchange mechanisms in these materials, the total exchange coupling strength (hence the size of $T_{\rm C}$) is relatively weak\cite{Huang_JACS2018}. A comprehensive study for $T_{\rm C}$ of few layers of CrX$_3$ has been carried out by Kim et al.\cite{Kim_PNAS2019}.  Cr$_2$X$_2$Te$_6$ (X=Ge,Si) exhibits semiconductor/insulator properties due to their electronic bands with about a 0.4 eV gap\cite{GongNature2017,Lin2016}. Very much like the case of CrX$_3$, a honeycomb Cr$^{3+}$ network stabilizes their ferromagnetic order owing to the super-exchange interaction via Te atoms. Another key group of magnetic vdW insulators is those of {\blue M}XPS$_3$ (M = V, Mn, Fe, Co, Ni, or Zn)\cite{Joy_PRB1992}, with a bandgap of 1.3 - 3.5 eV\cite{Wang_AdvFuncMater2018,Rehman_Micromach2018,Kim_NComm2019}. Transition metals X host magnetic moments within the honeycomb lattice, showing either antiferromagnetic or ferromagnetic order.

\subsection*{The double-exchange coupling in semiconductors}
When magnetic ions in a lattice have two types of ionic states, e.g. Mn$^{3+}$ and Mn$^{4+}$, an exchange of the two ionic states depends on the spin state of two magnetic ions. This exchange is {\blue generated by} %effectively 
one mobile carrier (either electron or hole) hopping between the two ions. This hopping is possible when spin states of the two magnetic ions are parallel, leading to delocalisation of the mobile carriers and reduction of their total kinetic energy. This is the double-exchange mechanism \cite{ZenerPhysRev1951} that stabilises a long-range magnetic order. A prototypical example of this is (La$_{1-x}$Ca$_x$)(Mn$^{3+}_{1-x}$Mn$^{4+}_{x}$)O$_3$\cite{Jonker_Physica1950}. In this material series, only one type of Mn ionic states exists for both ends of their compositions ($x=0,1$), which are good insulators, and the super-exchange interaction  stabilises antiferromagnetism in these limits. In their intermediate states, Mn$^{3+}$ and Mn$^{4+}$ coexist (effectively Mn$^{4+}$ being considered as Mn$^{3+}$ and one hole) and a ferromagnetic order outperforms the previously favored antiferromagnetic one via the double-exchange mechanism. In this regime, the presence of mobile carriers is confirmed by the conductivity which becomes several orders greater than those of the insulating compositions. 

The double-exchange mechanism has not been discussed extensively in 2D vdW magnetic systems so far, although this mechanism should be commonly present for magnets with moderate electric conduction. As discussed later, Cr$_2$Ge$_2$Te$_6$ with chemical and electric doping\cite{Wang_JACS2019,verzhbitskiy2020controlling} is one example where the exchange coupling strength is modified by the double-exchange mechanism, exhibiting large enhancements of the Curie temperature ($T_{\rm C}$) compared to undoped Cr$_2$Ge$_2$Te$_6$.

\subsection*{The band magnetism in metals}

When conduction electrons are fully delocalized and well hybridized with magnetic ions, we enter the itinerant regime where band magnetism is used to describe the magnetic order. In this mechanism, the total energy of the whole system with respect to magnetic order is discussed by the energy gain/loss of electrons close to the Fermi energy with the Pauli paramagnetism picture \cite{Blundell2001}. Based on this, together with the mean-field model \cite{Weiss1907}, the Stoner criterion $D(E_F)U\geq1$ (here $D(E_F)$ and $U$ are the density of states (DOS) at the Fermi level in a non-spin-polarized state and the electron correlation energy) can be {\blue applied}\cite{Stoner1,Stoner2}. This essentially means that when the Coulomb energy associated with the electron exchange is strong enough, spontaneous exchange-energy splitting takes place to drive ferromagnetism. This model is often employed to discuss the emergence of a magnetic order in metallic ferromagnets\cite{Blundell2001}.  

\begin{figure}[t!]
\centering
\includegraphics[width=12cm]{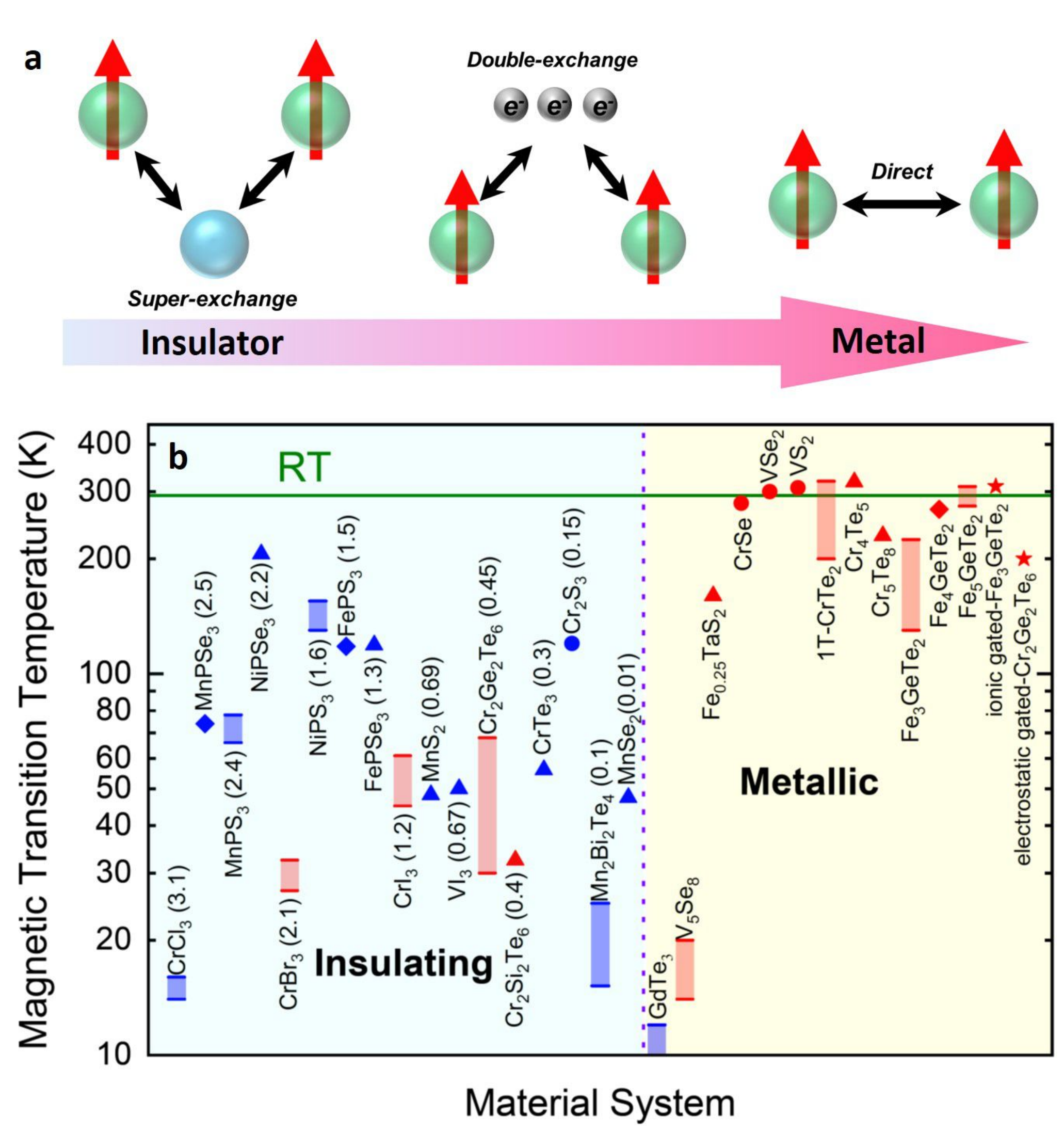}
\caption{Experimentally validated critical magnetic transition temperature fingerprints for various 2D vdWs magnetic systems. \textbf{a} Three different exchange coupling mechanisms of two individual moments as an underlying origin of the long-range magnetic ordering for material systems in the insulating and metallic phases going from left to right. \textbf{b} Two-dimensional vdW magnetic materials library showing experimentally determined magnetic transition temperature for materials in insulating and metallic phases. In the insulating phase, number in brackets corresponds to the bandgap value. The dataset is characterised in two-ways using the colours and symbols. Colours: Red (ferromagnetic ordering) and Blue (anti-ferromagnetic ordering). Symbols: Bar (top and bottom points in the bar highlight the transition temperature for bulk and mono/few-layers, respectively), $\bigtriangleup$ ($T_{\rm C}$ is reported for the bulk case), $\bigcirc$ ($T_{\rm C}$ is identified for mono/few-layers), $\Diamond$ ($T_{\rm C}$ is independent of material thickness), and $\bigstar$ ($T_{\rm C}$ is enhanced by the external techniques). The data points in the plot are taken from Refs. \cite{McGuire2017,cai2019atomically, wiedenmann1981neutron,Rehman_Micromach2018,long2017isolation,Wang_AdvFuncMater2018,le1982magnetic,ghazaryan2018magnon,samuelsen1971spin,Kim_NComm2019,lee2016ising,liu2020exploring,Huang_NNano2018,HuangNature2017,zhong2017van,chattopadhyay1991spin,Son_PRB2019,GongNature2017,khan2019spin,Lin2016b,McGuire_PRB2017,lv2015strain,sun2020room,OHara2018,chu2019sub,xie2020atomically,zeugner2019chemical,kan2014ferromagnetism,itoh1977magnetic,lei2020high,Nakano_NanoLett2019,Morosan_PRB2007,zhang2019ultrathin,Bonilla2018,zhang2013dimension,zhang2020critical,liu2019magnetic,seo2020nearly,May_ACSNano2019,Zhuang_PRB2016,Deng2018,verzhbitskiy2020controlling}}
\label{fig2}
\end{figure}

Magnetic vdW materials with electron conductivity are also available, such as VX$_2$ (X=Se,Te)\cite{Bonilla2018}, MnSe$_2$\cite{OHara2018}, V$_5$Se$_8$\cite{Nakano_NanoLett2019}, Fe-Ge-Te compounds\cite{fei2018two,Deng2018,May_PRB2016,May_ACSNano2019}, Fe$_{0.25}$TaS$_2$\cite{Morosan_PRB2007} and Cr-X (X=Te,Se) compounds\cite{McGuire_PRB2017,Wang_PRB2019,Yan_EPL2019}. As shown in Fig. \ref{fig2}, these metallic 2D vdW materials tend to display higher $T_{\rm C}$ due to the direct, stronger exchange interaction between individual moments. Some of them already possess an above-RT $T_{\rm C}$ which is relevant to spintronic device applications, although a much higher $T_{\rm C}$ than RT is welcomed for stable device operation in any of these applications. The Stoner criterion is indeed discussed for some of these materials, for example Fe$_3$GeTe$_2$\cite{Zhuang_PRB2016} was calculated to satisfy the inequality by $D(E_F)U=1.56\times0.71\geq1$.
 
A final remark of this section is on the Mermin-Wagner theorem \cite{Mermin1966} in which an absence of long range magnetic order in spin-rotational invariant systems is introduced. This could be statistically accurate but in reality it is difficult to achieve this since as soon as we make a real magnet, it automatically possesses any sort of magnetic anisotropies, be it from the spin-orbit interaction, anisotropic exchange interaction or magnetic dipole interaction, all of which opens up a gap for magnon excitations that suppress thermal fluctuations. 
 
\section*{Control of magnetism and magnetic states by external stimuli}

\begin{figure}[ht!]
\centering
\includegraphics[width=17cm]{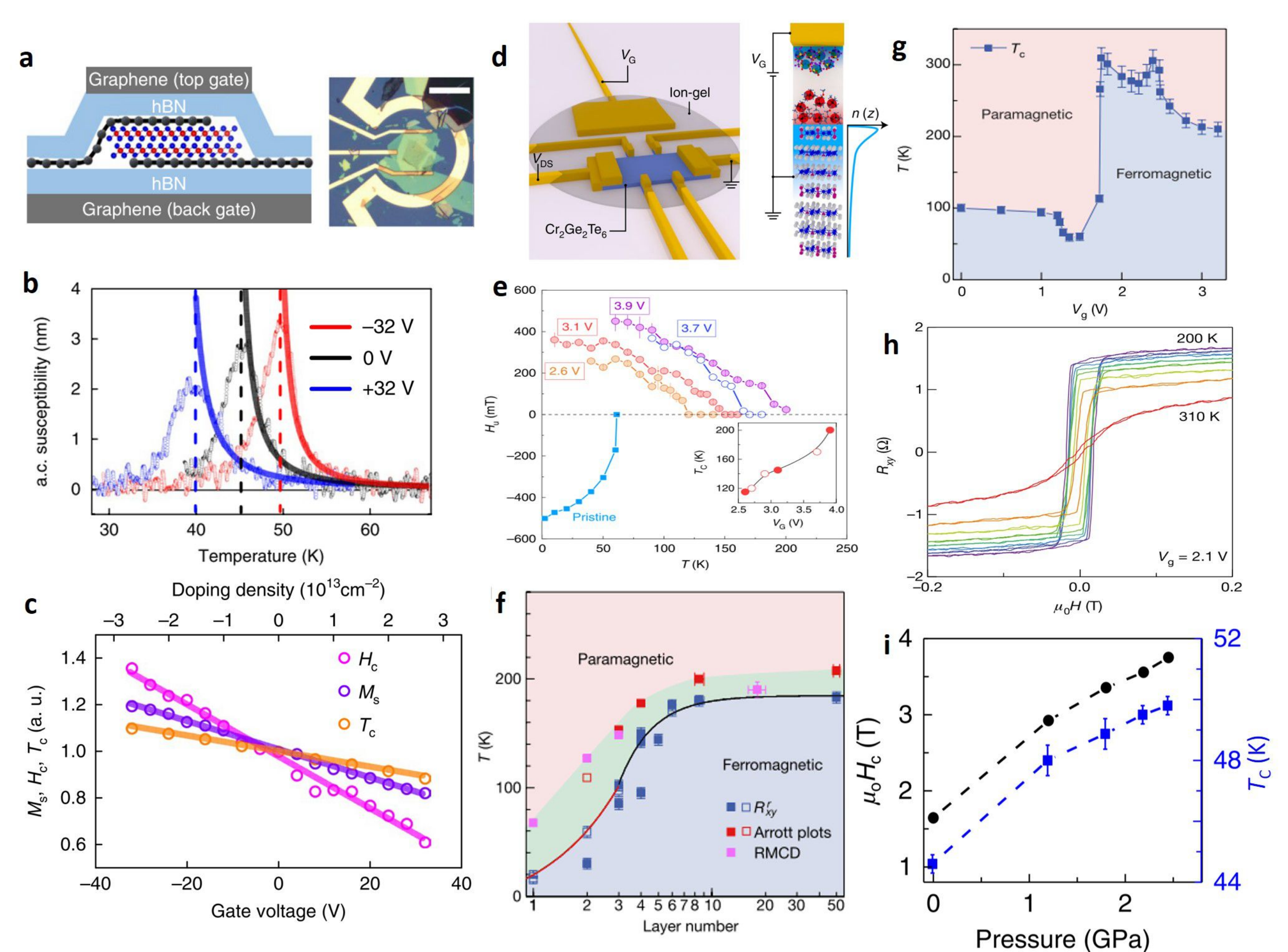}
\caption{Experimental observation of electric field control of magnetism in different vdW magnets. \textbf{a} A schematic side view of a field-effect device with a bilayer CrI$_3$ encapsulated in few-layer graphene and hBN. The left panel is an optical microscope image for their device with a scale bar of 20 $\mu$m. \textbf{b} Electric field control of magnetism in measured by ac susceptibility with different gate voltages as shown. \textbf{c}  Coercive force (magenta), saturation magnetization (purple) (both at 4 K) and Curie temperature (orange) normalized by their values at zero gate voltage as a function of gate voltage and induced doping density. \textbf{d} Schematics of a field-effect-transistor device with Cr$_2$Ge$_2$Te$_6$ and side view of charge accumulation by electric field. \textbf{e} Experimentally-deduced out-of-plane anisotropy field in Cr$_2$Ge$_2$Te$_6$ as a function of temperature for different gate voltages, with $T_{\rm C}$ with gate voltages (inset). \textbf{f} $T_{\rm C}$ of Fe$_3$GeTe$_2$ with different layer thicknesses measured by different analysis methods shown. \textbf{g} Gate voltage dependence of $T_{\rm C}$ in a trilayer Fe$_3$GeTe$_2$ and \textbf{h} their hysteresis loops for different temperatures. \textbf{i} the extracted critical field for spin-flip transition (black circles) and critical temperature (blue squares) as a function of pressure in a trilayer CrI$_3$. Figures are adapted from Ref. \cite{JiangNNano2018} (Fig. 3a-c), Ref. \cite{verzhbitskiy2020controlling} (Fig. 3d-e), Ref. \cite{Deng2018} (Fig. 3f-h) and Ref. \cite{song2019switching} (Fig. 3i) respectively. }
\label{fig3}
\end{figure}

As already summarized in the aforementioned section, the transport nature in magnets is strongly associated with the type of leading exchange interaction terms and their strength. The exchange interaction can be modified by the carrier concentration as well as the distance between magnetic ions. Here we summarize two external control of magnetism and magnetic states, electric field and strain, applied to 2D vdW magnetic systems. 

\subsection*{Electric field}

By fabricating a field-effect-transistor (FET)-type device with a thin-film magnet, an external control of magnetism by the charge accumulation at the surface is possible, and leads to a modulation of the Fermi level as well as surface carrier density. Therefore, this approach is a promising way of controlling the magnetic properties $in$-$situ$, which has clear relevance to not only achieving high-$T_{\rm C}$ materials but also gate-tunable spintronic devices. Electric field control of magnetism and magnetic states has been of interest within the field of spintronics where atomically-thin magnetic metals, dilute magnetic semiconductors and magnetic oxides are fundamental and instrumental materials \cite{Matsukura_NNano2015}. 2D vdW magnetic systems can offer added values to this already-developed research domain as summarised below. 

A dramatic change of magnetism by electric field, probed by modification of $T_{\rm C}$, has been observed in several 2D vdW magnets. Jiang et al.\cite{JiangNNano2018} fabricated parallel-plate capacitor devices with monolayer or bilayer CrI$_3$ together with a ring-shaped local ac-excitation metal patterns (schematic and microscopy images in Fig. \ref{fig3}a). Magnetic responses from the CrI$_3$ were measured by magnetic circular dichroism as ac susceptibility (Fig. \ref{fig3}b).  A clear change of $T_{\rm C}$ by applying electrostatic voltages was observed, together with other sizable changes of the saturation magnetization ($M_{\rm S}$) and coercive field ($H_{\rm C}$), as summarized in Fig. \ref{fig3}c. Bi-layer CrI$_3$ also exhibits a linear magneto-electric (ME) effect due to their magnetic point group with broken spatial-inversion and time-reversal symmetries\cite{McGuire_Crystals2017,Sivadas2016,Jiang2018,Huang_NNano2018} . Together with a relatively weak spin-flip transition in CrI$_3$, the linear ME effect can shift the transition field and demonstrate pure electric switch between ferromagnetic and anti-ferromagnetic states\cite{Jiang2018,Huang_NNano2018}  when the applied magnetic field is set at the point of transition while the gate voltage is swept. Song et al.\cite{Song_NanoLett2019} demonstrated that tunneling magnetoresistance (TMR) can be electrically modified by up to one order of magnitude in four-layer CrI$_3$ by using bistable magnetic states with the same net magnetization, yet showing different TMR values. 

An electrostatic modulation of magnetism has been also demonstrated in Cr$_2$Ge$_2$Te$_6$ by Wang et al.\cite{wang2018electric} and Verzhbitskiy et al. \cite{verzhbitskiy2020controlling} By applying the gate voltage to a 3.5 nm Cr$_2$Ge$_2$Te$_6$ film via a SiO$_2$ solid-state dielectric, Wang et al. achieved a carrier density of order of 10$^{12}$ cm$^{-2}$ for both electron and hole doing regimes, and observed a moderate change of $M_{\rm S}$ without any significant change of $T_{\rm C}$. Verzhbitskiy et al.\cite{verzhbitskiy2020controlling} attempted a similar FET-type device with 20 nm thick Cr$_2$Ge$_2$Te$_6$ with an ionic gel as a dielectric as shown in Fig. \ref{fig3}d. With this scheme, a much stronger doping action has been observed, at the level of 10$^{14}$ cm$^{-2}$ and $T_{\rm C}$ of Cr$_2$Ge$_2$Te$_6$ is enhanced from 64 K (bulk)\cite{GongNature2017} to 200 K. This enhancement is explained by the activation of the double-exchange interaction mechanism by introducing a significant number of electrons. In this regime, about one quarter of Cr$^{3+}$ magnetic ions become  Cr$^{2+}$, where hoping of electrons via the Te atoms contributes to the additional exchange mechanism, enhancing the exchange coefficient as well as $T_{\rm C}$. With this charge accumulation at the interface, the width of the region where electrons are doped electrostatically becomes comparable to the vdW gap ($\sim$ 1 nm). Therefore, we can create an artificial two-dimensional electron gas together with magnetism within the 20 nm thick Cr$_2$Ge$_2$Te$_6$. Furthermore, the magnetic easy axis of Cr$_2$Ge$_2$Te$_6$ is switched from out-of-plane to in-plane by this carrier doping\cite{verzhbitskiy2020controlling}. 

While CrI$_3$ and Cr$_2$Ge$_2$Te$_6$ are magnetic insulators in their pristine forms, the electric field modulation of magnetism has been also explored in metallic itinerant vdW magnets such as Fe$_3$GeTe$_2$ \cite{Deng2018}. Deng et al. performed a detailed study of Fe$_3$GeTe$_2$'s $T_{\rm C}$ with different layer numbers (Fig.\ref{fig3}f) and applied electric fields onto a trilayer Fe$_3$GeTe$_2$ film using solid electrolyte (LiClO$_4$ dissolved in polyethylene oxide matrix). $T_{\rm C}$ is greatly enhanced from 100 K to above 300 K by application of around 2 V, as shown in Fig. \ref{fig3}g\&h. This gate-tunability of $T_{\rm C}$ can be understood by the Stoner model\cite{Stoner1,Stoner2} since the energy gain by forming a ferromagnetic phase depends on the size of DOS at the Fermi level. Theoretical calculations suggest that by applying the electric field, the Fermi level can be lifted up to a large DOS peak area in Fe$_3$GeTe$_2$, mainly consisting of Fe $d_{z^2}$, d$_{xz}$ and d$_{yz}$ orbitals\cite{Deng2018}. The non-monotonic behaviour of $T_{\rm C}$ as a function of gate voltage may imply this Fermi level passing by the peak points.

\subsection*{Strain}

Strain is also a good external knob to modify the magnetic ground state of 2D vdW magnets. In some of these magnetic materials, the interlayer exchange coupling governs the ground-state spin orientation, leading for instance to antiferromagnetism in CrI$_3$. Since the exchange coupling is very sensitive to the distance between two spins (in this case, the vdW gap distance), any external perturbation modifying this length enables the control of magnetism, as shown with hydrostatic pressure of few GPa which can induce change of $T_{\rm C}$ of bulk vdW crystals Cr$_2$Ge$_2$Te$_6$\cite{Sun_APL2018}, CrI$_3$\cite{Mondal_PRB2019} and VI$_3$\cite{Son_PRB2019} as well as the sign change of the two-fold magnetic anisotropy in Cr$_2$Ge$_2$Te$_6$\cite{Lin_PRM2018}. In the thin-film regime, Song et al.\cite{song2019switching} and Li et al.\cite{Li_NMater2019} both independently applied hydrostatic pressure of a few GPa on a few monolayer of CrI$_3$ and observed a monoclinic-to-rhombohedral structural transition, driven by interlayer antiferromagnetic or ferromagnetic couplings \cite{Jiang_PRB2019,Jang_PRM2019,Sivadas_NanoL2019,SORIANO_SolidStateCom2019}. Similarly, a ferromagnetic-antiferromagnetic transition under applied pressure, as well as pressure-controlled magnetic state switching between antiferromagnetism to ferromagnetism have been also demonstrated \cite{song2019switching}. Such monoclinic-to-rhombohedral structural transition can be also observed in exfoliated CrCl$_3$ crystals with a variation of the interlayer exchange coupling by up to one order of magnitude, when compared to bulk samples\cite{klein2019enhancement}. This result highlights that sample fabrication processes unintentionally produce strain that greatly perturbs the magnetic properties of exfoliated magnets. Whether this monoclinic phase is thermodynamic ground states or meta-stable states induced by their exfoliation processes is yet to be revealed \cite{klein2019enhancement}. Furthermore, $T_{\rm C}$ can be modified by application of strain e.g. to a trilayer of CrI$_3$ as shown in Fig. \ref{fig3}i.

\section*{Spin-orbit torques in two-dimensional materials}

While the magnetic properties have been successfully controlled in a certain number of 2D vdW magnets, the access to out-of-equilibrium regimes in such materials by using electric currents and other means presents formidable challenges, but also holds paramount potential for future spintronic memory and logic applications\cite{Kent_Nnano2015,Dieny_NElec2020,BHATTI2017530,Grollier_NPhysRev2020}. This area of research is growing fast but to date only a very limited number of reports are available in the literature \cite{MacNeill_NPhys2017,Guimaraes_NanoLett2018,Shi_Nnano2019,Alghamdi_NanoLett2019,Wang_AdvSci2019,Ostwal_AdvMater2020}. Using electric fields, induced non-equilibrium spin polarization can result in substantial angular momentum transfer and magnetization control to a certain degree. This has been achieved using non-collinearity between two magnets\cite{RALPH_JMMM2008} in magnetic tunnel junctions, or via the intrinsic SOC of materials \cite{Manchon2019}. In this section, we start with an overview of the latest experimental demonstrations of magnetization control by electric currents with 2D vdW materials, and then present a theoretical analysis of spin torque physics.

TMDs with high spin-orbit coupling (SOC), such as WTe$_2$ and NbSe$_2$, have already been employed to generate spin torques in ferromagnetic metals\cite{MacNeill_NPhys2017,Guimaraes_NanoLett2018,Shi_Nnano2019}. In particular, MacNeill and coworkers observed that the torque symmetry arising in NiFe/WTe$_2$ bi-layers results from the symmetry lowering at the interface\cite{MacNeill_NPhys2017}, exhibiting the damping-like symmetry with the spin-polarization along the out-of-plane direction which cannot be generated by the spin-Hall effect\cite{Manchon2019}. Magnetic moments in vdW ferromagnets have been also manipulated by current-induced spin torques\cite{Alghamdi_NanoLett2019,Wang_AdvSci2019,Ostwal_AdvMater2020}. Interestingly, the spin-Hall effect in Pt was used to generate spin currents into $10\sim 20$-nm-thick exfoliated Fe$_3$GeTe$_2$ thin-films, and induce magnetization switching for low current density of 2.5 x 10$^{11}$ A/m$^2$\cite{Alghamdi_NanoLett2019,Wang_AdvSci2019}. This current density is actually comparable to the one needed to switch standard 3d transition metal alloys\cite{Manchon2019}. A magnetization switching by the spin-Hall effect has been also discussed in the insulator-Cr$_2$Ge$_2$Te$_6$/Ta system, requiring current density of $\sim$ 5 x 10$^{11}$ A/m$^2$, comparable to the one necessary for Tm$_3$Fe$_5$O$_{12}$\cite{Avci_NMat2017}. Finally very recently, using current density of about 3.9 x 10$^{12}$ A/m$^2$, a spin-orbit torque induced magnetization switching in an all vdW heterostructure was also reported in Fe$_3$GeTe$_2$/WTe$_2$ stacks \cite{shin2021spinorbit}. 

So far, the majority of studies on current-induced spin-torque manipulations in 2D vdW magnets exploit the properties of conventional heavy metals to generate the spin source. Due to the intrinsic nature of vdW materials, some 2D vdW magnets lack an inversion center in their crystalline structures, hence allowing for spin-orbit Hamiltonian with linear and/or cubic in momentum. This fundamentally means that these materials possess spin textures in momentum space which can be tailored by electric currents for magnetisation control\cite{Edelstein1990,Manchon2019}. While spin-orbit torque arising from spin-textures in momentum space has been demonstrated and studied in great detail in 3D crystals \cite{Chernyshov2009,Fang_NNano2011,Ciccarelli_NPhys2016,Yoshimi_SciAdv2018}, including the demonstration of damping-like torques due to the Berry-curvature \cite{Kurebayashi_NNano2014}, to date no such result have been reported in 2D vdW magnets. Since electronic states in 2D vdW materials are inherently very sensitive to any perturbation caused at interfaces, the 2D vdW material family offers a fascinating and promising playground to explore novel and efficient current-induced spin torques with a large number of possible heterostructure combinations. To guide this research development, we here present a summary of possible spin-orbit torque symmetries for different point group crystals, including the contributions emerging at interfaces. 
{\blue We remind the reader that following Neuman's principle, it is the point group which governs the allowed types of linear response tensors, 
being themselves translation invariant, and that it is the space group of the material which conditions  the band structure, which can have an effect on the magnitude of the response \cite{BradleyCracknellSymmBook}}.
We begin with basic descriptions of the underlying spin physics and then elaborate a broader picture of torque categories, followed by a discussion about the theoretical methodologies needed to explore the nature and strength of the torque phenomena. 

\subsection*{Basics of magnetic torques and spin-orbit interaction}
\begin{figure}[t!]
\centering
\includegraphics[width=16cm]{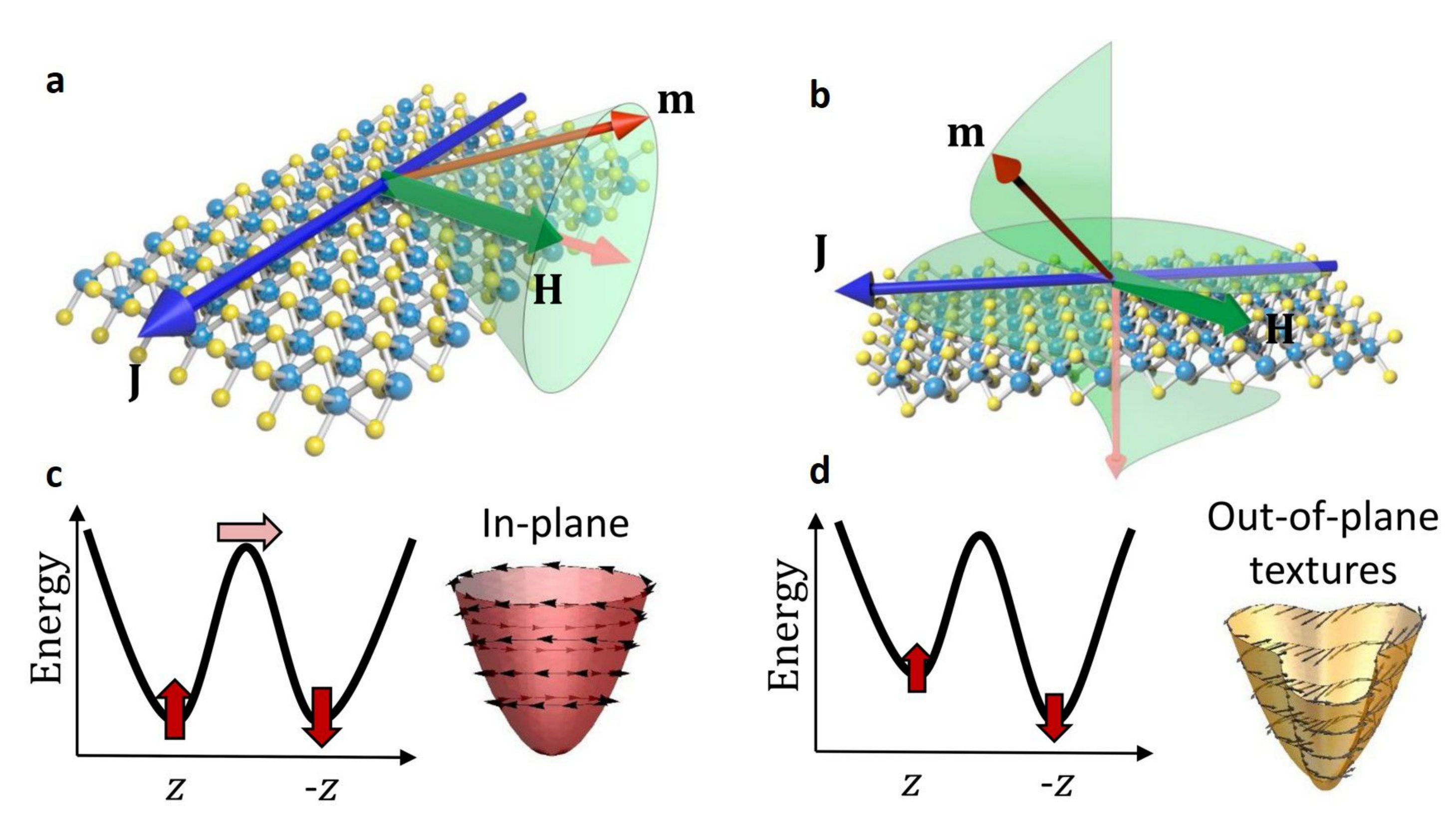}
\caption{Illustration of spin-orbit torque mechanisms for different spin-textures and their effects on the magnetic energy profile. \textbf{a} Magnetization dynamics due to a typical effective SOT field for systems with multiple orthogonal mirror planes. \textbf{b} Magnetization dynamics for effective SOC fields with a component parallel to the anisotropy axis. \textbf{c} Magnetic energy density for a system with perpendicular magnetic anisotropy and a weak in-plane effective SOC field produced an in-plane spin texture. \textbf{d} Magnetic energy density for a system with perpendicular magnetic anisotropy and an effective field with a component parallel to the anisotropy axis, and the spin-textures that enable these kind of fields.}
\label{fig4}
\end{figure}

The magnetization in ferromagnetic materials $\MM$ originates from the spin angular momentum $\bm{s}$ of lattice atoms. The exchange interaction enforces the magnetic order but since it is generally isotropic, a source of magnetic anisotropy is required to dictate the direction of the magnetization at equilibrium, and to stabilize the magnetic phase. Typically, this interaction defines two equilibrium positions along the positive and negative directions of the anisotropy axis, which is characterized by two symmetric minima in the system's energy. An external magnetic field with a component parallel to such an axis will then favor a particular minimum and translates into a torque $\TT$ on the magnetization, which will align it along the field direction. Still, large fields are required to induce stable and reversible switching, a condition which remains a significant technological challenge. An alternative way is provided by manipulating the SOC to produce electrically driven torques on the magnetization, known as SOT \cite{Manipatruni2019}.

SOC is the consequence of the periodic electrical potential due to the nuclei within the crystal, which is felt by the electrons as an effective magnetic field ${\bf{B}}_{\rm soc}$ proportional to its angular momentum {\bf{L}}\cite{Bandyopadhyay2008}. The characteristics of the coupling between the spin and this effective field dictate the resulting strength and symmetry properties of the dominant SOC terms. Additionally, the spin of the conduction electrons propagating through a magnetic material will couple to its total magnetization through a Zeeman-like interaction $-(2\Delta_{\rm ex}/\hbar)\bm{s}\cdot \MD$, where $\hbar$ the reduced Planck's constant\cite{Fo2004}, $\MD\equiv \MM/|\MM|$ the normalized magnetization, and $\Delta_{\rm ex}$ the effective exchange interaction between the localized and itinerant spins. Therefore, in magnetic systems with SOC, the spin of conduction electrons will be driven by the torque produced by a combination of the exchange field $\bm{B}_{\rm ex}\equiv -(2\Delta_{\rm ex}/\hbar)\MD$\cite{Fo2004,PRBFreimuth2014} and the effective momentum-dependent spin-orbit field $\bm{B}_{\rm soc}(\bm{p})$ \cite{Bandyopadhyay2008}. Conversely, the action-reaction law states that the magnetization will feel an opposite momentum-dependent torque due to the electronic spin $\TT_{\rm soc}(\bm{p})$ and will be then indirectly coupled to spins of the itinerant electrons which are also affected by the SOC\cite{Haney2007}.  We should remember however that since ${\bf{B}}_{\rm soc}$ is a relativistic effect, it does not exists in the laboratory frame. As a result, any non-zero torque induced by this field will cancel in equilibrium owing to time-reversal symmetry invariance. Nevertheless, under the action of an external electrical bias $\Delta V$, electrons around the Fermi level $E_F$ are bestowed with extra energy that promotes a reorganization of charge favoring electrons with momentum parallel to the electric field. This non-equilibrium state opens the possibility of a finite {\blue non-equilibirum} spin density, an effect first discovered by Edelstein \cite{Edelstein1990} and referred as inverse spin galvanic \cite{Manchon2019}. From a different point of view, such a non-equilibrium spin density can be understood as stemming from an effective macroscopic SOC field ${\bf{H}}_{\rm soc}(\Delta V)$, that is a function of the bias and promotes a macroscopic torque $\TT_{\rm soc} = \MM \times   {\bf H}_{\rm soc}$. In this sense, the bias provides the energy to maintain spins alignment with ${\bf H}_{\rm soc}(\Delta V)$, and it will also affect the magnetization direction through the exchange interaction. The direction and strength of this field will actually depend on the average of the momentum-dependent spin polarization, which is usually termed as spin texture and is computed as the expectation value of the spin operators in Bloch's states. In Fig.\ref{fig4}c, a sketch of this phenomenon is shown, where the surface represents the allowed energies ($\,E(\bm{p},\bm{s})\,$) and  the arrows indicate the spin texture pattern. It is clear from the figure that in the chosen situation, the spin texture compensate after summation over all allowed momentum. Alternatively, we could also consider the field ${\bf H}_{\rm soc}$ as an electrically induced {\blue non-equilibrium} magnetic anisotropy that promotes the alignment along a different orientation \cite{GaratPRB2009}. 

In systems with an inversion center, the Bloch's states are doubly degenerated for any given momentum, leading to two superimposed and opposite spin textures that will suppress any possible spin torque\cite{DresselhausM.S.andDresselhausG.andJorio2008}. This result implies that the SOTs requires a broken inversion symmetry, as typically achieved through the creation of an interface. In layered materials though, the inversion symmetry can be lifted by multilayer stacking \cite{Zollner2020,Zihlmann2018}, strain \cite{Guimaraes_NanoLett2018}, or even the interaction with the substrate \cite{Alghamdi_NanoLett2019}. The second class of highly symmetric 2D systems are those possessing a 4-fold rotation axis or higher. At the $\Gamma$ point of those systems, only in-plane fields perpendicular to the layer and the current such as the one illustrated in Fig.\ref{fig4}a are allowed by the in-plane spin textures (Fig.\ref{fig4}c) enforced by the symmetries \cite{Manchon2015}. In systems with  perpendicular magnetic anisotropy, such a field will gradually reduce the barrier in Fig.\ref{fig4}c without favouring any particular minima, and therefore, cannot produce deterministic switching. However spin-pseudospin coupling \cite{vantuan2014np} and other internal degrees of freedom could lead to nontrinvial out-of-plane textures\cite{Ok2019,Garcia2020}, a possibility that remains largely unexplored in the past. In fact, currently the role of symmetries is under extensive experimental study due to its potential for technological applications \cite{LiuNatNat2021,MacNeill_NPhys2017}. Exploiting this out-of-plane spin texture (as depicted in Fig. \ref{fig4}d) is particularly rewarding since it breaks the energy degeneracy of the moment pointing $\pm$ $z$ directions by an electric current as shown in Fig. \ref{fig4}b\&d. This leads to deterministic control of the magnetization switching by an electric current without any needs of external magnetic fields in such spintronic switching devices.

\subsection*{Symmetries of the spin-orbit torques in layered materials}

The quantitative determination of spin-textures in momentum space requires knowledge of the microscopic interactions at play, which are complicated to measure. Still, they lead to the SOT which is a macroscopic observable, or equivalently, an effective magnetic field ${\bf H}_{\rm soc}$. The underlying lattice, equilibrium magnetic symmetries and atomic structure fully define the SOC and, as consequence, will also determine the allowed SOT symmetries. Moreover, crystal symmetries will also manifests in the torque's dependence on the enforced current $\JJ$ and the magnetization direction $\MD$ \cite{PRBFreimuth2014}. In general, ${\bf H}_{\rm soc}$ can be {\blue written in terms of the} current and magnetization \cite{Garello2013,Belashchenko2019,Dolui2020,PRBFreimuth2014}. 

To illustrate the method, we will limit ourselves only to terms linear in the current and the magnetization. We already determined that a non-equilibrium state is mandatory to produce the spin-orbit fields forcing it to be proportional to current density at lowest order. Its dependence against the magnetization direction is more complicated since there are two possible contributions, 
\begin{equation}
 {\bf H}^e \equiv \Upsilon \JJ   ,\label{even_field}
\end{equation}
independent of magnetization and linked to the current density through the  $3\times3$ matrix $\Upsilon$, and
\begin{equation}
{\bf H}^o \equiv \MD\Lambda \JJ,\label{odd_field}
\end{equation}
 which is  proportional to the magnetization and is mixed linearly with the current density through the rank-3 tensor $\Lambda$. The superscript $e/o$ are included to highlight the even and odd dependence of these fields against magnetization reversal. These two objects comprises over 36 terms that need to be determined to fully describe the torque's symmetries. Through symmetry analysis, these elements can be further reduced to up to two parameters.

To outline how this symmetry analysis work, we will first start by defining a point symmetry $\mathcal{R}$, which is a geometric operation that leaves the positions of the system invariant with respect to a given point, and are representable as $3\times3$ unitary matrices if we represent positions as three-dimensional vectors. The main assumptions of symmetry analysis is that a representation of these symmetries will also act on all crystal's physical properties, which is known as Neumann's principle\cite{DresselhausM.S.andDresselhausG.andJorio2008,PRBFreimuth2014}. In this sense, both the current density and magnetization  must remain invariant against the action of all system's symmetries $\{\mathcal{R}_1,\mathcal{R}_2,\dots, \mathcal{R}_N\}$, where $N$ the total number of symmetry operations. This will lead to a set of coupled equations that will restrict the number of allowed components. In general, one must evaluate the $N$ symmetries present in the material, but by using group theory it is possible to reduce it to a much smaller minimal set of symmetries  known as the generating set,  that can be combined to span the whole group\cite{DresselhausM.S.andDresselhausG.andJorio2008}, and it is sufficient to impose invariance only in that set.

To wrap up and settle the concepts presented here, we provide an example in {\bf Textbox}, which illustrates how to use the symmetry analysis to determine the allowed torques of the $C_{3v}$ point group. This group possesses six symmetry elements but a generating set of only two, which we choose to be a rotation of $120^\circ$ and an reflection on the $zy$ plane, which are generic for 2D honeycomb structures and serve as a starting point to describe many different materials such as: graphene heterostructures \cite{KochanPRB2017}, hexagonal transition metal dichalcogenides, Fe$_3$GeTe$_2$\cite{PRLJohansen2019}, CrI$_{3}$. Materials belonging to higher symmetry groups such as $D_{6h}$ or $C_{6v}$ will
possess fewer allowed torques.

\begin{figure}[htpb]
\textbf{Textbox}
\begin{framed}
\begin{minipage}[c]{\textwidth}
The $C_{3v}$ point group of symmetries possess a generating set given by a $120^\circ$ rotation around the $z$-axis and a reflection on the $yz$ plane. For any possible symmetries $\mathcal{R}$, the matrices $\Upsilon$ and $\Lambda$ must satisfy the following set of equations \cite{PRBSeemann2015,PRBFreimuth2014} $\Upsilon = {\rm det}(\mathcal{R})\,\mathcal{R} \Upsilon \mathcal{R}^T$ and  $\Lambda^{i} = \sum_{i'=x,y,z}\mathcal{R}_{i,i'} \mathcal{R}\Lambda^{i'}\mathcal{R}^T$, where $i,j,k\in (x,y,z)$ defining the direction of the axes of a chosen Cartesian system. Below we show how  these tensors change after successive application of the two symmetries. 
\includegraphics[width=\columnwidth]{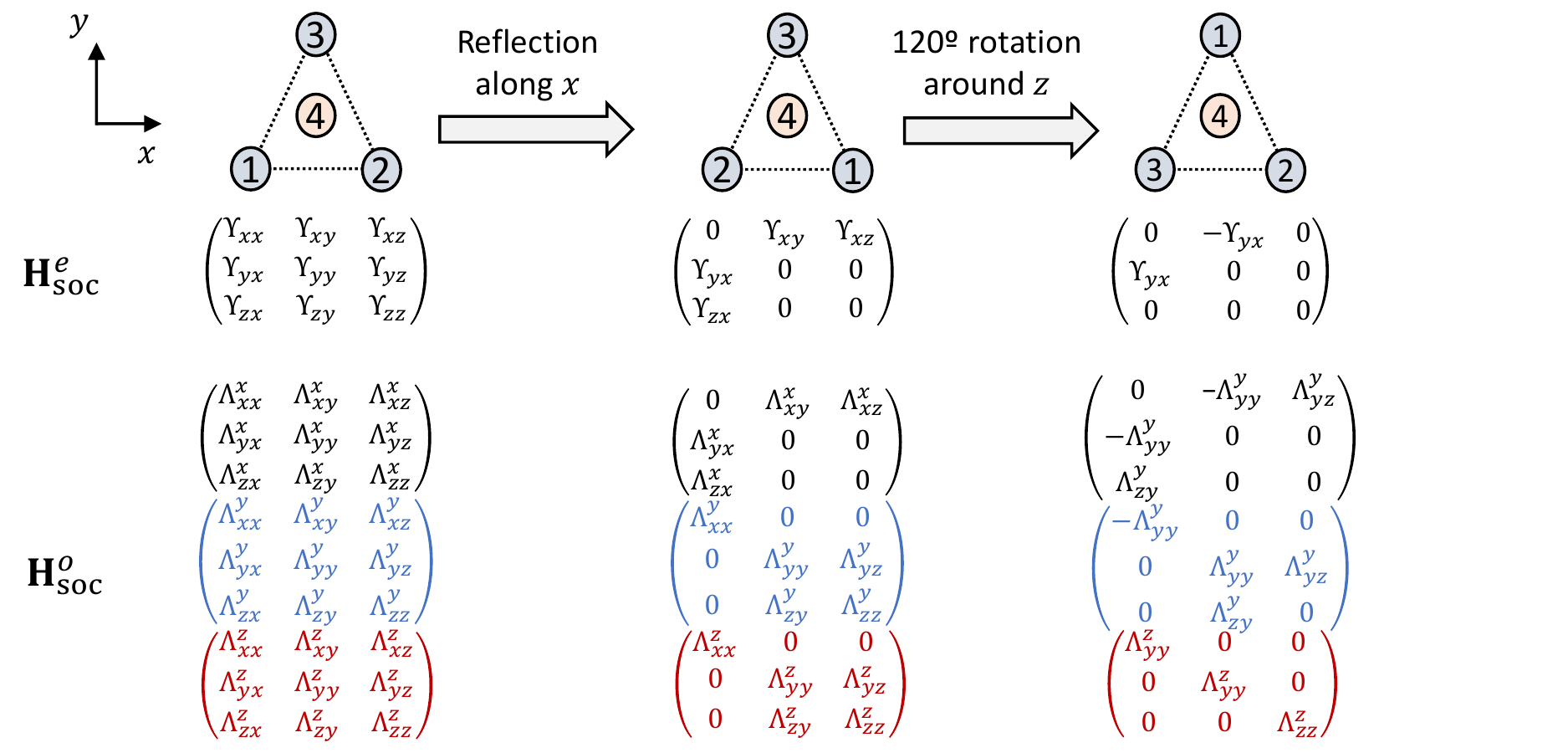}
The matrix $\Upsilon$ is antisymmetric and translate into a cross product between the current and the rotation axis  ${\bf H}^{e}= \Upsilon_{yx}\JJ\times \hat{\bm{z}}$. The second term is less straightforward but it is, still, fully defined by just five parameters, that reduces to three for 2D materials. This is a substantial reduction in the number of degree of freedom {\blue that one would, e.g., use in a phenomenological fit of the experiment. Also, note that a microscopic calculation would automatically give  such reduction, which can be used as a check, or simply one can use the symmetry to reduced calculation cost.}  
 \end{minipage}
\end{framed}
 \end{figure} 
 
A key feature of 2D materials and their combination is the possible variability of symmetry situations, including interface-induced symmetry breaking \cite{KochanPRB2017,MacneillPRB2017,Gupta2020,Cheng2016,Alghamdi_NanoLett2019}. For instance, stacking different 2D materials lift the inversion symmetry and can artificially enhance the SOC and exchange interaction \cite{Shao2016,Zhang2016,Dolui2020,Husain2020b,HiddingFM_2020,Zollner2020}. Accordingly, it is possible to design and tailor the SOT symmetry by the choice of materials and the engineering of 2D material stacking. There are several studies that show how symmetries affect the magnetization dynamics through a 2D layer inserted at an interface \cite{MacNeill_NPhys2017,Guimaraes_NanoLett2018,Lv2018,StiehlPRB2019,Xie2019,HiddingFM_2020,Ostwal_AdvMater2020,Liang2020}. However, 2D magnets can also display an internally generated SOT due to their combined SOC, broken inversion symmetry, and exchange interactions, i.e. not needing an interface or spin-injection from an adjacent layer, opening the door for \emph{all-in-one} SOT-memories.
To date, the number of synthesized 2D magnetic materials is still very limited, but novel materials and combination emerge regularly. Most of the presently known systems belong to the same spatial group and hence will display the same kind of torques. 
  \begin{figure}
  \centering
	\includegraphics[width=0.7\columnwidth]{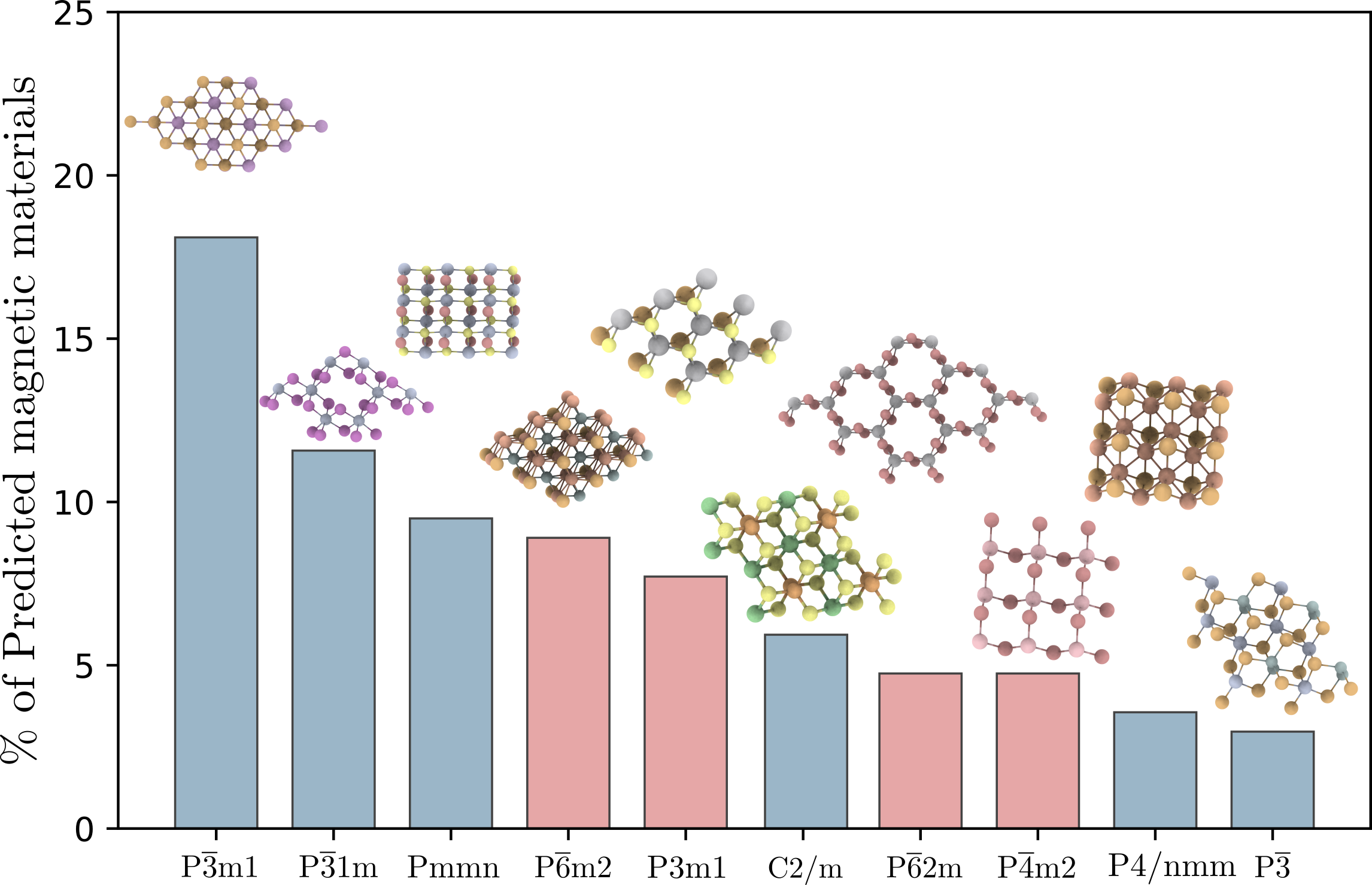}
 	\caption{Distributions of magnetic materials predicted using first principle calculations as a function of their Symmetry. The materials family, which requires a substrate-induced inversion symmetry breaking to generate torque components, is identified by the filled blue histograms. Non-centrosymmetric materials which may manifest SOT are outlined by filled red histograms.}
 	\label{fig5}
 \end{figure}
 
 \subsection*{Material screening and grouping}
    We here perform a brute-force screening of all possible 2D magnetic materials in the C2DB database \cite{Haastrup_2018} and group them according to their symmetry. To do this, we selected only those that were predicted to be stable in their freestanding form according to the database's stability criteria \cite{Haastrup_2018}. We also discarded all materials without a point group symmetry since we will restrict the discussion to those materials where symmetry analysis is able to provide physical insights of the torques based on its point group. The final result is presented in Fig. \ref{fig5}, where it is clear that hexagonal and trigonal systems are ubiquitous, in agreement with the trends observed in experimental synthesis. Table I provides allowed torques for the most common space groups and gives examples of representative materials that could display them. The space group is shown in the international notation, and we set the $z$-axis as the high-symmetry rotational axis and aligned the $x$-axis to one of the lattice vectors. The table is ordered based on the number of symmetry operators in each group, or equivalently, from the highest-to-lowest symmetric groups. In this sense, it is clear that the lowest symmetry group C2/m requires, in general, 12 parameters to define all possible torques, whenever the highest symmetry group is limited to three parameters. The table also reinforces the result presented in Textbox 1 regarding the stringent role of symmetries concerning the allowed torque components. For example, when there is a two-fold rotation, it is clear there will be different torques pinned to the two perpendicular directions that we choose to be the $x$ ad $y$ axes, and in general, one would expect all torques to be defined by a  $2\times 2$ matrix. The presence of vertical mirror planes or in-plane rotational axes further halves the torque components quantity. Systems with three-fold rotational symmetry are the best examples to understand the emergence of novel torques upon symmetry breaking. To better appreciate this mechanism, let us focus on the torques odd against magnetization reversals of materials with the spacegroup P$\bar{3}$, which are the simplest to analyse although the result extends to the those with even dependence. These materials are centrosymmetric but lose this property and belong to the reduced P3 group when deposited onto a substrate. The remaining three-fold rotational symmetry then permits only two kinds of odd torques,  one perpendicular and another parallel to the current. The trigonal structure which only maintains a single vertical mirror plane will then suppress the torque component parallel to the current. Similar ideas apply to the odd components of the torques. Therefore, the main message here is that within a given class of crystals, new torques will appear when the interaction with the environment starts to disrupt internal crystalline symmetries. These new torque components will be superimposed to the intrinsic ones robust to symmetry breaking effects providing a way to optimize the characteristics of the torque by symmetry engineering of materials. Another interesting result is that following this approach, we recover most of the synthetized 2D magnets such as CrI$_3$, CrGeTe$_3$ and Fe$_3$GeTe$_2$, but also some unexplored families such as the trihalides and the cubic magnets such as FeTe. In the same table we show the SOT directions allowed by the symmetries for all materials in the freestanding form (blue color) and under a substrate (red color). Most of the system present inversion symmetry and as a result, do not show any torque in their freestanding form and require the presence of a substrate (modelled here by either removing the mirror plane parallel to the layer or the inversion operator out of the generating set). Interestingly, we also identify a set of materials with intrinsic crystal asymmetry leading to an \emph{intrinsic} torque in  Fe$_3$GeTe$_2$ (P$\bar{6}$m2) and certain magnetic trihalides like CrBr$_3$ (P$\bar{4}$m2)\footnote{ Display intrinsic anisotropy-like SOT}, and magnetic Janus structures such as MnSTe (P$\bar{3}$m1)\footnote{Display intrinsic field-like and damping-like SOT}. Additionally, we confirm that the angular dependence of this internally generated torque changes once the substrate is incorporated and breaks additional symmetries, with the only exception of the magnetic Janus TMD \cite{Lu2017,Zhang2017a
}, for which the angular dependence remains the same. 

\subsection*{Isotropic and anisotropic torques}
The Eqs.(\ref{even_field},\ref{odd_field}) allow us to consider the SOT as a phenomenon produced by a current-induced effective magnetic field prescribed by the symmetries and the SOC. However, despite the apparent simplicity of these expressions, the fields are function of both the magnetization direction and the applied current, and can take very complex functional forms as it is evident from the less symmetric materials reported in Table 1. Nevertheless, given that the functional form of the allowed torques in isotropic systems (i.e, two-dimensional electron gas) resembles the one produced by the external effective magnetic given by ${\textbf H}_{\rm eff} \propto \JJ\times \hat{z}$, early studies classified them in two components, a field-like torque that produces magnetization precession and is characterized by the parameter $\tau_{\rm FL}$, and a damping-like torque that aligns the magnetization along the field direction and is proportional to  $\tau_{\rm DL}$\cite{Gambardella2011}. These two torques defines what we term as \emph{isotropic} torques 
 \begin{equation}
     {\TT}_{\rm iso} =\MD\times\left[\, \tau_{\rm FL} \JJ\times \hat{{\bf z}} +\tau_{\rm DL}\MD  \times(\JJ\times \hat{{\bf z}})+\tau_{\rm z} {\rm m}_z \JJ \,  \right], \label{eq:isotropicT}
 \end{equation}
given that the strength depends solely on the current magnitude $J$, allowing the definition of even and odd torques using a single parameter for each. Here, we use three parameters $\tau_{\rm FL}$, $\tau_{\rm DL}$ and $\tau_{\rm z}$ to represent the field strength for each symmetry. In fact, by following the procedure outlined in the Textbox 1, it is straightforward to show that such a simple description, where the torque is fully characterized by a \emph{single} effective field, becomes characteristic of systems with two orthogonal vertical mirror planes, including totally isotropic systems.
In the less symmetrical 2D systems without these two mirror planes, there will be a structural anisotropy between two orthogonal directions which would translate into an anisotropic electrical response. As a consequence, the current response will be characterized by the two diagonal elements of the conductivity tensor (i.e $\sigma_{xx}$ and $\sigma_{yy}$), while the torque response will be generally described by two or more effective magnetic fields. This means that the strength of the total torque will depend both on the current orientation and magnitude. 

To better grasp these concepts, let us consider first the case of system possessing a three-fold rotational symmetry and at least a mirror plane (trigonal). In Texbox 1 we demonstrated how to compute the torque matrices for the $C_{3v}$ point group symmetry and in Table 1 we combined it with Eq.(\ref{even_field}) and Eq.(\ref{odd_field}) to express it in vectorial form, leading to Eq.(\ref{eq:isotropicT}). The first two torques are torques produced by the current-induced spin-polarization along $\JJ\times \hat{{\bf z}}$ with field-like and damping-like mechanisms respectively. Both are described by a single parameter indicating that the strength of the effective field does not depend on the current orientation. The last torque term is parallel to the current, becoming non-negligible only when there is an out-of-plane component of the magnetization ${\rm m}_z$. When the mirror plane is removed, a novel field-like torque emerges that is perpendicular to both the current and the conventional field-like torque
\begin{equation}
     {\TT}_{\rm FL}^{\rm ani} =\MD\times\left[\, \tau_{\rm FL}^{I} \JJ\times \hat{{\bf z}} +\tau_{\rm FL}^{II} \JJ \right].
     \label{eq:anisotropicT}
 \end{equation}
As a result, this acts as a superposition of two perpendicular in-plane effective magnetic fields and thus leads to an anisotropic response as a function of the current direction.

Lower symmetrical systems, such as those with two-fold rotational symmetry, possess a structural anisotropy between two orthogonal crystalline directions. Therefore, the most general relation between the current and the torques should appear in the form of a $2\times2$ matrix, which is indeed the case reported in Table 1, 
\begin{equation}
     {\TT}^{\rm ani}_{\rm FL} = \MD\times \left( \begin{array}{cc}
     0 & \tau_{\rm FL}^{yx}\\
     \tau_{\rm FL}^{xy}& 0  \\
     \end{array}\right) {\textbf J}
\end{equation}
where $\tau_{\rm FL}^{yx}$ and $\tau_{\rm FL}^{xy}$ two independent parameters. Since this field is invariant against two-fold rotation, it is fundamentally different to that in Eq. (\ref{eq:anisotropicT}). For example, for $\tau_{\rm FL}^{xy}=0$, the field will vanish if the current is applied in the $y$ direction, which will not happen for systems with three-fold symmetry. Therefore, it is evident here that the existence of two-parameters will change the effective field strength with the current direction enabling a general case of anisotropic torques.

\subsection*{Anisotropy-like spin-orbit torques and low symmetry cases}
Other systems such as Fe$_3$GeTe$_2$ do not have inversion symmetry due to the lack of two orthogonal mirror planes. In these systems, the spin-orbit interaction allows for a SOT that behaves similarly to the one produced by a magnetic anisotropy, which is the reason we call it an anisotropy-like SOT. The axis of the anisotropy is defined by the remaining mirror axis and the direction of the current. It is similar to a magnetic anisotropy in the sense that one can associate the following energy density \cite{PRLJohansen2019} 
\begin{equation}
E_{\rm AL}=J_x m_x m_y - \frac{1}{2} J_y ( m_x^2 -m_y^2),
\end{equation} 
which is quadratic in $\MD$, and it produces the effective torque through its gradient with respect to the magnetization  ${\TT}_{\rm AL} =-\tau_{\rm AL}\bm{\nabla}_{\MD} E_{\rm AL}$. 

Certain systems such as monoclinic crystals with just one mirror axis and a two-fold rotation, facilitate the formation of torques possessing components from effective magnetic field proportionals to $\JJ\times \hat{{\bf y}}$ and $\JJ\times \hat{{\bf x}}$, which are favorable conditions for switching magnets with an out-of-plane perpendicular magnetic anisotropy. 

\begin{figure}
\begin{minipage}[c]{1.0\textwidth}
\textbf{Table 1}\\
%\emph{Even and odd torques' field for different %groups of two-dimensional magnetic materials %based on their point group. The blue color %identify those torques observable in %freestanding materials that posses an inherent %broken inversion symmetry, while in red we %considered the system under a substrate. The %allowed torques are also valid for %semiconductor materials when doped or gated %into the conductive states. } 
\includegraphics[width=0.95\columnwidth]{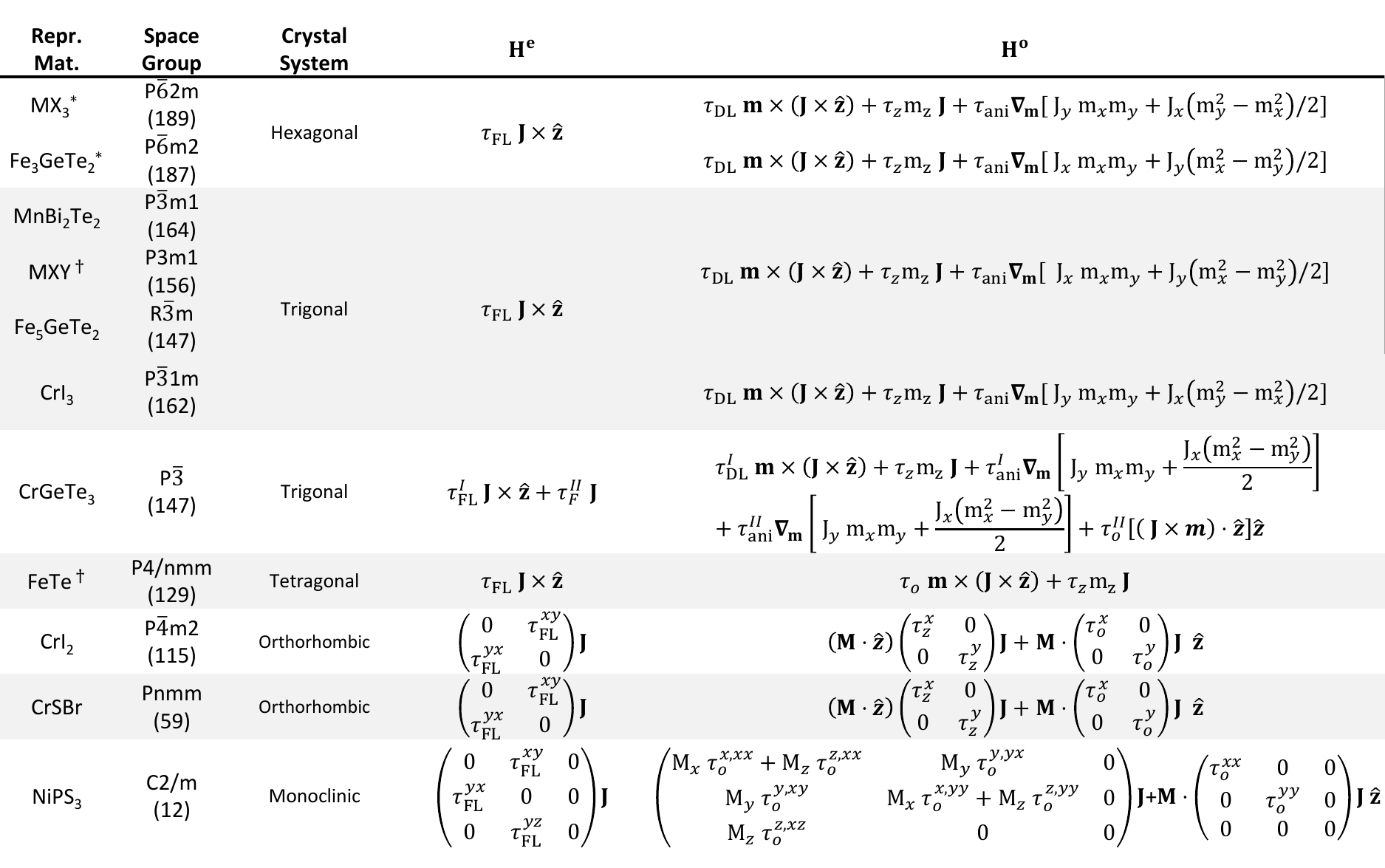} 
 \end{minipage}
 \end{figure} 

\subsection*{Measurement symmetry with the predicted ${\bf H}_{\rm SOC}$}

Experimentally, we are able to detect these predicted torques via low-frequency second-harmonic measurements\cite{Hayashi_PRB2014} and current-induced FMR experiments\cite{Liu_PRL2011,Fang_NNano2011}. In both detection schemes, we can consider the current-induced effective fields that tilt the static magnetization direction by a free energy change of ${\bf M} \cdot {\bf H}_{\rm SOC}$, or that drive magnetic precession with a magnetic torque of ${\bf M} \times {\bf H}_{\rm SOC}$. We can substitute the vector expressions of ${\bf H}_{\rm SOC}$ fields in Table 1 to predict the torque symmetry for future magnetization-control experiments. For example, by using $\JJ\times \hat{{\bf z}}$ as ${\bf H}_{\rm SOC}$, we have the torque expression of ${\bf M} \times( \JJ\times \hat{{\bf z}})$ for spin dynamics excitation. Assuming that a driving current is flowing along the $x$ direction, i.e. $\JJ \parallel \hat{{\bf x}}$, the torque expression reduces into ${\bf M} \times\hat{{\bf y}}$. This suggests that FMR amplitudes excited by this ${\bf H}_{\rm SOC}$ display $\cos{\phi}$ dependence when the magnetization is rotated in-plane with $\phi$ defined by the angle between $\JJ$ and ${\bf M}$. In general, measured FMR voltages also have $\sin{2\phi}$ dependence due to a time-dependent change of resistance arising from anisotropic magnetoresistance and magnetic precession\cite{Mecking_PRB2007}. As a result, we expect ${\bf H}_{\rm SOC}(\propto \JJ\times \hat{{\bf z}})$ produces $\sin{2\phi}\cos{\phi}$ dependence when we perform FMR voltage measurements for in-plane angle rotation of ${\bf M}$. We can apply the same vector analysis for other effective fields for FMR experiments as well as the low-frequency second-harmonic experiments. 

%We note that ${\bf H}_{\rm SOC}$ along the $z$ direction does not provide any in-plane angle dependence since ${\bf M} \times {\bf H}_{\rm SOC}$ is constant.  

%via  ST-FMR experiments, the fieldlike and dampinglike torques will appear as the anti-symmetric and symmetric contributions respectively and will provide an angular dependence given by $\cos\phi\equiv (\MD \cdot \JJ)$. The in-plane torque proportional to the out-of-plane component of the magnetization will appear as a constant contribution. Differently, the magnetic anisotropic torque exhibits two behaviors; when the current is aligned with the inversion symmetry axis, the torque will have an angular dependence given by $|\sin(\phi)|$, whereas for incoming current perpendicular to the mirror axis, a constant shift will result for the torque. 

\subsection*{Strength of the torques}

An important point for comparison with experimental data is how to obtain theoretically the strength of the SOT parameters. In 2D materials, the SOT originates solely from the current-induced spin density in the material. Therefore, in the limit of small currents, the best approach to compute these parameters is via the Kubo formula within the linear-response approximation \cite{FAN20211}. This approach requires knowledge of the  magnetization-dependent Hamiltonian $H(\MD)$ and an effective way to compute its associated retarded Green's functions $G^+(H,E)$. For non-interacting electrons, the spin-orbit torque is determined by the following expression
\begin{equation}
   \TT^\alpha(\mu,T ) =  -\frac{ 2\pi\Delta_{\rm ex} {\rm J}_\alpha}{\sigma_{\alpha\alpha}}\int dE\, f(E-\mu,T)\, {\rm Im Tr}\left[\,{\rm Im}G^+\,\hat{\rm J}_\alpha\, \frac{d G^+}{dE}( \hat{\bm{\sigma}}\times \MD)\right],\label{kubo_bastin}
\end{equation}
where {\rm Im Tr} represents the imaginary part of the trace of the operators in the brackets, $f(E-\mu,T)$ the Fermi-Dirac distribution for a given chemical potential $\mu$ and temperature $T$, $\hat{\JJ}$ the current-density operator and $\JJ$ its macroscopic average, $\sigma_{\alpha\alpha}$ the conductivity, and $\hat{\bm{\sigma}}$ the Pauli matrices operator. This approach gives access to the SOTs by making no assumptions on their dependence against magnetization. Furthermore, Eq. (\ref{kubo_bastin}) can be further sliced into two contributions, 
{\blue those resulting from from scattering mechanisms of electrons at the Fermi level (proportional to the conductivity), and those arising from a Fermi-sea contribution, }
%those resulting from the electrons with around the Fermi level which can only provide dissipative torques like the damping-like, and those arising from electrons deep in the valence band (Fermi-sea contribution), 
which in principle can provide both dissipative and non-dissipative SOTs, and {\blue are primarily scattering independent contributions}
%have a connection with a Berry-like phase 
\cite{SinovaRMP2015}. The existence of a %dissipative torque 
{\blue SOT proportional to scattering times}
also evidences that disorder plays a major role to determine the system's behavior. 
In simple systems, the Kubo formula can be evaluated following a semi-analytic solutions based on perturbation theory \cite{LeeKiSeungPRB2015,Sousa2020}, where the disorder is usually described by point impurities. For low concentration of impurities and when the single scattering approximation is valid, the Kubo formula reduces to the Boltzmann equation, which offers a simple way to estimate the behavior of the torque parameters \cite{}. However, a quantitative determination for a generic disorder or a structurally complex system is only attainable through linear scaling quantum transport methodologies \cite{FAN20211}, that enable reaching the required size scales to study the torques in the diffusive and localized regimes. For ferromagnetic materials, the magnetization can be instead replaced by the exchange-correlation field, facilitating the use of the Kubo formula to determine the torques from first principles simulations \cite{Xue2020,Mahfouzi2020}. 

 Beyond the Kubo formula, nonadiabatic time-dependent evolution in realistic systems is also a great challenge, although there are some recent efforts along this direction \cite{Nikolic2018}. 

The Kubo formula provides a method to compute the torque parameters in units of a torque density (eV/m$^3$). However, since these parameters depend directly on the driven current density, a current-to-torque conversion efficiency is usually defined. Such an efficiency definition is quite straightforward in systems where the torque is mostly due to the spin Hall effect, so it can be connected to the spin Hall angle which is given by the ratio between the spin and charge conductivities \cite{SinovaRMP2015}. Following similar considerations, the torque conductivity $\sigma_{\tau}$ can be defined as the angular momentum absorbed by the magnet per second per unit interface area per unit applied electric field\cite{MacneillPRB2017}, from which the torque efficiency $\chi_{\alpha}^{X}$ is derived as:  
\begin{equation}
\chi_{\alpha}^{X}\equiv  \frac{2e}{\hbar} \frac{M_s t}{\gamma} \frac{\tau_{X}}{  {\rm J}_\alpha}, 
\end{equation}
where $M_s$ the saturation magnetization, $X$ the torque label (FL,DL, etc), $t$ the thickness of the magnet, which is taken as the Van der Waal gap in 2D materials. The former expression defines an efficiency for different kind of torques classified according to their symmetries, but there is an alternative dimensionless expression which describes the efficiency according to the torque's directions expressed as\cite{PRBFreimuth2014}
\begin{equation}
\chi_{\alpha \beta}\equiv  \frac{e \Delta_{\rm ex  }t}{\hbar}\frac{{\rm T}_\beta}{  {\rm J}_\alpha}.
\end{equation}
Both expressions allow to quantify a torque efficiency due to a source of spins independently of geometric factors. For a Cobalt/Platinum system $\chi_{\alpha} \approx 0.0075$ for a $t\approx 0.85$ nm \cite{Fan2014c}, while in experiments considering TMD/Py heterostructures the efficiency only reaches values between $10^{-5}-10^{-3}\,\,\,$  \cite{HiddingFM_2020}. Another important parameter is given by $\chi_{\alpha}^{X}/\chi_{\alpha}^{\rm FL}= \tau_X/ \tau_{\rm FL}$ which measures the relative value of all torques against the dissipative field-like contribution and generally ranges from 1-10\cite{}. Notice that in 2D materials M$_{s}t$ is actually the magnetization per unit area or sheet magnetization \cite{Jiang2018}. From all this discussion, it becomes clear that a lot of work remains to be done to obtain realistic and quantitative information about spin torque efficiency in vdW heterostructures combining 2D magnets with strong SOC materials. Beyond well-controlled ab-initio simulations and the elaboration of properly designed SOC-dependent tight-binding models, the use of computational methodologies that can numerically solve the Kubo formula Eq. (\ref{kubo_bastin}) for large scale models is a grand challenge, particularly when disorder effects affecting the spin physics demand non-perturbative treatments. Brute-force simulations using linear scaling methods demand further developments and implementations, as well as comparison of their predictions with future experiments.

As a final remark, the spin-Hall effect\cite{SinovaRMP2015} should exist in a variety of vdW material systems, both magnetic and non-magnetic, except in the single monolayer limit. In order to exploit such spin-charge phenomena, vdW heterostructures are usually fabricated with one magnetic layer in direct contact with a spin current source, so as to have a close and efficient magnetization manipulation. The symmetry of spin-Hall effect is well-established and the spin-current source layer could be also magnetic, in which case additional magnetization-dependent anomalous components \cite{TaniguchiPRAppl2015,BaekNatMater2018,Iihama_NatElec2018} might appear.

\section*{Future Outlook}

The present review has put in perspective the formidable playground and plethora of possibilities to create new hybrid materials with emerging spin-dependent interface properties. Among the vast zoo of novel structures, theory and modelling based on symmetry features of interfaces bring a pathfinder to find the most suitable conditions for best controlling spin transport through spin-orbit or exchange coupling fields. One most urgent quest is the search for the upper limit of the achievable spin-orbit torque efficiency (at RT and with scalable materials), which could ultimately enable low-energy and fast switching of the material magnetization in the frame of SOT memory technologies\cite{Dieny_NElec2020,KAWAHARA2012613,BHATTI_MatToday2017,Cubukcu2018}. Such is a challenging task which will require proper combination of theoretical and experimental research within a large community of material grower, advanced characterization and simulation tools, as well as optimized device fabrication approaches. The search for materials with high Curie or N\'eel temperatures is also a key aspect of this research, and efforts to guide experiments such as the recently studied monolayer Fe$_3$P with predicted T$_C$ of 420 K\cite{Zheng_JPhysChemLett2019} should be followed. In this regard, the exchange coupling strength by controlling molecular orbital alignments in 2D magnetic semiconductors\cite{Huang_JACS2018} could be an interesting approach where CrWI$_6$ and CrWGe$_2$Te$_6$ are predicted to display ferromagnetic order above RT via the purely super-exchange interaction. Furthermore, high-throughput computation screening using an existing materials database\cite{Torelli_2DMater2019} and other material informatics methods could be an effective approach to guide us towards discoveries of novel high-$T_{\rm C}$ vdW magnetic materials.
Here we have focused our attention to the more familiar ferromagnetic cases, but  this approach is valid for predicting torques in non-magnetic vdW systems by using the same space group analysis as well as for novel types of torques in antiferromagnetic systems, such as the Néel SOT\cite{WadleyScience2016} and other antiferromagnetic specific SOC effects\cite{JungwirthNPhys2018} that are ripe for exploration in these exciting 2D magnetic materials class.
Looking further for more exotic 2D vdW materials with order parameters, multiferroics is predicted in a certain class of 2D materials\cite{Seixas_PRL2016} such as in a doped CrBr$_3$ monolayer\cite{Huang_PRL2018}, in a bilayer of halogen-intercalated phosphorene\cite{Yang_JACS2017} as well as in a monolayer ReWCl$_6$ with magnetic transitions induced by electric field\cite{Xu_PRL2020}. Some vdW magnetic materials, such as MnBi$_2$Te$_4$\cite{Li_SciAdv2020,Wu_SciAdv2020,Hu_NComm2020}, host topological electronic states and offer unique topologically-protected transport phenomena\cite{Deng2020}. We anticipate that this research domain will be rapidly expanding to explore the topological transport as well as its use in spintronic applications\cite{Mellnik_Nature2014}. Besides, nontrivial spin transport physics is also expected to develop at complex interfaces of vdW heterostructures, including the generation of skyrmions\cite{Yang_SciAdv2020,Wu_NComm2020,Park_PRB2021}, the excitation and detection of magnons at the 2D limit\cite{Zhang_NMater2020,McCreary_NComm2020,Cenker_NPhys2021} and their interaction with other spin-dependent characteristics of interfaced materials, as well as a wealth of proximity effects which could reveal unprecedented manifestations of entanglement and many-body physics, as recently discussed in the context of topological superconductivity or unconventional ferroelectricity in magic-angle twisted bi/trilayer graphene and related moir\'e superlattices \cite{Huang2020}. In conclusion, there is a lot of rooms for further exploration of spin transport at the frontier of layered magnetism and SOC materials, and likely many exciting discoveries waiting for emerging to the surface.

\bibliography{2DspinReview}

\begin{thebibliography}{100}
\urlstyle{rm}
\expandafter\ifx\csname url\endcsname\relax
  \def\url#1{\texttt{#1}}\fi
\expandafter\ifx\csname urlprefix\endcsname\relax\def\urlprefix{URL }\fi
\expandafter\ifx\csname doiprefix\endcsname\relax\def\doiprefix{DOI: }\fi
\providecommand{\bibinfo}[2]{#2}
\providecommand{\eprint}[2][]{\url{#2}}

\bibitem{HuangNature2017}
\bibinfo{author}{Huang, B.} \emph{et~al.}
\newblock \bibinfo{journal}{\bibinfo{title}{{Layer-dependent ferromagnetism in
  a van der Waals crystal down to the monolayer limit}}}.
\newblock {\emph{\JournalTitle{Nature}}} \textbf{\bibinfo{volume}{546}},
  \bibinfo{pages}{270--273}, \doiprefix\url{10.1038/nature22391}
  (\bibinfo{year}{2017}).

\bibitem{GongNature2017}
\bibinfo{author}{Gong, C.} \emph{et~al.}
\newblock \bibinfo{journal}{\bibinfo{title}{{Discovery of intrinsic
  ferromagnetism in two-dimensional van der Waals crystals}}}.
\newblock {\emph{\JournalTitle{Nature}}} \textbf{\bibinfo{volume}{546}},
  \bibinfo{pages}{265--269}, \doiprefix\url{10.1038/nature22060}
  (\bibinfo{year}{2017}).

\bibitem{Burch2018a}
\bibinfo{author}{Burch, K.~S.}, \bibinfo{author}{Mandrus, D.} \&
  \bibinfo{author}{Park, J.~G.}
\newblock \bibinfo{journal}{\bibinfo{title}{{Magnetism in two-dimensional van
  der Waals materials}}}.
\newblock {\emph{\JournalTitle{Nature}}} \textbf{\bibinfo{volume}{563}},
  \bibinfo{pages}{47--52}, \doiprefix\url{10.1038/s41586-018-0631-z}
  (\bibinfo{year}{2018}).

\bibitem{fei2018two}
\bibinfo{author}{Fei, Z.} \emph{et~al.}
\newblock \bibinfo{journal}{\bibinfo{title}{{Two-dimensional itinerant
  ferromagnetism in atomically thin Fe$_3$GeTe$_2$}}}.
\newblock {\emph{\JournalTitle{Nature Materials}}}
  \textbf{\bibinfo{volume}{17}}, \bibinfo{pages}{778--782},
  \doiprefix\url{10.1038/s41563-018-0149-7} (\bibinfo{year}{2018}).

\bibitem{Li2019b}
\bibinfo{author}{Li, H.}, \bibinfo{author}{Ruan, S.} \& \bibinfo{author}{Zeng,
  Y.~J.}
\newblock \bibinfo{journal}{\bibinfo{title}{{Intrinsic Van Der Waals Magnetic
  Materials from Bulk to the 2D Limit: New Frontiers of Spintronics}}}.
\newblock {\emph{\JournalTitle{Advanced Materials}}}
  \textbf{\bibinfo{volume}{31}}, \bibinfo{pages}{1--34},
  \doiprefix\url{10.1002/adma.201900065} (\bibinfo{year}{2019}).

\bibitem{Gong2019b}
\bibinfo{author}{Gong, Y.} \emph{et~al.}
\newblock \bibinfo{journal}{\bibinfo{title}{{Experimental Realization of an
  Intrinsic Magnetic Topological Insulator *}}}.
\newblock {\emph{\JournalTitle{Chinese Physics Letters}}}
  \textbf{\bibinfo{volume}{36}}, \bibinfo{pages}{076801},
  \doiprefix\url{10.1088/0256-307X/36/7/076801} (\bibinfo{year}{2019}).

\bibitem{GeimNature2013}
\bibinfo{author}{Geim, A.~K.} \& \bibinfo{author}{Grigorieva, I.~V.}
\newblock \bibinfo{journal}{\bibinfo{title}{Van der waals heterostructures}}.
\newblock {\emph{\JournalTitle{Nature}}} \textbf{\bibinfo{volume}{499}},
  \bibinfo{pages}{419--425}, \doiprefix\url{10.1038/nature12385}
  (\bibinfo{year}{2013}).

\bibitem{LiuNRMater2016}
\bibinfo{author}{Liu, Y.} \emph{et~al.}
\newblock \bibinfo{journal}{\bibinfo{title}{Van der waals heterostructures and
  devices}}.
\newblock {\emph{\JournalTitle{Nature Reviews Materials}}}
  \textbf{\bibinfo{volume}{1}}, \bibinfo{pages}{16042},
  \doiprefix\url{10.1038/natrevmats.2016.42} (\bibinfo{year}{2016}).

\bibitem{AndreiNRMater2021}
\bibinfo{author}{Andrei, E.~Y.} \emph{et~al.}
\newblock \bibinfo{journal}{\bibinfo{title}{The marvels of moir{\'e}
  materials}}.
\newblock {\emph{\JournalTitle{Nature Reviews Materials}}}
  \textbf{\bibinfo{volume}{6}}, \bibinfo{pages}{201--206},
  \doiprefix\url{10.1038/s41578-021-00284-1} (\bibinfo{year}{2021}).

\bibitem{Freitas2015}
\bibinfo{author}{Freitas, D.~C.} \emph{et~al.}
\newblock \bibinfo{journal}{\bibinfo{title}{{Ferromagnetism in layered
  metastable 1T-CrTe$_2$}}}.
\newblock {\emph{\JournalTitle{Journal of Physics Condensed Matter}}}
  \textbf{\bibinfo{volume}{27}}, \bibinfo{pages}{176002},
  \doiprefix\url{10.1088/0953-8984/27/17/176002} (\bibinfo{year}{2015}).

\bibitem{Freitas2013}
\bibinfo{author}{Freitas, D.~C.} \emph{et~al.}
\newblock \bibinfo{journal}{\bibinfo{title}{{Antiferromagnetism and
  ferromagnetism in layered 1T-CrSe$_2$ with v and Ti replacements}}}.
\newblock {\emph{\JournalTitle{Physical Review B - Condensed Matter and
  Materials Physics}}} \textbf{\bibinfo{volume}{87}}, \bibinfo{pages}{014420},
  \doiprefix\url{10.1103/PhysRevB.87.014420} (\bibinfo{year}{2013}).

\bibitem{Zhang2015j}
\bibinfo{author}{Zhang, W.-B.}, \bibinfo{author}{Qu, Q.}, \bibinfo{author}{Zhu,
  P.} \& \bibinfo{author}{Lam, C.-H.}
\newblock \bibinfo{journal}{\bibinfo{title}{{Robust intrinsic ferromagnetism
  and half semiconductivity in stable two-dimensional single-layer chromium
  trihalides}}}.
\newblock {\emph{\JournalTitle{Journal of Materials Chemistry C}}}
  \textbf{\bibinfo{volume}{3}}, \bibinfo{pages}{12457--12468},
  \doiprefix\url{10.1039/C5TC02840J} (\bibinfo{year}{2015}).
\newblock \eprint{1507.07275}.

\bibitem{McGuire2014}
\bibinfo{author}{McGuire, M.~A.}, \bibinfo{author}{Dixit, H.},
  \bibinfo{author}{Cooper, V.~R.} \& \bibinfo{author}{Sales, B.~C.}
\newblock \bibinfo{journal}{\bibinfo{title}{{Coupling of Crystal Structure and
  Magnetism in the Layered, Ferromagnetic Insulator CrI$_3$}}}.
\newblock {\emph{\JournalTitle{Chemistry of Materials}}}
  \textbf{\bibinfo{volume}{27}}, \bibinfo{pages}{612--620},
  \doiprefix\url{10.1021/cm504242t} (\bibinfo{year}{2015}).

\bibitem{McGuire2017}
\bibinfo{author}{McGuire, M.~A.} \emph{et~al.}
\newblock \bibinfo{journal}{\bibinfo{title}{{Magnetic behavior and spin-lattice
  coupling in cleavable van der Waals layered CrCl$_3$ crystals}}}.
\newblock {\emph{\JournalTitle{Physical Review Materials}}}
  \textbf{\bibinfo{volume}{1}}, \bibinfo{pages}{14001},
  \doiprefix\url{10.1103/PhysRevMaterials.1.014001} (\bibinfo{year}{2017}).

\bibitem{Joy_PRB1992}
\bibinfo{author}{Joy, P.~A.} \& \bibinfo{author}{Vasudevan, S.}
\newblock \bibinfo{journal}{\bibinfo{title}{{Magnetism in the layered
  transition-metal thiophosphates MPS$_3$ (M=Mn, Fe, and Ni)}}}.
\newblock {\emph{\JournalTitle{Physical Review B}}}
  \textbf{\bibinfo{volume}{46}}, \bibinfo{pages}{5425--5433},
  \doiprefix\url{10.1103/PhysRevB.46.5425} (\bibinfo{year}{1992}).

\bibitem{Lin2016}
\bibinfo{author}{Lin, W.}, \bibinfo{author}{Chen, K.}, \bibinfo{author}{Zhang,
  S.} \& \bibinfo{author}{Chien, C.~L.}
\newblock \bibinfo{journal}{\bibinfo{title}{{Enhancement of Thermally Injected
  Spin Current through an Antiferromagnetic Insulator}}}.
\newblock {\emph{\JournalTitle{Physical Review Letters}}}
  \textbf{\bibinfo{volume}{116}}, \bibinfo{pages}{186601},
  \doiprefix\url{10.1103/PhysRevLett.116.186601} (\bibinfo{year}{2016}).
\newblock \eprint{1603.00931}.

\bibitem{Bonilla2018}
\bibinfo{author}{Bonilla, M.} \emph{et~al.}
\newblock \bibinfo{journal}{\bibinfo{title}{{Strong room-temperature
  ferromagnetism in VSe$_2$ monolayers on van der Waals substrates}}}.
\newblock {\emph{\JournalTitle{Nature Nanotechnology}}}
  \textbf{\bibinfo{volume}{13}}, \bibinfo{pages}{289--293},
  \doiprefix\url{10.1038/s41565-018-0063-9} (\bibinfo{year}{2018}).

\bibitem{Deng2018}
\bibinfo{author}{Deng, Y.} \emph{et~al.}
\newblock \bibinfo{journal}{\bibinfo{title}{{Gate-tunable room-temperature
  ferromagnetism in two-dimensional Fe$_3$GeTe$_2$}}}.
\newblock {\emph{\JournalTitle{Nature}}} \textbf{\bibinfo{volume}{563}},
  \bibinfo{pages}{94--99}, \doiprefix\url{10.1038/s41586-018-0626-9}
  (\bibinfo{year}{2018}).

\bibitem{OHara2018}
\bibinfo{author}{O'Hara, D.~J.} \emph{et~al.}
\newblock \bibinfo{journal}{\bibinfo{title}{{Room Temperature Intrinsic
  Ferromagnetism in Epitaxial Manganese Selenide Films in the Monolayer
  Limit}}}.
\newblock {\emph{\JournalTitle{Nano Letters}}} \textbf{\bibinfo{volume}{18}},
  \bibinfo{pages}{3125--3131}, \doiprefix\url{10.1021/acs.nanolett.8b00683}
  (\bibinfo{year}{2018}).
\newblock \eprint{1802.08152}.

\bibitem{OHara2018b}
\bibinfo{author}{O'Hara, D.~J.}, \bibinfo{author}{Zhu, T.} \&
  \bibinfo{author}{Kawakami, R.~K.}
\newblock \bibinfo{journal}{\bibinfo{title}{{Importance of Paramagnetic
  Background Subtraction for Determining the Magnetic Moment in Epitaxially
  Grown Ultrathin van der Waals Magnets}}}.
\newblock {\emph{\JournalTitle{IEEE Magnetics Letters}}}
  \textbf{\bibinfo{volume}{9}}, \bibinfo{pages}{1--5},
  \doiprefix\url{10.1109/LMAG.2018.2867339} (\bibinfo{year}{2018}).

\bibitem{Walker1983}
\bibinfo{author}{Walker, M.~B.} \& \bibinfo{author}{Withers, R.~L.}
\newblock \bibinfo{journal}{\bibinfo{title}{{Stacking of charge-density waves
  in 1T transition-metal dichalcogenides}}}.
\newblock {\emph{\JournalTitle{Physical Review B}}}
  \textbf{\bibinfo{volume}{28}}, \bibinfo{pages}{2766--2774},
  \doiprefix\url{10.1103/PhysRevB.28.2766} (\bibinfo{year}{1983}).

\bibitem{Eaglesham1986}
\bibinfo{author}{Eaglesham, D.~J.}, \bibinfo{author}{Withers, R.~L.} \&
  \bibinfo{author}{Bird, D.~M.}
\newblock \bibinfo{journal}{\bibinfo{title}{{Charge-density-wave transitions in
  1T-VSe$_2$}}}.
\newblock {\emph{\JournalTitle{Journal of Physics C: Solid State Physics}}}
  \textbf{\bibinfo{volume}{19}}, \bibinfo{pages}{359--367},
  \doiprefix\url{10.1088/0022-3719/19/3/006} (\bibinfo{year}{1986}).

\bibitem{Feng2018}
\bibinfo{author}{Feng, J.} \emph{et~al.}
\newblock \bibinfo{journal}{\bibinfo{title}{{Electronic Structure and Enhanced
  Charge-Density Wave Order of Monolayer VSe$_2$}}}.
\newblock {\emph{\JournalTitle{Nano Letters}}} \textbf{\bibinfo{volume}{18}},
  \bibinfo{pages}{4493--4499}, \doiprefix\url{10.1021/acs.nanolett.8b01649}
  (\bibinfo{year}{2018}).

\bibitem{Yang2014b}
\bibinfo{author}{Yang, J.} \emph{et~al.}
\newblock \bibinfo{journal}{\bibinfo{title}{{Thickness dependence of the
  charge-density-wave transition temperature in VSe$_2$}}}.
\newblock {\emph{\JournalTitle{Applied Physics Letters}}}
  \textbf{\bibinfo{volume}{105}}, \bibinfo{pages}{063109},
  \doiprefix\url{10.1063/1.4893027} (\bibinfo{year}{2014}).

\bibitem{Coelho2019}
\bibinfo{author}{Coelho, P.~M.} \emph{et~al.}
\newblock \bibinfo{journal}{\bibinfo{title}{{Charge Density Wave State
  Suppresses Ferromagnetic Ordering in VSe$_2$ Monolayers}}}.
\newblock {\emph{\JournalTitle{The Journal of Physical Chemistry C}}}
  \textbf{\bibinfo{volume}{123}}, \bibinfo{pages}{14089--14096},
  \doiprefix\url{10.1021/acs.jpcc.9b04281} (\bibinfo{year}{2019}).

\bibitem{HIROHATAJMMM2020}
\bibinfo{author}{Hirohata, A.} \emph{et~al.}
\newblock \bibinfo{journal}{\bibinfo{title}{Review on spintronics: Principles
  and device applications}}.
\newblock {\emph{\JournalTitle{Journal of Magnetism and Magnetic Materials}}}
  \textbf{\bibinfo{volume}{509}}, \bibinfo{pages}{166711},
  \doiprefix\url{https://doi.org/10.1016/j.jmmm.2020.166711}
  (\bibinfo{year}{2020}).

\bibitem{GrollierNElec2020}
\bibinfo{author}{Grollier, J.} \emph{et~al.}
\newblock \bibinfo{journal}{\bibinfo{title}{Neuromorphic spintronics}}.
\newblock {\emph{\JournalTitle{Nature Electronics}}}
  \textbf{\bibinfo{volume}{3}}, \bibinfo{pages}{360--370},
  \doiprefix\url{10.1038/s41928-019-0360-9} (\bibinfo{year}{2020}).

\bibitem{LinNElec2019}
\bibinfo{author}{Lin, X.}, \bibinfo{author}{Yang, W.}, \bibinfo{author}{Wang,
  K.~L.} \& \bibinfo{author}{Zhao, W.}
\newblock \bibinfo{journal}{\bibinfo{title}{Two-dimensional spintronics for
  low-power electronics}}.
\newblock {\emph{\JournalTitle{Nature Electronics}}}
  \textbf{\bibinfo{volume}{2}}, \bibinfo{pages}{274--283},
  \doiprefix\url{10.1038/s41928-019-0273-7} (\bibinfo{year}{2019}).

\bibitem{KAWAHARA2012613}
\bibinfo{author}{Kawahara, T.}, \bibinfo{author}{Ito, K.},
  \bibinfo{author}{Takemura, R.} \& \bibinfo{author}{Ohno, H.}
\newblock \bibinfo{journal}{\bibinfo{title}{Spin-transfer torque ram
  technology: Review and prospect}}.
\newblock {\emph{\JournalTitle{Microelectronics Reliability}}}
  \textbf{\bibinfo{volume}{52}}, \bibinfo{pages}{613--627},
  \doiprefix\url{https://doi.org/10.1016/j.microrel.2011.09.028}
  (\bibinfo{year}{2012}).
\newblock \bibinfo{note}{Advances in non-volatile memory technology}.

\bibitem{BHATTI2017530}
\bibinfo{author}{Bhatti, S.} \emph{et~al.}
\newblock \bibinfo{journal}{\bibinfo{title}{Spintronics based random access
  memory: a review}}.
\newblock {\emph{\JournalTitle{Materials Today}}}
  \textbf{\bibinfo{volume}{20}}, \bibinfo{pages}{530--548},
  \doiprefix\url{https://doi.org/10.1016/j.mattod.2017.07.007}
  (\bibinfo{year}{2017}).

\bibitem{Chernyshov2009}
\bibinfo{author}{Chernyshov, A.} \emph{et~al.}
\newblock \bibinfo{journal}{\bibinfo{title}{{Evidence for reversible control of
  magnetization in a ferromagnetic material by means of spin–orbit magnetic
  field}}}.
\newblock {\emph{\JournalTitle{Nature Physics}}} \textbf{\bibinfo{volume}{5}},
  \bibinfo{pages}{656--659}, \doiprefix\url{10.1038/nphys1362}
  (\bibinfo{year}{2009}).
\newblock \eprint{0812.3160}.

\bibitem{Miron2011b}
\bibinfo{author}{Miron, I.~M.} \emph{et~al.}
\newblock \bibinfo{journal}{\bibinfo{title}{{Perpendicular switching of a
  single ferromagnetic layer induced by in-plane current injection}}}.
\newblock {\emph{\JournalTitle{Nature}}} \textbf{\bibinfo{volume}{476}},
  \bibinfo{pages}{189--193}, \doiprefix\url{10.1038/nature10309}
  (\bibinfo{year}{2011}).

\bibitem{Liu2012}
\bibinfo{author}{Liu, L.} \emph{et~al.}
\newblock \bibinfo{journal}{\bibinfo{title}{{Spin-Torque Switching with the
  Giant Spin Hall Effect of Tantalum}}}.
\newblock {\emph{\JournalTitle{Science}}} \textbf{\bibinfo{volume}{336}},
  \bibinfo{pages}{555--558}, \doiprefix\url{10.1126/science.1218197}
  (\bibinfo{year}{2012}).
\newblock \eprint{1203.2875}.

\bibitem{Cubukcu2018}
\bibinfo{author}{Cubukcu, M.} \emph{et~al.}
\newblock \bibinfo{journal}{\bibinfo{title}{{Ultra-Fast Perpendicular
  Spin-Orbit Torque MRAM}}}.
\newblock {\emph{\JournalTitle{IEEE Transactions on Magnetics}}}
  \textbf{\bibinfo{volume}{54}}, \bibinfo{pages}{9300204},
  \doiprefix\url{10.1109/TMAG.2017.2772185} (\bibinfo{year}{2018}).
\newblock \eprint{1509.02375}.

\bibitem{SinovaRMP2015}
\bibinfo{author}{Sinova, J.}, \bibinfo{author}{Valenzuela, S.~O.},
  \bibinfo{author}{Wunderlich, J.}, \bibinfo{author}{Back, C.~H.} \&
  \bibinfo{author}{Jungwirth, T.}
\newblock \bibinfo{journal}{\bibinfo{title}{{Spin Hall effects}}}.
\newblock {\emph{\JournalTitle{Reviews of Modern Physics}}}
  \textbf{\bibinfo{volume}{87}}, \bibinfo{pages}{1213--1260},
  \doiprefix\url{10.1103/RevModPhys.87.1213} (\bibinfo{year}{2015}).

\bibitem{Edelstein1990}
\bibinfo{author}{Edelstein, V.~M.}
\newblock \bibinfo{journal}{\bibinfo{title}{{Spin polarization of conduction
  electrons induced by electric current in two-dimensional asymmetric electron
  systems}}}.
\newblock {\emph{\JournalTitle{Solid State Communications}}}
  \textbf{\bibinfo{volume}{73}}, \bibinfo{pages}{233--235},
  \doiprefix\url{10.1016/0038-1098(90)90963-C} (\bibinfo{year}{1990}).

\bibitem{Manchon2019}
\bibinfo{author}{Manchon, A.} \emph{et~al.}
\newblock \bibinfo{journal}{\bibinfo{title}{{Current-induced spin-orbit torques
  in ferromagnetic and antiferromagnetic systems}}}.
\newblock {\emph{\JournalTitle{Reviews of Modern Physics}}}
  \textbf{\bibinfo{volume}{91}}, \bibinfo{pages}{035004},
  \doiprefix\url{10.1103/RevModPhys.91.035004} (\bibinfo{year}{2019}).
\newblock \eprint{1801.09636}.

\bibitem{Burch_Nature2018}
\bibinfo{author}{Burch, K.~S.}, \bibinfo{author}{Mandrus, D.} \&
  \bibinfo{author}{Park, J.-G.}
\newblock \bibinfo{journal}{\bibinfo{title}{Magnetism in two-dimensional van
  der waals materials}}.
\newblock {\emph{\JournalTitle{Nature}}} \textbf{\bibinfo{volume}{563}},
  \bibinfo{pages}{47--52}, \doiprefix\url{10.1038/s41586-018-0631-z}
  (\bibinfo{year}{2018}).

\bibitem{Gibertini_NatNano2019}
\bibinfo{author}{Gibertini, M.}, \bibinfo{author}{Koperski, M.},
  \bibinfo{author}{Morpurgo, A.~F.} \& \bibinfo{author}{Novoselov, K.~S.}
\newblock \bibinfo{journal}{\bibinfo{title}{Magnetic 2d materials and
  heterostructures}}.
\newblock {\emph{\JournalTitle{Nature Nanotechnology}}}
  \textbf{\bibinfo{volume}{14}}, \bibinfo{pages}{408--419},
  \doiprefix\url{10.1038/s41565-019-0438-6} (\bibinfo{year}{2019}).

\bibitem{Gong_eaav4450}
\bibinfo{author}{Gong, C.} \& \bibinfo{author}{Zhang, X.}
\newblock \bibinfo{journal}{\bibinfo{title}{Two-dimensional magnetic crystals
  and emergent heterostructure devices}}.
\newblock {\emph{\JournalTitle{Science}}} \textbf{\bibinfo{volume}{363}},
  \doiprefix\url{10.1126/science.aav4450} (\bibinfo{year}{2019}).
\newblock
  \eprint{https://science.sciencemag.org/content/363/6428/eaav4450.full.pdf}.

\bibitem{Mak_NatRevPhys2019}
\bibinfo{author}{Mak, K.~F.}, \bibinfo{author}{Shan, J.} \&
  \bibinfo{author}{Ralph, D.~C.}
\newblock \bibinfo{journal}{\bibinfo{title}{Probing and controlling magnetic
  states in 2d layered magnetic materials}}.
\newblock {\emph{\JournalTitle{Nature Reviews Physics}}}
  \textbf{\bibinfo{volume}{1}}, \bibinfo{pages}{646--661},
  \doiprefix\url{10.1038/s42254-019-0110-y} (\bibinfo{year}{2019}).

\bibitem{Huang_NatMater2020}
\bibinfo{author}{Huang, B.} \emph{et~al.}
\newblock \bibinfo{journal}{\bibinfo{title}{Emergent phenomena and proximity
  effects in two-dimensional magnets and heterostructures}}.
\newblock {\emph{\JournalTitle{Nature Materials}}}
  \textbf{\bibinfo{volume}{19}}, \bibinfo{pages}{1276--1289},
  \doiprefix\url{10.1038/s41563-020-0791-8} (\bibinfo{year}{2020}).

\bibitem{Huang_Nanoscale2020}
\bibinfo{author}{Huang, P.} \emph{et~al.}
\newblock \bibinfo{journal}{\bibinfo{title}{Recent advances in two-dimensional
  ferromagnetism: materials synthesis{,} physical properties and device
  applications}}.
\newblock {\emph{\JournalTitle{Nanoscale}}} \textbf{\bibinfo{volume}{12}},
  \bibinfo{pages}{2309--2327}, \doiprefix\url{10.1039/C9NR08890C}
  (\bibinfo{year}{2020}).

\bibitem{Wei_2DMater2020}
\bibinfo{author}{Wei, S.} \emph{et~al.}
\newblock \bibinfo{journal}{\bibinfo{title}{Emerging intrinsic magnetism in
  two-dimensional materials: theory and applications}}.
\newblock {\emph{\JournalTitle{2D Materials}}} \textbf{\bibinfo{volume}{8}},
  \bibinfo{pages}{012005}, \doiprefix\url{10.1088/2053-1583/abc8cb}
  (\bibinfo{year}{2020}).

\bibitem{Han_APLMater2016}
\bibinfo{author}{Han, W.}
\newblock \bibinfo{journal}{\bibinfo{title}{Perspectives for spintronics in 2d
  materials}}.
\newblock {\emph{\JournalTitle{APL Materials}}} \textbf{\bibinfo{volume}{4}},
  \bibinfo{pages}{032401}, \doiprefix\url{10.1063/1.4941712}
  (\bibinfo{year}{2016}).
\newblock \eprint{https://doi.org/10.1063/1.4941712}.

\bibitem{Avsar_RevModPhys2020}
\bibinfo{author}{Avsar, A.} \emph{et~al.}
\newblock \bibinfo{journal}{\bibinfo{title}{Colloquium: Spintronics in graphene
  and other two-dimensional materials}}.
\newblock {\emph{\JournalTitle{Rev. Mod. Phys.}}}
  \textbf{\bibinfo{volume}{92}}, \bibinfo{pages}{021003},
  \doiprefix\url{10.1103/RevModPhys.92.021003} (\bibinfo{year}{2020}).

\bibitem{Husain_APRev2020}
\bibinfo{author}{Husain, S.} \emph{et~al.}
\newblock \bibinfo{journal}{\bibinfo{title}{Emergence of spinâ€“orbit
  torques in 2d transition metal dichalcogenides: A status update}}.
\newblock {\emph{\JournalTitle{Applied Physics Reviews}}}
  \textbf{\bibinfo{volume}{7}}, \bibinfo{pages}{041312},
  \doiprefix\url{10.1063/5.0025318} (\bibinfo{year}{2020}).
\newblock \eprint{https://doi.org/10.1063/5.0025318}.

\bibitem{Ahn_npj2020}
\bibinfo{author}{Ahn, E.~C.}
\newblock \bibinfo{journal}{\bibinfo{title}{2d materials for spintronic
  devices}}.
\newblock {\emph{\JournalTitle{npj 2D Materials and Applications}}}
  \textbf{\bibinfo{volume}{4}}, \bibinfo{pages}{17},
  \doiprefix\url{10.1038/s41699-020-0152-0} (\bibinfo{year}{2020}).

\bibitem{Anderson_PR1950}
\bibinfo{author}{Anderson, P.~W.}
\newblock \bibinfo{journal}{\bibinfo{title}{{Antiferromagnetism. Theory of
  superexchange interaction}}}.
\newblock {\emph{\JournalTitle{Physical Review}}}
  \textbf{\bibinfo{volume}{79}}, \bibinfo{pages}{350--356},
  \doiprefix\url{10.1103/PhysRev.79.350} (\bibinfo{year}{1950}).

\bibitem{Lado_2017}
\bibinfo{author}{Lado, J.~L.} \& \bibinfo{author}{Fern{\'{a}}ndez-Rossier, J.}
\newblock \bibinfo{journal}{\bibinfo{title}{{On the origin of magnetic
  anisotropy in two dimensional CrI$_3$}}}.
\newblock {\emph{\JournalTitle{2D Materials}}} \textbf{\bibinfo{volume}{4}},
  \bibinfo{pages}{35002}, \doiprefix\url{10.1088/2053-1583/aa75ed}
  (\bibinfo{year}{2017}).
\newblock \eprint{1704.03849}.

\bibitem{KANAMORI1959}
\bibinfo{author}{Kanamori, J.}
\newblock \bibinfo{journal}{\bibinfo{title}{{Superexchange interaction and
  symmetry properties of electron orbitals}}}.
\newblock {\emph{\JournalTitle{Journal of Physics and Chemistry of Solids}}}
  \textbf{\bibinfo{volume}{10}}, \bibinfo{pages}{87--98},
  \doiprefix\url{10.1016/0022-3697(59)90061-7} (\bibinfo{year}{1959}).

\bibitem{Goodenough_PR1955}
\bibinfo{author}{Goodenough, J.~B.}
\newblock \bibinfo{journal}{\bibinfo{title}{{Theory of the role of covalence in
  the perovskite-type manganites [La,M(II)]MnO$_3$}}}.
\newblock {\emph{\JournalTitle{Physical Review}}}
  \textbf{\bibinfo{volume}{100}}, \bibinfo{pages}{564--573},
  \doiprefix\url{10.1103/PhysRev.100.564} (\bibinfo{year}{1955}).

\bibitem{Huang_JACS2018}
\bibinfo{author}{Huang, C.} \emph{et~al.}
\newblock \bibinfo{journal}{\bibinfo{title}{{Toward Intrinsic Room-Temperature
  Ferromagnetism in Two-Dimensional Semiconductors}}}.
\newblock {\emph{\JournalTitle{Journal of the American Chemical Society}}}
  \textbf{\bibinfo{volume}{140}}, \bibinfo{pages}{11519--11525},
  \doiprefix\url{10.1021/jacs.8b07879} (\bibinfo{year}{2018}).

\bibitem{Kim_PNAS2019}
\bibinfo{author}{Kim, H.~H.} \emph{et~al.}
\newblock \bibinfo{journal}{\bibinfo{title}{{Evolution of interlayer and
  intralayer magnetism in three atomically thin chromium trihalides}}}.
\newblock {\emph{\JournalTitle{Proceedings of the National Academy of Sciences
  of the United States of America}}} \textbf{\bibinfo{volume}{166}},
  \bibinfo{pages}{11131--11136}, \doiprefix\url{10.1073/pnas.1902100116}
  (\bibinfo{year}{2019}).
\newblock \eprint{1903.01409}.

\bibitem{Wang_AdvFuncMater2018}
\bibinfo{author}{Wang, F.} \emph{et~al.}
\newblock \bibinfo{journal}{\bibinfo{title}{{New Frontiers on van der Waals
  Layered Metal Phosphorous Trichalcogenides}}}.
\newblock {\emph{\JournalTitle{Advanced Functional Materials}}}
  \textbf{\bibinfo{volume}{28}}, \bibinfo{pages}{1802151},
  \doiprefix\url{10.1002/adfm.201802151} (\bibinfo{year}{2018}).

\bibitem{Rehman_Micromach2018}
\bibinfo{author}{{Ur Rehman}, Z.} \emph{et~al.}
\newblock \bibinfo{journal}{\bibinfo{title}{{Magnetic isotropy/anisotropy in
  layered metal phosphorous Trichalcogenide MPS$_3$ (M = Mn, Fe)single
  crystals}}}.
\newblock {\emph{\JournalTitle{Micromachines}}} \textbf{\bibinfo{volume}{9}},
  \doiprefix\url{10.3390/mi9060292} (\bibinfo{year}{2018}).

\bibitem{Kim_NComm2019}
\bibinfo{author}{Kim, K.} \emph{et~al.}
\newblock \bibinfo{journal}{\bibinfo{title}{{Suppression of magnetic ordering
  in XXZ-type antiferromagnetic monolayer NiPS$_3$}}}.
\newblock {\emph{\JournalTitle{Nature Communications}}}
  \textbf{\bibinfo{volume}{10}}, \bibinfo{pages}{345},
  \doiprefix\url{10.1038/s41467-018-08284-6} (\bibinfo{year}{2019}).

\bibitem{ZenerPhysRev1951}
\bibinfo{author}{Zener, C.}
\newblock \bibinfo{journal}{\bibinfo{title}{{Interaction between the d-shells
  in the transition metals. II. Ferromagnetic compounds of manganese with
  Perovskite structure}}}.
\newblock {\emph{\JournalTitle{Physical Review}}}
  \textbf{\bibinfo{volume}{82}}, \bibinfo{pages}{403--405},
  \doiprefix\url{10.1103/PhysRev.82.403} (\bibinfo{year}{1951}).

\bibitem{Jonker_Physica1950}
\bibinfo{author}{Jonker, G.} \& \bibinfo{author}{{Van Santen}, J.}
\newblock \bibinfo{journal}{\bibinfo{title}{{Ferromagnetic compounds of
  manganese with perovskite structure}}}.
\newblock {\emph{\JournalTitle{Physica}}} \textbf{\bibinfo{volume}{16}},
  \bibinfo{pages}{337--349}, \doiprefix\url{10.1016/0031-8914(50)90033-4}
  (\bibinfo{year}{1950}).

\bibitem{Wang_JACS2019}
\bibinfo{author}{Wang, N.} \emph{et~al.}
\newblock \bibinfo{journal}{\bibinfo{title}{Transition from ferromagnetic
  semiconductor to ferromagnetic metal with enhanced curie temperature in
  cr$_2$ge$_2$te$_6$ via organic ion intercalation}}.
\newblock {\emph{\JournalTitle{Journal of the American Chemical Society}}}
  \textbf{\bibinfo{volume}{141}}, \bibinfo{pages}{17166--17173},
  \doiprefix\url{10.1021/jacs.9b06929} (\bibinfo{year}{2019}).
\newblock \bibinfo{note}{PMID: 31599579},
  \eprint{https://doi.org/10.1021/jacs.9b06929}.

\bibitem{verzhbitskiy2020controlling}
\bibinfo{author}{Verzhbitskiy, I.~A.} \emph{et~al.}
\newblock \bibinfo{journal}{\bibinfo{title}{{Controlling the magnetic
  anisotropy in Cr$_2$Ge$_2$Te$_6$ by electrostatic gating}}}.
\newblock {\emph{\JournalTitle{Nature Electronics}}}
  \textbf{\bibinfo{volume}{3}}, \bibinfo{pages}{460--465},
  \doiprefix\url{10.1038/s41928-020-0427-7} (\bibinfo{year}{2020}).

\bibitem{Blundell2001}
\bibinfo{author}{Blundell, S.} \& \bibinfo{author}{Thouless, D.}
\newblock \emph{\bibinfo{title}{{Magnetism in Condensed Matter}}},
  vol.~\bibinfo{volume}{71} of \emph{\bibinfo{series}{Oxford master series in
  condensed matter physics}} (\bibinfo{publisher}{Oxford University Press},
  \bibinfo{address}{Oxford}, \bibinfo{year}{2003}).

\bibitem{Weiss1907}
\bibinfo{author}{Weiss, P.}
\newblock \bibinfo{journal}{\bibinfo{title}{{L'hypoth{\`{e}}se du champ
  mol{\'{e}}culaire et la propri{\'{e}}t{\'{e}} ferromagn{\'{e}}tique}}}.
\newblock {\emph{\JournalTitle{Journal de Physique Th{\'{e}}orique et
  Appliqu{\'{e}}e}}} \textbf{\bibinfo{volume}{6}}, \bibinfo{pages}{661--690},
  \doiprefix\url{10.1051/jphystap:019070060066100} (\bibinfo{year}{1907}).

\bibitem{Stoner1}
\bibinfo{author}{Watanabe, H.}
\newblock \bibinfo{journal}{\bibinfo{title}{{Collective Electron
  Ferromagnetism, II}}}.
\newblock {\emph{\JournalTitle{Journal of the Physical Society of Japan}}}
  \textbf{\bibinfo{volume}{3}}, \bibinfo{pages}{317--322},
  \doiprefix\url{10.1143/JPSJ.3.317} (\bibinfo{year}{1948}).

\bibitem{Stoner2}
\bibinfo{author}{Stoner, E.~C.}
\newblock \bibinfo{journal}{\bibinfo{title}{{Collective electron ferromagnetism
  II. Energy and specific heat}}}.
\newblock {\emph{\JournalTitle{Proceedings of the Royal Society of London.
  Series A. Mathematical and Physical Sciences}}}
  \textbf{\bibinfo{volume}{169}}, \bibinfo{pages}{339--371},
  \doiprefix\url{10.1098/rspa.1939.0003} (\bibinfo{year}{1939}).

\bibitem{Nakano_NanoLett2019}
\bibinfo{author}{Nakano, M.} \emph{et~al.}
\newblock \bibinfo{journal}{\bibinfo{title}{{Intrinsic 2D Ferromagnetism in
  V$_5$Se$_8$ Epitaxial Thin Films}}}.
\newblock {\emph{\JournalTitle{Nano Letters}}} \textbf{\bibinfo{volume}{19}},
  \bibinfo{pages}{8806--8810}, \doiprefix\url{10.1021/acs.nanolett.9b03614}
  (\bibinfo{year}{2019}).

\bibitem{May_PRB2016}
\bibinfo{author}{May, A.~F.}, \bibinfo{author}{Calder, S.},
  \bibinfo{author}{Cantoni, C.}, \bibinfo{author}{Cao, H.} \&
  \bibinfo{author}{McGuire, M.~A.}
\newblock \bibinfo{journal}{\bibinfo{title}{{Magnetic structure and phase
  stability of the van der Waals bonded ferromagnet Fe$_{3-x}$GeTe$_2$}}}.
\newblock {\emph{\JournalTitle{Physical Review B}}}
  \textbf{\bibinfo{volume}{93}}, \bibinfo{pages}{14411},
  \doiprefix\url{10.1103/PhysRevB.93.014411} (\bibinfo{year}{2016}).
\newblock \eprint{1508.06959}.

\bibitem{May_ACSNano2019}
\bibinfo{author}{May, A.~F.} \emph{et~al.}
\newblock \bibinfo{journal}{\bibinfo{title}{{Ferromagnetism Near Room
  Temperature in the Cleavable van der Waals Crystal Fe$_5$GeTe$_2$}}}.
\newblock {\emph{\JournalTitle{ACS Nano}}} \textbf{\bibinfo{volume}{13}},
  \bibinfo{pages}{4436--4442}, \doiprefix\url{10.1021/acsnano.8b09660}
  (\bibinfo{year}{2019}).

\bibitem{Morosan_PRB2007}
\bibinfo{author}{Morosan, E.} \emph{et~al.}
\newblock \bibinfo{journal}{\bibinfo{title}{{Sharp switching of the
  magnetization in Fe$_{0.25}$TaS$_2$}}}.
\newblock {\emph{\JournalTitle{Physical Review B}}}
  \textbf{\bibinfo{volume}{75}}, \bibinfo{pages}{104401},
  \doiprefix\url{10.1103/PhysRevB.75.104401} (\bibinfo{year}{2007}).

\bibitem{McGuire_PRB2017}
\bibinfo{author}{McGuire, M.~A.} \emph{et~al.}
\newblock \bibinfo{journal}{\bibinfo{title}{{Antiferromagnetism in the van der
  Waals layered spin-lozenge semiconductor CrTe$_3$}}}.
\newblock {\emph{\JournalTitle{Physical Review B}}}
  \textbf{\bibinfo{volume}{95}}, \bibinfo{pages}{144421},
  \doiprefix\url{10.1103/PhysRevB.95.144421} (\bibinfo{year}{2017}).
\newblock \eprint{1701.08621}.

\bibitem{Wang_PRB2019}
\bibinfo{author}{Wang, Y.} \emph{et~al.}
\newblock \bibinfo{journal}{\bibinfo{title}{{Magnetic anisotropy and
  topological Hall effect in the trigonal chromium tellurides Cr$_5$Te$_8$}}}.
\newblock {\emph{\JournalTitle{Physical Review B}}}
  \textbf{\bibinfo{volume}{100}}, \bibinfo{pages}{24434},
  \doiprefix\url{10.1103/PhysRevB.100.024434} (\bibinfo{year}{2019}).

\bibitem{Yan_EPL2019}
\bibinfo{author}{Yan, J.} \emph{et~al.}
\newblock \bibinfo{journal}{\bibinfo{title}{Anomalous hall effect of the
  quasi-two-dimensional weak itinerant ferromagnet cr 4.14 te 8}}.
\newblock {\emph{\JournalTitle{{EPL} (Europhysics Letters)}}}
  \textbf{\bibinfo{volume}{124}}, \bibinfo{pages}{67005},
  \doiprefix\url{10.1209/0295-5075/124/67005} (\bibinfo{year}{2019}).

\bibitem{Zhuang_PRB2016}
\bibinfo{author}{Zhuang, H.~L.}, \bibinfo{author}{Kent, P.~R.} \&
  \bibinfo{author}{Hennig, R.~G.}
\newblock \bibinfo{journal}{\bibinfo{title}{{Strong anisotropy and
  magnetostriction in the two-dimensional Stoner ferromagnet Fe$_3$GeTe$_2$}}}.
\newblock {\emph{\JournalTitle{Physical Review B}}}
  \textbf{\bibinfo{volume}{93}}, \bibinfo{pages}{134407},
  \doiprefix\url{10.1103/PhysRevB.93.134407} (\bibinfo{year}{2016}).

\bibitem{cai2019atomically}
\bibinfo{author}{Cai, X.} \emph{et~al.}
\newblock \bibinfo{journal}{\bibinfo{title}{Atomically thin {C}r{C}l$_3$: an
  in-plane layered antiferromagnetic insulator}}.
\newblock {\emph{\JournalTitle{Nano letters}}} \textbf{\bibinfo{volume}{19}},
  \bibinfo{pages}{3993--3998} (\bibinfo{year}{2019}).

\bibitem{wiedenmann1981neutron}
\bibinfo{author}{Wiedenmann, A.}, \bibinfo{author}{Rossat-Mignod, J.},
  \bibinfo{author}{Louisy, A.}, \bibinfo{author}{Brec, R.} \&
  \bibinfo{author}{Rouxel, J.}
\newblock \bibinfo{journal}{\bibinfo{title}{Neutron diffraction study of the
  layered compounds {M}n{PS}e$_3$ and {F}e{PS}e$_3$}}.
\newblock {\emph{\JournalTitle{Solid State Communications}}}
  \textbf{\bibinfo{volume}{40}}, \bibinfo{pages}{1067--1072}
  (\bibinfo{year}{1981}).

\bibitem{long2017isolation}
\bibinfo{author}{Long, G.} \emph{et~al.}
\newblock \bibinfo{journal}{\bibinfo{title}{Isolation and characterization of
  few-layer manganese thiophosphite}}.
\newblock {\emph{\JournalTitle{ACS nano}}} \textbf{\bibinfo{volume}{11}},
  \bibinfo{pages}{11330--11336} (\bibinfo{year}{2017}).

\bibitem{le1982magnetic}
\bibinfo{author}{Le~Flem, G.}, \bibinfo{author}{Brec, R.},
  \bibinfo{author}{Ouvard, G.}, \bibinfo{author}{Louisy, A.} \&
  \bibinfo{author}{Segransan, P.}
\newblock \bibinfo{journal}{\bibinfo{title}{Magnetic interactions in the layer
  compounds {MPX}$_3$ ({M}= {M}n, {F}e, {N}i; {X}= {S}, {S}e)}}.
\newblock {\emph{\JournalTitle{Journal of Physics and Chemistry of Solids}}}
  \textbf{\bibinfo{volume}{43}}, \bibinfo{pages}{455--461}
  (\bibinfo{year}{1982}).

\bibitem{ghazaryan2018magnon}
\bibinfo{author}{Ghazaryan, D.} \emph{et~al.}
\newblock \bibinfo{journal}{\bibinfo{title}{Magnon-assisted tunnelling in van
  der waals heterostructures based on {C}r{B}r$_3$}}.
\newblock {\emph{\JournalTitle{Nature Electronics}}}
  \textbf{\bibinfo{volume}{1}}, \bibinfo{pages}{344--349}
  (\bibinfo{year}{2018}).

\bibitem{samuelsen1971spin}
\bibinfo{author}{Samuelsen, E.}, \bibinfo{author}{Silberglitt, R.},
  \bibinfo{author}{Shirane, G.} \& \bibinfo{author}{Remeika, J.}
\newblock \bibinfo{journal}{\bibinfo{title}{Spin waves in ferromagnetic
  {C}r{B}r$_3$ studied by inelastic neutron scattering}}.
\newblock {\emph{\JournalTitle{Physical Review B}}}
  \textbf{\bibinfo{volume}{3}}, \bibinfo{pages}{157} (\bibinfo{year}{1971}).

\bibitem{lee2016ising}
\bibinfo{author}{Lee, J.-U.} \emph{et~al.}
\newblock \bibinfo{journal}{\bibinfo{title}{Ising-type magnetic ordering in
  atomically thin {F}e{PS}$_3$}}.
\newblock {\emph{\JournalTitle{Nano letters}}} \textbf{\bibinfo{volume}{16}},
  \bibinfo{pages}{7433--7438} (\bibinfo{year}{2016}).

\bibitem{liu2020exploring}
\bibinfo{author}{Liu, P.} \emph{et~al.}
\newblock \bibinfo{journal}{\bibinfo{title}{Exploring the magnetic ordering in
  atomically thin antiferromagnetic {M}n{PS}e$_3$ by raman spectroscopy}}.
\newblock {\emph{\JournalTitle{Journal of Alloys and Compounds}}}
  \textbf{\bibinfo{volume}{828}}, \bibinfo{pages}{154432}
  (\bibinfo{year}{2020}).

\bibitem{Huang_NNano2018}
\bibinfo{author}{Huang, B.} \emph{et~al.}
\newblock \bibinfo{journal}{\bibinfo{title}{{Electrical control of 2D magnetism
  in bilayer CrI$_3$}}}.
\newblock {\emph{\JournalTitle{Nature Nanotechnology}}}
  \textbf{\bibinfo{volume}{13}}, \bibinfo{pages}{544--548},
  \doiprefix\url{10.1038/s41565-018-0121-3} (\bibinfo{year}{2018}).
\newblock \eprint{1802.06979}.

\bibitem{zhong2017van}
\bibinfo{author}{Zhong, D.} \emph{et~al.}
\newblock \bibinfo{journal}{\bibinfo{title}{Van der waals engineering of
  ferromagnetic semiconductor heterostructures for spin and valleytronics}}.
\newblock {\emph{\JournalTitle{Science advances}}}
  \textbf{\bibinfo{volume}{3}}, \bibinfo{pages}{e1603113}
  (\bibinfo{year}{2017}).

\bibitem{chattopadhyay1991spin}
\bibinfo{author}{Chattopadhyay, T.}, \bibinfo{author}{Br{\"u}ckel, T.} \&
  \bibinfo{author}{Burlet, P.}
\newblock \bibinfo{journal}{\bibinfo{title}{Spin correlation in the frustrated
  antiferromagnet {M}n{S}$_2$ above the n{\'e}el temperature}}.
\newblock {\emph{\JournalTitle{Physical Review B}}}
  \textbf{\bibinfo{volume}{44}}, \bibinfo{pages}{7394} (\bibinfo{year}{1991}).

\bibitem{Son_PRB2019}
\bibinfo{author}{Son, S.} \emph{et~al.}
\newblock \bibinfo{journal}{\bibinfo{title}{{Bulk properties of the van der
  Waals hard ferromagnet VI$_3$}}}.
\newblock {\emph{\JournalTitle{Physical Review B}}}
  \textbf{\bibinfo{volume}{99}}, \bibinfo{pages}{41402},
  \doiprefix\url{10.1103/PhysRevB.99.041402} (\bibinfo{year}{2019}).
\newblock \eprint{1812.05284}.

\bibitem{khan2019spin}
\bibinfo{author}{Khan, S.} \emph{et~al.}
\newblock \bibinfo{journal}{\bibinfo{title}{Spin dynamics study in layered van
  der waals single-crystal {C}r$_2${G}e$_2${T}e$_6$}}.
\newblock {\emph{\JournalTitle{Physical Review B}}}
  \textbf{\bibinfo{volume}{100}}, \bibinfo{pages}{134437}
  (\bibinfo{year}{2019}).

\bibitem{Lin2016b}
\bibinfo{author}{Lin, M.-W.} \emph{et~al.}
\newblock \bibinfo{journal}{\bibinfo{title}{{Ultrathin nanosheets of CrSiTe$_3$
  : a semiconducting two-dimensional ferromagnetic material}}}.
\newblock {\emph{\JournalTitle{Journal of Materials Chemistry C}}}
  \textbf{\bibinfo{volume}{4}}, \bibinfo{pages}{315--322},
  \doiprefix\url{10.1039/C5TC03463A} (\bibinfo{year}{2016}).

\bibitem{lv2015strain}
\bibinfo{author}{Lv, H.}, \bibinfo{author}{Lu, W.}, \bibinfo{author}{Shao, D.},
  \bibinfo{author}{Liu, Y.} \& \bibinfo{author}{Sun, Y.}
\newblock \bibinfo{journal}{\bibinfo{title}{Strain-controlled switch between
  ferromagnetism and antiferromagnetism in 1{T}-{C}r{X}$_2$ ({X}= {S}e, {T}e)
  monolayers}}.
\newblock {\emph{\JournalTitle{Physical Review B}}}
  \textbf{\bibinfo{volume}{92}}, \bibinfo{pages}{214419}
  (\bibinfo{year}{2015}).

\bibitem{sun2020room}
\bibinfo{author}{Sun, X.} \emph{et~al.}
\newblock \bibinfo{journal}{\bibinfo{title}{Room temperature ferromagnetism in
  ultra-thin van der waals crystals of 1{T}-{C}r{T}e$_2$}}.
\newblock {\emph{\JournalTitle{Nano Research}}} \textbf{\bibinfo{volume}{13}},
  \bibinfo{pages}{3358--3363} (\bibinfo{year}{2020}).

\bibitem{chu2019sub}
\bibinfo{author}{Chu, J.} \emph{et~al.}
\newblock \bibinfo{journal}{\bibinfo{title}{Sub-millimeter-scale growth of
  one-unit-cell-thick ferrimagnetic {C}r$_2${S}$_3$ nanosheets}}.
\newblock {\emph{\JournalTitle{Nano letters}}} \textbf{\bibinfo{volume}{19}},
  \bibinfo{pages}{2154--2161} (\bibinfo{year}{2019}).

\bibitem{xie2020atomically}
\bibinfo{author}{Xie, L.} \emph{et~al.}
\newblock \bibinfo{journal}{\bibinfo{title}{An atomically thin air-stable
  narrow-gap semiconductor {C}r$_2${S}$_3$ for broadband photodetection with
  high responsivity}}.
\newblock {\emph{\JournalTitle{Advanced Electronic Materials}}}
  \bibinfo{pages}{2000962} (\bibinfo{year}{2020}).

\bibitem{zeugner2019chemical}
\bibinfo{author}{Zeugner, A.} \emph{et~al.}
\newblock \bibinfo{journal}{\bibinfo{title}{Chemical aspects of the candidate
  antiferromagnetic topological insulator {M}n{B}i$_2${T}e$_4$}}.
\newblock {\emph{\JournalTitle{Chemistry of Materials}}}
  \textbf{\bibinfo{volume}{31}}, \bibinfo{pages}{2795--2806}
  (\bibinfo{year}{2019}).

\bibitem{kan2014ferromagnetism}
\bibinfo{author}{Kan, M.}, \bibinfo{author}{Adhikari, S.} \&
  \bibinfo{author}{Sun, Q.}
\newblock \bibinfo{journal}{\bibinfo{title}{Ferromagnetism in {M}n{X}$_2$ ({X}=
  {S}, {S}e) monolayers}}.
\newblock {\emph{\JournalTitle{Physical Chemistry Chemical Physics}}}
  \textbf{\bibinfo{volume}{16}}, \bibinfo{pages}{4990--4994}
  (\bibinfo{year}{2014}).

\bibitem{itoh1977magnetic}
\bibinfo{author}{Itoh, H.} \& \bibinfo{author}{Miyahara, S.}
\newblock \bibinfo{journal}{\bibinfo{title}{Magnetic susceptibility and thermal
  expansion of {M}n{S}e$_2$ with pyrite structure}}.
\newblock {\emph{\JournalTitle{Journal of the Physical Society of Japan}}}
  \textbf{\bibinfo{volume}{42}}, \bibinfo{pages}{470--472}
  (\bibinfo{year}{1977}).

\bibitem{lei2020high}
\bibinfo{author}{Lei, S.} \emph{et~al.}
\newblock \bibinfo{journal}{\bibinfo{title}{High mobility in a van der waals
  layered antiferromagnetic metal}}.
\newblock {\emph{\JournalTitle{Science advances}}}
  \textbf{\bibinfo{volume}{6}}, \bibinfo{pages}{eaay6407}
  (\bibinfo{year}{2020}).

\bibitem{zhang2019ultrathin}
\bibinfo{author}{Zhang, Y.} \emph{et~al.}
\newblock \bibinfo{journal}{\bibinfo{title}{Ultrathin magnetic 2d
  single-crystal {C}r{S}e}}.
\newblock {\emph{\JournalTitle{Advanced Materials}}}
  \textbf{\bibinfo{volume}{31}}, \bibinfo{pages}{1900056}
  (\bibinfo{year}{2019}).

\bibitem{zhang2013dimension}
\bibinfo{author}{Zhang, H.}, \bibinfo{author}{Liu, L.-M.} \&
  \bibinfo{author}{Lau, W.-M.}
\newblock \bibinfo{journal}{\bibinfo{title}{Dimension-dependent phase
  transition and magnetic properties of {VS}$_2$}}.
\newblock {\emph{\JournalTitle{Journal of Materials Chemistry A}}}
  \textbf{\bibinfo{volume}{1}}, \bibinfo{pages}{10821--10828}
  (\bibinfo{year}{2013}).

\bibitem{zhang2020critical}
\bibinfo{author}{Zhang, L.-Z.} \emph{et~al.}
\newblock \bibinfo{journal}{\bibinfo{title}{Critical behavior and
  magnetocaloric effect of the quasi-two-dimensional room-temperature
  ferromagnet {C}r$_4${T}e$_5$}}.
\newblock {\emph{\JournalTitle{Physical Review B}}}
  \textbf{\bibinfo{volume}{101}}, \bibinfo{pages}{214413}
  (\bibinfo{year}{2020}).

\bibitem{liu2019magnetic}
\bibinfo{author}{Liu, Y.} \emph{et~al.}
\newblock \bibinfo{journal}{\bibinfo{title}{Magnetic anisotropy and entropy
  change in trigonal {C}r$_5${T}e$_8$}}.
\newblock {\emph{\JournalTitle{Physical Review B}}}
  \textbf{\bibinfo{volume}{100}}, \bibinfo{pages}{245114}
  (\bibinfo{year}{2019}).

\bibitem{seo2020nearly}
\bibinfo{author}{Seo, J.} \emph{et~al.}
\newblock \bibinfo{journal}{\bibinfo{title}{Nearly room temperature
  ferromagnetism in a magnetic metal-rich van der waals metal}}.
\newblock {\emph{\JournalTitle{Science advances}}}
  \textbf{\bibinfo{volume}{6}}, \bibinfo{pages}{eaay8912}
  (\bibinfo{year}{2020}).

\bibitem{Mermin1966}
\bibinfo{author}{Mermin, N.~D.} \& \bibinfo{author}{Wagner, H.}
\newblock \bibinfo{journal}{\bibinfo{title}{{Absence of ferromagnetism or
  antiferromagnetism in one- or two-dimensional isotropic Heisenberg models}}}.
\newblock {\emph{\JournalTitle{Physical Review Letters}}}
  \textbf{\bibinfo{volume}{17}}, \bibinfo{pages}{1133--1136},
  \doiprefix\url{10.1103/PhysRevLett.17.1133} (\bibinfo{year}{1966}).

\bibitem{JiangNNano2018}
\bibinfo{author}{Jiang, S.}, \bibinfo{author}{Li, L.}, \bibinfo{author}{Wang,
  Z.}, \bibinfo{author}{Mak, K.~F.} \& \bibinfo{author}{Shan, J.}
\newblock \bibinfo{journal}{\bibinfo{title}{Controlling magnetism in 2d cri$_3$
  by electrostatic doping}}.
\newblock {\emph{\JournalTitle{Nature Nanotechnology}}}
  \textbf{\bibinfo{volume}{13}}, \bibinfo{pages}{549--553},
  \doiprefix\url{10.1038/s41565-018-0135-x} (\bibinfo{year}{2018}).

\bibitem{song2019switching}
\bibinfo{author}{Song, T.} \emph{et~al.}
\newblock \bibinfo{journal}{\bibinfo{title}{{Switching 2D magnetic states via
  pressure tuning of layer stacking}}}.
\newblock {\emph{\JournalTitle{Nature Materials}}}
  \textbf{\bibinfo{volume}{18}}, \bibinfo{pages}{1298--1302},
  \doiprefix\url{10.1038/s41563-019-0505-2} (\bibinfo{year}{2019}).
\newblock \eprint{1905.10860}.

\bibitem{Matsukura_NNano2015}
\bibinfo{author}{Matsukura, F.}, \bibinfo{author}{Tokura, Y.} \&
  \bibinfo{author}{Ohno, H.}
\newblock \bibinfo{journal}{\bibinfo{title}{{Control of magnetism by electric
  fields}}}.
\newblock {\emph{\JournalTitle{Nature Nanotechnology}}}
  \textbf{\bibinfo{volume}{10}}, \bibinfo{pages}{209--220},
  \doiprefix\url{10.1038/nnano.2015.22} (\bibinfo{year}{2015}).

\bibitem{McGuire_Crystals2017}
\bibinfo{author}{McGuire, M.~A.}
\newblock \bibinfo{journal}{\bibinfo{title}{{Crystal and magnetic structures in
  layered, transition metal dihalides and trihalides}}}.
\newblock {\emph{\JournalTitle{Crystals}}} \textbf{\bibinfo{volume}{7}},
  \doiprefix\url{10.3390/cryst7050121} (\bibinfo{year}{2017}).
\newblock \eprint{1704.08225}.

\bibitem{Sivadas2016}
\bibinfo{author}{Sivadas, N.}, \bibinfo{author}{Okamoto, S.} \&
  \bibinfo{author}{Xiao, D.}
\newblock \bibinfo{journal}{\bibinfo{title}{{Gate-Controllable Magneto-optic
  Kerr Effect in Layered Collinear Antiferromagnets}}}.
\newblock {\emph{\JournalTitle{Physical Review Letters}}}
  \textbf{\bibinfo{volume}{117}}, \bibinfo{pages}{267203},
  \doiprefix\url{10.1103/PhysRevLett.117.267203} (\bibinfo{year}{2016}).
\newblock \eprint{1607.02156}.

\bibitem{Jiang2018}
\bibinfo{author}{Jiang, S.}, \bibinfo{author}{Shan, J.} \&
  \bibinfo{author}{Mak, K.~F.}
\newblock \bibinfo{journal}{\bibinfo{title}{{Electric-field switching of
  two-dimensional van der Waals magnets}}}.
\newblock {\emph{\JournalTitle{Nature Materials}}}
  \textbf{\bibinfo{volume}{17}}, \bibinfo{pages}{406--410},
  \doiprefix\url{10.1038/s41563-018-0040-6} (\bibinfo{year}{2018}).

\bibitem{Song_NanoLett2019}
\bibinfo{author}{Song, T.} \emph{et~al.}
\newblock \bibinfo{journal}{\bibinfo{title}{{Voltage Control of a van der Waals
  Spin-Filter Magnetic Tunnel Junction}}}.
\newblock {\emph{\JournalTitle{Nano Letters}}} \textbf{\bibinfo{volume}{19}},
  \bibinfo{pages}{915--920}, \doiprefix\url{10.1021/acs.nanolett.8b04160}
  (\bibinfo{year}{2019}).

\bibitem{wang2018electric}
\bibinfo{author}{Wang, Z.} \emph{et~al.}
\newblock \bibinfo{journal}{\bibinfo{title}{{Electric-field control of
  magnetism in a few-layered van der Waals ferromagnetic semiconductor}}}.
\newblock {\emph{\JournalTitle{Nature Nanotechnology}}}
  \textbf{\bibinfo{volume}{13}}, \bibinfo{pages}{554--559},
  \doiprefix\url{10.1038/s41565-018-0186-z} (\bibinfo{year}{2018}).

\bibitem{Sun_APL2018}
\bibinfo{author}{Sun, Y.} \emph{et~al.}
\newblock \bibinfo{journal}{\bibinfo{title}{{Effects of hydrostatic pressure on
  spin-lattice coupling in two-dimensional ferromagnetic Cr$_2$Ge$_2$Te$_6$}}}.
\newblock {\emph{\JournalTitle{Applied Physics Letters}}}
  \textbf{\bibinfo{volume}{112}}, \bibinfo{pages}{72409},
  \doiprefix\url{10.1063/1.5016568} (\bibinfo{year}{2018}).

\bibitem{Mondal_PRB2019}
\bibinfo{author}{Mondal, S.} \emph{et~al.}
\newblock \bibinfo{journal}{\bibinfo{title}{{Effect of hydrostatic pressure on
  ferromagnetism in two-dimensional CrI$_3$}}}.
\newblock {\emph{\JournalTitle{Physical Review B}}}
  \textbf{\bibinfo{volume}{99}}, \bibinfo{pages}{180407},
  \doiprefix\url{10.1103/PhysRevB.99.180407} (\bibinfo{year}{2019}).
\newblock \eprint{1901.00706}.

\bibitem{Lin_PRM2018}
\bibinfo{author}{Lin, Z.} \emph{et~al.}
\newblock \bibinfo{journal}{\bibinfo{title}{{Pressure-induced spin
  reorientation transition in layered ferromagnetic insulator
  Cr$_2$Ge$_2$Te$_6$}}}.
\newblock {\emph{\JournalTitle{Physical Review Materials}}}
  \textbf{\bibinfo{volume}{2}}, \bibinfo{pages}{51004},
  \doiprefix\url{10.1103/PhysRevMaterials.2.051004} (\bibinfo{year}{2018}).

\bibitem{Li_NMater2019}
\bibinfo{author}{Li, T.} \emph{et~al.}
\newblock \bibinfo{journal}{\bibinfo{title}{{Pressure-controlled interlayer
  magnetism in atomically thin CrI$_3$}}}.
\newblock {\emph{\JournalTitle{Nature Materials}}}
  \textbf{\bibinfo{volume}{18}}, \bibinfo{pages}{1303--1308},
  \doiprefix\url{10.1038/s41563-019-0506-1} (\bibinfo{year}{2019}).
\newblock \eprint{1905.10905}.

\bibitem{Jiang_PRB2019}
\bibinfo{author}{Jiang, P.} \emph{et~al.}
\newblock \bibinfo{journal}{\bibinfo{title}{{Stacking tunable interlayer
  magnetism in bilayer CrI$_3$}}}.
\newblock {\emph{\JournalTitle{Physical Review B}}}
  \textbf{\bibinfo{volume}{99}}, \bibinfo{pages}{144401},
  \doiprefix\url{10.1103/PhysRevB.99.144401} (\bibinfo{year}{2019}).
\newblock \eprint{1806.09274}.

\bibitem{Jang_PRM2019}
\bibinfo{author}{Jang, S.~W.}, \bibinfo{author}{Jeong, M.~Y.},
  \bibinfo{author}{Yoon, H.}, \bibinfo{author}{Ryee, S.} \&
  \bibinfo{author}{Han, M.~J.}
\newblock \bibinfo{journal}{\bibinfo{title}{{Microscopic understanding of
  magnetic interactions in bilayer CrI$_3$}}}.
\newblock {\emph{\JournalTitle{Physical Review Materials}}}
  \textbf{\bibinfo{volume}{3}}, \bibinfo{pages}{31001},
  \doiprefix\url{10.1103/PhysRevMaterials.3.031001} (\bibinfo{year}{2019}).
\newblock \eprint{1809.01388}.

\bibitem{Sivadas_NanoL2019}
\bibinfo{author}{Sivadas, N.}, \bibinfo{author}{Okamoto, S.},
  \bibinfo{author}{Xu, X.}, \bibinfo{author}{Fennie, C.~J.} \&
  \bibinfo{author}{Xiao, D.}
\newblock \bibinfo{journal}{\bibinfo{title}{{Stacking-Dependent Magnetism in
  Bilayer CrI$_3$}}}.
\newblock {\emph{\JournalTitle{Nano Letters}}} \textbf{\bibinfo{volume}{18}},
  \bibinfo{pages}{7658--7664}, \doiprefix\url{10.1021/acs.nanolett.8b03321}
  (\bibinfo{year}{2018}).
\newblock \eprint{1808.06559}.

\bibitem{SORIANO_SolidStateCom2019}
\bibinfo{author}{Soriano, D.}, \bibinfo{author}{Cardoso, C.} \&
  \bibinfo{author}{Fern{\'{a}}ndez-Rossier, J.}
\newblock \bibinfo{journal}{\bibinfo{title}{{Interplay between interlayer
  exchange and stacking in CrI$_3$ bilayers}}}.
\newblock {\emph{\JournalTitle{Solid State Communications}}}
  \textbf{\bibinfo{volume}{299}}, \bibinfo{pages}{113662},
  \doiprefix\url{10.1016/j.ssc.2019.113662} (\bibinfo{year}{2019}).
\newblock \eprint{1807.00357}.

\bibitem{klein2019enhancement}
\bibinfo{author}{Klein, D.~R.} \emph{et~al.}
\newblock \bibinfo{journal}{\bibinfo{title}{{Enhancement of interlayer exchange
  in an ultrathin two-dimensional magnet}}}.
\newblock {\emph{\JournalTitle{Nature Physics}}} \textbf{\bibinfo{volume}{15}},
  \bibinfo{pages}{1255--1260}, \doiprefix\url{10.1038/s41567-019-0651-0}
  (\bibinfo{year}{2019}).

\bibitem{Kent_Nnano2015}
\bibinfo{author}{Kent, A.~D.} \& \bibinfo{author}{Worledge, D.~C.}
\newblock \bibinfo{journal}{\bibinfo{title}{{A new spin on magnetic
  memories}}}.
\newblock {\emph{\JournalTitle{Nature Nanotechnology}}}
  \textbf{\bibinfo{volume}{10}}, \bibinfo{pages}{187--191},
  \doiprefix\url{10.1038/nnano.2015.24} (\bibinfo{year}{2015}).

\bibitem{Dieny_NElec2020}
\bibinfo{author}{Dieny, B.} \emph{et~al.}
\newblock \bibinfo{journal}{\bibinfo{title}{{Opportunities and challenges for
  spintronics in the microelectronics industry}}}.
\newblock {\emph{\JournalTitle{Nature Electronics}}}
  \textbf{\bibinfo{volume}{3}}, \bibinfo{pages}{446--459},
  \doiprefix\url{10.1038/s41928-020-0461-5} (\bibinfo{year}{2020}).
\newblock \eprint{1908.10584}.

\bibitem{Grollier_NPhysRev2020}
\bibinfo{author}{Grollier, J.} \emph{et~al.}
\newblock \bibinfo{journal}{\bibinfo{title}{{Neuromorphic spintronics}}}.
\newblock {\emph{\JournalTitle{Nature Electronics}}}
  \textbf{\bibinfo{volume}{3}}, \bibinfo{pages}{360--370},
  \doiprefix\url{10.1038/s41928-019-0360-9} (\bibinfo{year}{2020}).

\bibitem{MacNeill_NPhys2017}
\bibinfo{author}{MacNeill, D.} \emph{et~al.}
\newblock \bibinfo{journal}{\bibinfo{title}{{Control of spin-orbit torques
  through crystal symmetry in WTe$_2$ /ferromagnet bilayers}}}.
\newblock {\emph{\JournalTitle{Nature Physics}}} \textbf{\bibinfo{volume}{13}},
  \bibinfo{pages}{300--305}, \doiprefix\url{10.1038/nphys3933}
  (\bibinfo{year}{2017}).
\newblock \eprint{1605.02712}.

\bibitem{Guimaraes_NanoLett2018}
\bibinfo{author}{Guimar{\~{a}}es, M.~H.}, \bibinfo{author}{Stiehl, G.~M.},
  \bibinfo{author}{MacNeill, D.}, \bibinfo{author}{Reynolds, N.~D.} \&
  \bibinfo{author}{Ralph, D.~C.}
\newblock \bibinfo{journal}{\bibinfo{title}{{Spin-Orbit Torques in
  NbSe$_2$/Permalloy Bilayers}}}.
\newblock {\emph{\JournalTitle{Nano Letters}}} \textbf{\bibinfo{volume}{18}},
  \bibinfo{pages}{1311--1316}, \doiprefix\url{10.1021/acs.nanolett.7b04993}
  (\bibinfo{year}{2018}).
\newblock \eprint{1801.07281}.

\bibitem{Shi_Nnano2019}
\bibinfo{author}{Shi, S.} \emph{et~al.}
\newblock \bibinfo{journal}{\bibinfo{title}{{All-electric magnetization
  switching and Dzyaloshinskii–Moriya interaction in WTe$_2$/ferromagnet
  heterostructures}}}.
\newblock {\emph{\JournalTitle{Nature Nanotechnology}}}
  \textbf{\bibinfo{volume}{14}}, \bibinfo{pages}{945--949},
  \doiprefix\url{10.1038/s41565-019-0525-8} (\bibinfo{year}{2019}).

\bibitem{Alghamdi_NanoLett2019}
\bibinfo{author}{Alghamdi, M.} \emph{et~al.}
\newblock \bibinfo{journal}{\bibinfo{title}{{Highly Efficient Spin-Orbit Torque
  and Switching of Layered Ferromagnet Fe$_3$GeTe$_2$}}}.
\newblock {\emph{\JournalTitle{Nano Letters}}} \textbf{\bibinfo{volume}{19}},
  \bibinfo{pages}{4400--4405}, \doiprefix\url{10.1021/acs.nanolett.9b01043}
  (\bibinfo{year}{2019}).
\newblock \eprint{1903.00571}.

\bibitem{Wang_AdvSci2019}
\bibinfo{author}{Wang, X.} \emph{et~al.}
\newblock \bibinfo{journal}{\bibinfo{title}{{Current-driven magnetization
  switching in a van der Waals ferromagnet Fe$_3$GeTe$_2$}}}.
\newblock {\emph{\JournalTitle{Science Advances}}}
  \textbf{\bibinfo{volume}{5}}, \doiprefix\url{10.1126/sciadv.aaw8904}
  (\bibinfo{year}{2019}).

\bibitem{Ostwal_AdvMater2020}
\bibinfo{author}{Ostwal, V.}, \bibinfo{author}{Shen, T.} \&
  \bibinfo{author}{Appenzeller, J.}
\newblock \bibinfo{journal}{\bibinfo{title}{{Efficient Spin-Orbit Torque
  Switching of the Semiconducting Van Der Waals Ferromagnet
  Cr$_2$Ge$_2$Te$_6$}}}.
\newblock {\emph{\JournalTitle{Advanced Materials}}}
  \textbf{\bibinfo{volume}{32}}, \bibinfo{pages}{1--7},
  \doiprefix\url{10.1002/adma.201906021} (\bibinfo{year}{2020}).

\bibitem{RALPH_JMMM2008}
\bibinfo{author}{Ralph, D.} \& \bibinfo{author}{Stiles, M.}
\newblock \bibinfo{journal}{\bibinfo{title}{{Spin transfer torques}}}.
\newblock {\emph{\JournalTitle{Journal of Magnetism and Magnetic Materials}}}
  \textbf{\bibinfo{volume}{320}}, \bibinfo{pages}{1190--1216},
  \doiprefix\url{10.1016/j.jmmm.2007.12.019} (\bibinfo{year}{2008}).

\bibitem{Avci_NMat2017}
\bibinfo{author}{Avci, C.~O.} \emph{et~al.}
\newblock \bibinfo{journal}{\bibinfo{title}{{Current-induced switching in a
  magnetic insulator}}}.
\newblock {\emph{\JournalTitle{Nature Materials}}}
  \textbf{\bibinfo{volume}{16}}, \bibinfo{pages}{309--314},
  \doiprefix\url{10.1038/nmat4812} (\bibinfo{year}{2017}).

\bibitem{shin2021spinorbit}
\bibinfo{author}{Shin, I.} \emph{et~al.}
\newblock \bibinfo{title}{{Spin-orbit Torque Switching in an All-Van der Waals
  Heterostructure}} (\bibinfo{year}{2021}).
\newblock \eprint{2102.09300}.

\bibitem{Fang_NNano2011}
\bibinfo{author}{Fang, D.} \emph{et~al.}
\newblock \bibinfo{journal}{\bibinfo{title}{{Spin-orbit-driven ferromagnetic
  resonance}}}.
\newblock {\emph{\JournalTitle{Nature Nanotechnology}}}
  \textbf{\bibinfo{volume}{6}}, \bibinfo{pages}{413--417},
  \doiprefix\url{10.1038/nnano.2011.68} (\bibinfo{year}{2011}).

\bibitem{Ciccarelli_NPhys2016}
\bibinfo{author}{Ciccarelli, C.} \emph{et~al.}
\newblock \bibinfo{journal}{\bibinfo{title}{{Room-temperature spin-orbit torque
  in NiMnSb}}}.
\newblock {\emph{\JournalTitle{Nature Physics}}} \textbf{\bibinfo{volume}{12}},
  \bibinfo{pages}{855--860}, \doiprefix\url{10.1038/nphys3772}
  (\bibinfo{year}{2016}).
\newblock \eprint{1510.03356}.

\bibitem{Yoshimi_SciAdv2018}
\bibinfo{author}{Yoshimi, R.} \emph{et~al.}
\newblock \bibinfo{journal}{\bibinfo{title}{{Current-driven magnetization
  switching in ferromagnetic bulk Rashba semiconductor (Ge,Mn)Te}}}.
\newblock {\emph{\JournalTitle{Science Advances}}}
  \textbf{\bibinfo{volume}{4}}, \doiprefix\url{10.1126/sciadv.aat9989}
  (\bibinfo{year}{2018}).

\bibitem{Kurebayashi_NNano2014}
\bibinfo{author}{Kurebayashi, H.} \emph{et~al.}
\newblock \bibinfo{journal}{\bibinfo{title}{{An antidamping spin-orbit torque
  originating from the Berry curvature}}}.
\newblock {\emph{\JournalTitle{Nature Nanotechnology}}}
  \textbf{\bibinfo{volume}{9}}, \bibinfo{pages}{211--217},
  \doiprefix\url{10.1038/nnano.2014.15} (\bibinfo{year}{2014}).

\bibitem{BradleyCracknellSymmBook}
\bibinfo{author}{Bradley, C.~J.} \& \bibinfo{author}{Cracknell, A.~P.}
\newblock \emph{\bibinfo{title}{{The Mathematical Theory of Symmetry in Solids:
  Representation theory for point groups and space groups}}}
  (\bibinfo{publisher}{Oxford University Press}, \bibinfo{year}{1972}).

\bibitem{Manipatruni2019}
\bibinfo{author}{Manipatruni, S.} \emph{et~al.}
\newblock \bibinfo{journal}{\bibinfo{title}{{Scalable energy-efficient
  magnetoelectric spin–orbit logic}}}.
\newblock {\emph{\JournalTitle{Nature}}} \textbf{\bibinfo{volume}{565}},
  \bibinfo{pages}{35--42}, \doiprefix\url{10.1038/s41586-018-0770-2}
  (\bibinfo{year}{2019}).

\bibitem{Bandyopadhyay2008}
\bibinfo{author}{Bandyopadhyay, S.} \& \bibinfo{author}{Cahay, M.}
\newblock \emph{\bibinfo{title}{{Introduction to Spintronics}}}
  (\bibinfo{publisher}{CRC Press}, \bibinfo{year}{2008}),
  \bibinfo{edition}{first} edn.

\bibitem{Fo2004}
\bibinfo{author}{Kurz, P.}, \bibinfo{author}{F{\"{o}}rster, F.},
  \bibinfo{author}{Nordstr{\"{o}}m, L.}, \bibinfo{author}{Bihlmayer, G.} \&
  \bibinfo{author}{Bl{\"{u}}gel, S.}
\newblock \bibinfo{journal}{\bibinfo{title}{{Ab initio treatment of
  noncollinear magnets with the full-potential linearized augmented plane wave
  method}}}.
\newblock {\emph{\JournalTitle{Physical Review B}}}
  \textbf{\bibinfo{volume}{69}}, \bibinfo{pages}{024415},
  \doiprefix\url{10.1103/PhysRevB.69.024415} (\bibinfo{year}{2004}).

\bibitem{PRBFreimuth2014}
\bibinfo{author}{Freimuth, F.}, \bibinfo{author}{Bl{\"{u}}gel, S.} \&
  \bibinfo{author}{Mokrousov, Y.}
\newblock \bibinfo{journal}{\bibinfo{title}{{Spin-orbit torques in Co/Pt(111)
  and Mn/W(001) magnetic bilayers from first principles}}}.
\newblock {\emph{\JournalTitle{Physical Review B}}}
  \textbf{\bibinfo{volume}{90}}, \bibinfo{pages}{174423},
  \doiprefix\url{10.1103/PhysRevB.90.174423} (\bibinfo{year}{2014}).
\newblock \eprint{1305.4873}.

\bibitem{Haney2007}
\bibinfo{author}{Haney, P.~M.} \emph{et~al.}
\newblock \bibinfo{journal}{\bibinfo{title}{{Current-induced order parameter
  dynamics: Microscopic theory applied to Co/Cu/Co}}}.
\newblock {\emph{\JournalTitle{Physical Review B}}}
  \textbf{\bibinfo{volume}{76}}, \bibinfo{pages}{024404},
  \doiprefix\url{10.1103/PhysRevB.76.024404} (\bibinfo{year}{2007}).

\bibitem{GaratPRB2009}
\bibinfo{author}{Garate, I.} \& \bibinfo{author}{MacDonald, A.~H.}
\newblock \bibinfo{journal}{\bibinfo{title}{{Influence of a transport current
  on magnetic anisotropy in gyrotropic ferromagnets}}}.
\newblock {\emph{\JournalTitle{Physical Review B}}}
  \textbf{\bibinfo{volume}{80}}, \bibinfo{pages}{134403},
  \doiprefix\url{10.1103/PhysRevB.80.134403} (\bibinfo{year}{2009}).
\newblock \eprint{0905.3856}.

\bibitem{DresselhausM.S.andDresselhausG.andJorio2008}
\bibinfo{author}{Dresselhaus, M.~S.}, \bibinfo{author}{Dresselhaus, G.} \&
  \bibinfo{author}{Jorio, A.}
\newblock \emph{\bibinfo{title}{{Group theory}}}, vol.~\bibinfo{volume}{53}
  (\bibinfo{publisher}{Springer Berlin Heidelberg}, \bibinfo{address}{Berlin,
  Heidelberg}, \bibinfo{year}{2008}).

\bibitem{Zollner2020}
\bibinfo{author}{Zollner, K.} \emph{et~al.}
\newblock \bibinfo{journal}{\bibinfo{title}{{Scattering-induced and highly
  tunable by gate damping-like spin-orbit torque in graphene doubly
  proximitized by two-dimensional magnet Cr$_2$Ge$_2$Te$_6$ and monolayer
  WS$_2$}}}.
\newblock {\emph{\JournalTitle{Physical Review Research}}}
  \textbf{\bibinfo{volume}{2}}, \bibinfo{pages}{1--12},
  \doiprefix\url{10.1103/physrevresearch.2.043057} (\bibinfo{year}{2020}).
\newblock \eprint{1910.08072}.

\bibitem{Zihlmann2018}
\bibinfo{author}{Zihlmann, S.} \emph{et~al.}
\newblock \bibinfo{journal}{\bibinfo{title}{{Large spin relaxation anisotropy
  and valley-Zeeman spin-orbit coupling in WSe$_2$/Gr/hBN heterostructures}}}.
\newblock {\emph{\JournalTitle{Physical Review B}}}
  \textbf{\bibinfo{volume}{97}}, \bibinfo{pages}{075434},
  \doiprefix\url{10.1103/PhysRevB.97.075434} (\bibinfo{year}{2018}).

\bibitem{Manchon2015}
\bibinfo{author}{Manchon, A.}, \bibinfo{author}{Koo, H.~C.},
  \bibinfo{author}{Nitta, J.}, \bibinfo{author}{Frolov, S.~M.} \&
  \bibinfo{author}{Duine, R.~A.}
\newblock \bibinfo{journal}{\bibinfo{title}{{New perspectives for Rashba
  spin–orbit coupling}}}.
\newblock {\emph{\JournalTitle{Nature Materials}}}
  \textbf{\bibinfo{volume}{14}}, \bibinfo{pages}{871--882},
  \doiprefix\url{10.1038/nmat4360} (\bibinfo{year}{2015}).
\newblock \eprint{1507.02408}.

\bibitem{vantuan2014np}
\bibinfo{author}{Tuan, D.~V.}, \bibinfo{author}{Ortmann, F.},
  \bibinfo{author}{Soriano, D.}, \bibinfo{author}{Valenzuela, S.~O.} \&
  \bibinfo{author}{Roche, S.}
\newblock \bibinfo{journal}{\bibinfo{title}{{Pseudospin-driven spin relaxation
  mechanism in graphene}}}.
\newblock {\emph{\JournalTitle{Nature Physics}}} \textbf{\bibinfo{volume}{10}},
  \bibinfo{pages}{857--863}, \doiprefix\url{10.1038/nphys3083}
  (\bibinfo{year}{2014}).

\bibitem{Ok2019}
\bibinfo{author}{Ok, S.} \emph{et~al.}
\newblock \bibinfo{journal}{\bibinfo{title}{{Custodial glide symmetry of
  quantum spin Hall edge modes in monolayer WTe$_2$}}}.
\newblock {\emph{\JournalTitle{Physical Review B}}}
  \textbf{\bibinfo{volume}{99}}, \bibinfo{pages}{121105},
  \doiprefix\url{10.1103/PhysRevB.99.121105} (\bibinfo{year}{2019}).

\bibitem{Garcia2020}
\bibinfo{author}{Garcia, J.~H.} \emph{et~al.}
\newblock \bibinfo{journal}{\bibinfo{title}{{Canted Persistent Spin Texture and
  Quantum Spin Hall Effect in WTe$_2$}}}.
\newblock {\emph{\JournalTitle{Physical Review Letters}}}
  \textbf{\bibinfo{volume}{125}}, \bibinfo{pages}{256603},
  \doiprefix\url{10.1103/PhysRevLett.125.256603} (\bibinfo{year}{2020}).

\bibitem{LiuNatNat2021}
\bibinfo{author}{Liu, L.} \emph{et~al.}
\newblock \bibinfo{journal}{\bibinfo{title}{{Symmetry-dependent field-free
  switching of perpendicular magnetization}}}.
\newblock {\emph{\JournalTitle{Nature Nanotechnology}}}
  \textbf{\bibinfo{volume}{16}}, \bibinfo{pages}{277--282},
  \doiprefix\url{10.1038/s41565-020-00826-8} (\bibinfo{year}{2021}).

\bibitem{Garello2013}
\bibinfo{author}{Garello, K.} \emph{et~al.}
\newblock \bibinfo{journal}{\bibinfo{title}{{Symmetry and magnitude of
  spin-orbit torques in ferromagnetic heterostructures}}}.
\newblock {\emph{\JournalTitle{Nature Nanotechnology}}}
  \textbf{\bibinfo{volume}{8}}, \bibinfo{pages}{587--593},
  \doiprefix\url{10.1038/nnano.2013.145} (\bibinfo{year}{2013}).

\bibitem{Belashchenko2019}
\bibinfo{author}{Belashchenko, K.~D.}, \bibinfo{author}{Kovalev, A.~A.} \&
  \bibinfo{author}{{Van Schilfgaarde}, M.}
\newblock \bibinfo{journal}{\bibinfo{title}{{First-principles calculation of
  spin-orbit torque in a Co/Pt bilayer}}}.
\newblock {\emph{\JournalTitle{Physical Review Materials}}}
  \textbf{\bibinfo{volume}{3}}, \bibinfo{pages}{11401},
  \doiprefix\url{10.1103/PhysRevMaterials.3.011401} (\bibinfo{year}{2019}).
\newblock \eprint{1810.11003}.

\bibitem{Dolui2020}
\bibinfo{author}{Dolui, K.} \emph{et~al.}
\newblock \bibinfo{journal}{\bibinfo{title}{{Proximity Spin-Orbit Torque on a
  Two-Dimensional Magnet within van der Waals Heterostructure: Current-Driven
  Antiferromagnet-to-Ferromagnet Reversible Nonequilibrium Phase Transition in
  Bilayer CrI$_3$}}}.
\newblock {\emph{\JournalTitle{Nano Letters}}} \textbf{\bibinfo{volume}{20}},
  \bibinfo{pages}{2288--2295}, \doiprefix\url{10.1021/acs.nanolett.9b04556}
  (\bibinfo{year}{2020}).

\bibitem{KochanPRB2017}
\bibinfo{author}{Kochan, D.}, \bibinfo{author}{Irmer, S.} \&
  \bibinfo{author}{Fabian, J.}
\newblock \bibinfo{journal}{\bibinfo{title}{{Model spin-orbit coupling
  Hamiltonians for graphene systems}}}.
\newblock {\emph{\JournalTitle{Physical Review B}}}
  \textbf{\bibinfo{volume}{95}}, \bibinfo{pages}{165415},
  \doiprefix\url{10.1103/PhysRevB.95.165415} (\bibinfo{year}{2017}).
\newblock \eprint{1610.08794}.

\bibitem{PRLJohansen2019}
\bibinfo{author}{Johansen, {\O}.}, \bibinfo{author}{Risingg{\aa}rd, V.},
  \bibinfo{author}{Sudb{\o}, A.}, \bibinfo{author}{Linder, J.} \&
  \bibinfo{author}{Brataas, A.}
\newblock \bibinfo{journal}{\bibinfo{title}{Current control of magnetism in
  two-dimensional fe$_3$gete$_2$}}.
\newblock {\emph{\JournalTitle{Physical Review Letters}}}
  \textbf{\bibinfo{volume}{122}}, \bibinfo{pages}{217203},
  \doiprefix\url{10.1103/PhysRevLett.122.217203} (\bibinfo{year}{2019}).
\newblock \eprint{1812.06096}.

\bibitem{PRBSeemann2015}
\bibinfo{author}{Seemann, M.}, \bibinfo{author}{K{\"{o}}dderitzsch, D.},
  \bibinfo{author}{Wimmer, S.} \& \bibinfo{author}{Ebert, H.}
\newblock \bibinfo{journal}{\bibinfo{title}{{Symmetry-imposed shape of linear
  response tensors}}}.
\newblock {\emph{\JournalTitle{Phys. Rev. B}}} \textbf{\bibinfo{volume}{92}},
  \bibinfo{pages}{155138}, \doiprefix\url{10.1103/PhysRevB.92.155138}
  (\bibinfo{year}{2015}).
\newblock \eprint{1507.04947}.

\bibitem{MacneillPRB2017}
\bibinfo{author}{MacNeill, D.} \emph{et~al.}
\newblock \bibinfo{journal}{\bibinfo{title}{{Thickness dependence of spin-orbit
  torques generated by WTe$_2$}}}.
\newblock {\emph{\JournalTitle{Physical Review B}}}
  \textbf{\bibinfo{volume}{96}}, \bibinfo{pages}{054450},
  \doiprefix\url{10.1103/PhysRevB.96.054450} (\bibinfo{year}{2017}).
\newblock \eprint{1707.03757}.

\bibitem{Gupta2020}
\bibinfo{author}{Gupta, V.} \emph{et~al.}
\newblock \bibinfo{journal}{\bibinfo{title}{{Manipulation of the van der Waals
  Magnet Cr$_2$Ge$_2$Te$_6$ by Spin-Orbit Torques}}}.
\newblock {\emph{\JournalTitle{Nano Letters}}} \textbf{\bibinfo{volume}{20}},
  \bibinfo{pages}{7482--7488}, \doiprefix\url{10.1021/acs.nanolett.0c02965}
  (\bibinfo{year}{2020}).

\bibitem{Cheng2016}
\bibinfo{author}{Cheng, C.}, \bibinfo{author}{Sun, J.~T.},
  \bibinfo{author}{Chen, X.~R.}, \bibinfo{author}{Fu, H.~X.} \&
  \bibinfo{author}{Meng, S.}
\newblock \bibinfo{journal}{\bibinfo{title}{{Nonlinear Rashba spin splitting in
  transition metal dichalcogenide monolayers}}}.
\newblock {\emph{\JournalTitle{Nanoscale}}} \textbf{\bibinfo{volume}{8}},
  \bibinfo{pages}{17854--17860}, \doiprefix\url{10.1039/c6nr04235j}
  (\bibinfo{year}{2016}).

\bibitem{Shao2016}
\bibinfo{author}{Shao, Q.} \emph{et~al.}
\newblock \bibinfo{journal}{\bibinfo{title}{{Strong Rashba-Edelstein
  Effect-Induced Spin–Orbit Torques in Monolayer Transition Metal
  Dichalcogenide/Ferromagnet Bilayers}}}.
\newblock {\emph{\JournalTitle{Nano Letters}}} \textbf{\bibinfo{volume}{16}},
  \bibinfo{pages}{7514--7520}, \doiprefix\url{10.1021/acs.nanolett.6b03300}
  (\bibinfo{year}{2016}).

\bibitem{Zhang2016}
\bibinfo{author}{Zhang, W.} \emph{et~al.}
\newblock \bibinfo{journal}{\bibinfo{title}{{Research Update: Spin transfer
  torques in permalloy on monolayer MoS$_2$}}}.
\newblock {\emph{\JournalTitle{APL Materials}}} \textbf{\bibinfo{volume}{4}},
  \bibinfo{pages}{32302}, \doiprefix\url{10.1063/1.4943076}
  (\bibinfo{year}{2016}).

\bibitem{Husain2020b}
\bibinfo{author}{Husain, S.} \emph{et~al.}
\newblock \bibinfo{journal}{\bibinfo{title}{{Emergence of spin–orbit torques
  in 2D transition metal dichalcogenides: A status update}}}.
\newblock {\emph{\JournalTitle{Applied Physics Reviews}}}
  \textbf{\bibinfo{volume}{7}}, \bibinfo{pages}{041312},
  \doiprefix\url{10.1063/5.0025318} (\bibinfo{year}{2020}).

\bibitem{HiddingFM_2020}
\bibinfo{author}{Hidding, J.} \& \bibinfo{author}{Guimar{\~{a}}es, M.~H.}
\newblock \bibinfo{journal}{\bibinfo{title}{{Spin-Orbit Torques in Transition
  Metal Dichalcogenide/Ferromagnet Heterostructures}}}.
\newblock {\emph{\JournalTitle{Frontiers in Materials}}}
  \textbf{\bibinfo{volume}{7}}, \doiprefix\url{10.3389/fmats.2020.594771}
  (\bibinfo{year}{2020}).

\bibitem{Lv2018}
\bibinfo{author}{Lv, W.} \emph{et~al.}
\newblock \bibinfo{journal}{\bibinfo{title}{{Electric-Field Control of
  Spin-Orbit Torques in WS$_2$/Permalloy Bilayers}}}.
\newblock {\emph{\JournalTitle{ACS Applied Materials and Interfaces}}}
  \textbf{\bibinfo{volume}{10}}, \bibinfo{pages}{2843--2849},
  \doiprefix\url{10.1021/acsami.7b16919} (\bibinfo{year}{2018}).

\bibitem{StiehlPRB2019}
\bibinfo{author}{Stiehl, G.~M.} \emph{et~al.}
\newblock \bibinfo{journal}{\bibinfo{title}{{Layer-dependent spin-orbit torques
  generated by the centrosymmetric transition metal dichalcogenide
  $\beta$-MoTe$_2$}}}.
\newblock {\emph{\JournalTitle{Physical Review B}}}
  \textbf{\bibinfo{volume}{100}}, \doiprefix\url{10.1103/PhysRevB.100.184402}
  (\bibinfo{year}{2019}).
\newblock \eprint{1906.01068}.

\bibitem{Xie2019}
\bibinfo{author}{Xie, Q.} \emph{et~al.}
\newblock \bibinfo{journal}{\bibinfo{title}{{Giant Enhancements of
  Perpendicular Magnetic Anisotropy and Spin-Orbit Torque by a MoS$_2$
  Layer}}}.
\newblock {\emph{\JournalTitle{Advanced Materials}}}
  \textbf{\bibinfo{volume}{31}}, \bibinfo{pages}{1--9},
  \doiprefix\url{10.1002/adma.201900776} (\bibinfo{year}{2019}).

\bibitem{Liang2020}
\bibinfo{author}{Liang, S.} \emph{et~al.}
\newblock \bibinfo{journal}{\bibinfo{title}{{Spin-Orbit Torque Magnetization
  Switching in MoTe$_2$/Permalloy Heterostructures}}}.
\newblock {\emph{\JournalTitle{Advanced Materials}}}
  \textbf{\bibinfo{volume}{32}}, \bibinfo{pages}{1--6},
  \doiprefix\url{10.1002/adma.202002799} (\bibinfo{year}{2020}).

\bibitem{Haastrup_2018}
\bibinfo{author}{Haastrup, S.} \emph{et~al.}
\newblock \bibinfo{journal}{\bibinfo{title}{{The Computational 2D Materials
  Database: High-throughput modeling and discovery of atomically thin
  crystals}}}.
\newblock {\emph{\JournalTitle{2D Materials}}} \textbf{\bibinfo{volume}{5}},
  \bibinfo{pages}{042002}, \doiprefix\url{10.1088/2053-1583/aacfc1}
  (\bibinfo{year}{2018}).
\newblock \eprint{1806.03173}.

\bibitem{Lu2017}
\bibinfo{author}{Lu, A.~Y.} \emph{et~al.}
\newblock \bibinfo{journal}{\bibinfo{title}{{Janus monolayers of transition
  metal dichalcogenides}}}.
\newblock {\emph{\JournalTitle{Nature Nanotechnology}}}
  \textbf{\bibinfo{volume}{12}}, \bibinfo{pages}{744--749},
  \doiprefix\url{10.1038/nnano.2017.100} (\bibinfo{year}{2017}).

\bibitem{Zhang2017a}
\bibinfo{author}{Zhang, J.} \emph{et~al.}
\newblock \bibinfo{journal}{\bibinfo{title}{{Janus Monolayer Transition-Metal
  Dichalcogenides}}}.
\newblock {\emph{\JournalTitle{ACS Nano}}} \textbf{\bibinfo{volume}{11}},
  \bibinfo{pages}{8192--8198}, \doiprefix\url{10.1021/acsnano.7b03186}
  (\bibinfo{year}{2017}).
\newblock \eprint{1704.06389}.

\bibitem{Gambardella2011}
\bibinfo{author}{Gambardella, P.} \& \bibinfo{author}{Miron, I.~M.}
\newblock \bibinfo{journal}{\bibinfo{title}{{Current-induced spin–orbit
  torques}}}.
\newblock {\emph{\JournalTitle{Philosophical Transactions of the Royal Society
  A: Mathematical, Physical and Engineering Sciences}}}
  \textbf{\bibinfo{volume}{369}}, \bibinfo{pages}{3175--3197},
  \doiprefix\url{10.1098/rsta.2010.0336} (\bibinfo{year}{2011}).

\bibitem{Hayashi_PRB2014}
\bibinfo{author}{Hayashi, M.}, \bibinfo{author}{Kim, J.},
  \bibinfo{author}{Yamanouchi, M.} \& \bibinfo{author}{Ohno, H.}
\newblock \bibinfo{journal}{\bibinfo{title}{{Quantitative characterization of
  the spin-orbit torque using harmonic Hall voltage measurements}}}.
\newblock {\emph{\JournalTitle{Physical Review B}}}
  \textbf{\bibinfo{volume}{89}}, \bibinfo{pages}{144425},
  \doiprefix\url{10.1103/PhysRevB.89.144425} (\bibinfo{year}{2014}).

\bibitem{Liu_PRL2011}
\bibinfo{author}{Liu, L.}, \bibinfo{author}{Moriyama, T.},
  \bibinfo{author}{Ralph, D.~C.} \& \bibinfo{author}{Buhrman, R.~A.}
\newblock \bibinfo{journal}{\bibinfo{title}{{Spin-torque ferromagnetic
  resonance induced by the spin Hall effect}}}.
\newblock {\emph{\JournalTitle{Physical Review Letters}}}
  \textbf{\bibinfo{volume}{106}}, \bibinfo{pages}{36601},
  \doiprefix\url{10.1103/PhysRevLett.106.036601} (\bibinfo{year}{2011}).
\newblock \eprint{1011.2788}.

\bibitem{Mecking_PRB2007}
\bibinfo{author}{Mecking, N.}, \bibinfo{author}{Gui, Y.~S.} \&
  \bibinfo{author}{Hu, C.-M.}
\newblock \bibinfo{journal}{\bibinfo{title}{{Microwave photovoltage and
  photoresistance effects in ferromagnetic microstrips}}}.
\newblock {\emph{\JournalTitle{Physical Review B}}}
  \textbf{\bibinfo{volume}{76}}, \bibinfo{pages}{224430},
  \doiprefix\url{10.1103/PhysRevB.76.224430} (\bibinfo{year}{2007}).

\bibitem{FAN20211}
\bibinfo{author}{Fan, Z.} \emph{et~al.}
\newblock \bibinfo{journal}{\bibinfo{title}{{Linear scaling quantum transport
  methodologies}}}.
\newblock {\emph{\JournalTitle{Physics Reports}}}
  \textbf{\bibinfo{volume}{903}}, \bibinfo{pages}{1--69},
  \doiprefix\url{10.1016/j.physrep.2020.12.001} (\bibinfo{year}{2021}).
\newblock \eprint{1811.07387}.

\bibitem{LeeKiSeungPRB2015}
\bibinfo{author}{Lee, K.-S.} \emph{et~al.}
\newblock \bibinfo{journal}{\bibinfo{title}{{Angular dependence of spin-orbit
  spin-transfer torques}}}.
\newblock {\emph{\JournalTitle{Physical Review B}}}
  \textbf{\bibinfo{volume}{91}}, \bibinfo{pages}{144401},
  \doiprefix\url{10.1103/PhysRevB.91.144401} (\bibinfo{year}{2015}).
\newblock \eprint{1409.5600}.

\bibitem{Sousa2020}
\bibinfo{author}{Sousa, F.}, \bibinfo{author}{Tatara, G.} \&
  \bibinfo{author}{Ferreira, A.}
\newblock \bibinfo{journal}{\bibinfo{title}{{Skew-scattering-induced giant
  antidamping spin-orbit torques: Collinear and out-of-plane Edelstein effects
  at two-dimensional material/ferromagnet interfaces}}}.
\newblock {\emph{\JournalTitle{Physical Review Research}}}
  \textbf{\bibinfo{volume}{2}}, \bibinfo{pages}{43401},
  \doiprefix\url{10.1103/physrevresearch.2.043401} (\bibinfo{year}{2020}).

\bibitem{Xue2020}
\bibinfo{author}{Xue, F.}, \bibinfo{author}{Rohmann, C.}, \bibinfo{author}{Li,
  J.}, \bibinfo{author}{Amin, V.} \& \bibinfo{author}{Haney, P.}
\newblock \bibinfo{journal}{\bibinfo{title}{{Unconventional spin-orbit torque
  in transition metal dichalcogenide-ferromagnet bilayers from first-principles
  calculations}}}.
\newblock {\emph{\JournalTitle{Physical Review B}}}
  \textbf{\bibinfo{volume}{102}}, \bibinfo{pages}{14401},
  \doiprefix\url{10.1103/PhysRevB.102.014401} (\bibinfo{year}{2020}).
\newblock \eprint{2005.01109}.

\bibitem{Mahfouzi2020}
\bibinfo{author}{Mahfouzi, F.}, \bibinfo{author}{Mishra, R.},
  \bibinfo{author}{Chang, P.~H.}, \bibinfo{author}{Yang, H.} \&
  \bibinfo{author}{Kioussis, N.}
\newblock \bibinfo{journal}{\bibinfo{title}{{Microscopic origin of spin-orbit
  torque in ferromagnetic heterostructures: A first-principles approach}}}.
\newblock {\emph{\JournalTitle{Physical Review B}}}
  \textbf{\bibinfo{volume}{101}}, \bibinfo{pages}{60405},
  \doiprefix\url{10.1103/PhysRevB.101.060405} (\bibinfo{year}{2020}).

\bibitem{Nikolic2018}
\bibinfo{author}{Nikoli{\'{c}}, B.~K.} \emph{et~al.}
\newblock \bibinfo{title}{{First-principles quantum transport modeling of
  spin-transfer and spin-orbit torques in magnetic multilayers}}.
\newblock In \emph{\bibinfo{booktitle}{arXiv}}, \bibinfo{pages}{1--35},
  \doiprefix\url{10.1007/978-3-319-50257-1_112-1} (\bibinfo{publisher}{Springer
  International Publishing}, \bibinfo{address}{Cham}, \bibinfo{year}{2018}).
\newblock \eprint{1801.05793}.

\bibitem{Fan2014c}
\bibinfo{author}{Fan, X.} \emph{et~al.}
\newblock \bibinfo{journal}{\bibinfo{title}{{Quantifying interface and bulk
  contributions to spin–orbit torque in magnetic bilayers}}}.
\newblock {\emph{\JournalTitle{Nature Communications}}}
  \textbf{\bibinfo{volume}{5}}, \bibinfo{pages}{3042},
  \doiprefix\url{10.1038/ncomms4042} (\bibinfo{year}{2014}).

\bibitem{TaniguchiPRAppl2015}
\bibinfo{author}{Taniguchi, T.}, \bibinfo{author}{Grollier, J.} \&
  \bibinfo{author}{Stiles, M.~D.}
\newblock \bibinfo{journal}{\bibinfo{title}{Spin-transfer torques generated by
  the anomalous hall effect and anisotropic magnetoresistance}}.
\newblock {\emph{\JournalTitle{Phys. Rev. Applied}}}
  \textbf{\bibinfo{volume}{3}}, \bibinfo{pages}{044001},
  \doiprefix\url{10.1103/PhysRevApplied.3.044001} (\bibinfo{year}{2015}).

\bibitem{BaekNatMater2018}
\bibinfo{author}{Baek, S.-h.~C.} \emph{et~al.}
\newblock \bibinfo{journal}{\bibinfo{title}{Spin currents and spin--orbit
  torques in ferromagnetic trilayers}}.
\newblock {\emph{\JournalTitle{Nature Materials}}}
  \textbf{\bibinfo{volume}{17}}, \bibinfo{pages}{509--513},
  \doiprefix\url{10.1038/s41563-018-0041-5} (\bibinfo{year}{2018}).

\bibitem{Iihama_NatElec2018}
\bibinfo{author}{Iihama, S.} \emph{et~al.}
\newblock \bibinfo{journal}{\bibinfo{title}{Spin-transfer torque induced by the
  spin anomalous hall effect}}.
\newblock {\emph{\JournalTitle{Nature Electronics}}}
  \textbf{\bibinfo{volume}{1}}, \bibinfo{pages}{120--123},
  \doiprefix\url{10.1038/s41928-018-0026-z} (\bibinfo{year}{2018}).

\bibitem{BHATTI_MatToday2017}
\bibinfo{author}{Bhatti, S.} \emph{et~al.}
\newblock \bibinfo{journal}{\bibinfo{title}{{Spintronics based random access
  memory: a review}}}.
\newblock {\emph{\JournalTitle{Materials Today}}}
  \textbf{\bibinfo{volume}{20}}, \bibinfo{pages}{530--548},
  \doiprefix\url{10.1016/j.mattod.2017.07.007} (\bibinfo{year}{2017}).

\bibitem{Zheng_JPhysChemLett2019}
\bibinfo{author}{Zheng, S.} \emph{et~al.}
\newblock \bibinfo{journal}{\bibinfo{title}{{High-Temperature Ferromagnetism in
  an Fe$_3$P Monolayer with a Large Magnetic Anisotropy}}}.
\newblock {\emph{\JournalTitle{Journal of Physical Chemistry Letters}}}
  \textbf{\bibinfo{volume}{10}}, \bibinfo{pages}{2733--2738},
  \doiprefix\url{10.1021/acs.jpclett.9b00970} (\bibinfo{year}{2019}).

\bibitem{Torelli_2DMater2019}
\bibinfo{author}{Torelli, D.}, \bibinfo{author}{Thygesen, K.~S.} \&
  \bibinfo{author}{Olsen, T.}
\newblock \bibinfo{journal}{\bibinfo{title}{High throughput computational
  screening for 2d ferromagnetic materials: the critical role of anisotropy and
  local correlations}}.
\newblock {\emph{\JournalTitle{2D Materials}}} \textbf{\bibinfo{volume}{6}},
  \bibinfo{pages}{045018}, \doiprefix\url{10.1088/2053-1583/ab2c43}
  (\bibinfo{year}{2019}).

\bibitem{WadleyScience2016}
\bibinfo{author}{Wadley, P.} \emph{et~al.}
\newblock \bibinfo{journal}{\bibinfo{title}{Electrical switching of an
  antiferromagnet}}.
\newblock {\emph{\JournalTitle{Science}}}
  \doiprefix\url{10.1126/science.aab1031} (\bibinfo{year}{2016}).
\newblock
  \eprint{https://science.sciencemag.org/content/early/2016/01/13/science.aab1031.full.pdf}.

\bibitem{JungwirthNPhys2018}
\bibinfo{author}{Jungwirth, T.} \emph{et~al.}
\newblock \bibinfo{journal}{\bibinfo{title}{The multiple directions of
  antiferromagnetic spintronics}}.
\newblock {\emph{\JournalTitle{Nature Physics}}} \textbf{\bibinfo{volume}{14}},
  \bibinfo{pages}{200--203}, \doiprefix\url{10.1038/s41567-018-0063-6}
  (\bibinfo{year}{2018}).

\bibitem{Seixas_PRL2016}
\bibinfo{author}{Seixas, L.}, \bibinfo{author}{Rodin, A.~S.},
  \bibinfo{author}{Carvalho, A.} \& \bibinfo{author}{{Castro Neto}, A.~H.}
\newblock \bibinfo{journal}{\bibinfo{title}{{Multiferroic Two-Dimensional
  Materials}}}.
\newblock {\emph{\JournalTitle{Physical Review Letters}}}
  \textbf{\bibinfo{volume}{116}}, \bibinfo{pages}{206803},
  \doiprefix\url{10.1103/PhysRevLett.116.206803} (\bibinfo{year}{2016}).
\newblock \eprint{1601.06438}.

\bibitem{Huang_PRL2018}
\bibinfo{author}{Huang, C.} \emph{et~al.}
\newblock \bibinfo{journal}{\bibinfo{title}{{Prediction of Intrinsic
  Ferromagnetic Ferroelectricity in a Transition-Metal Halide Monolayer}}}.
\newblock {\emph{\JournalTitle{Physical Review Letters}}}
  \textbf{\bibinfo{volume}{120}}, \bibinfo{pages}{147601},
  \doiprefix\url{10.1103/PhysRevLett.120.147601} (\bibinfo{year}{2018}).

\bibitem{Yang_JACS2017}
\bibinfo{author}{Yang, Q.}, \bibinfo{author}{Xiong, W.}, \bibinfo{author}{Zhu,
  L.}, \bibinfo{author}{Gao, G.} \& \bibinfo{author}{Wu, M.}
\newblock \bibinfo{journal}{\bibinfo{title}{{Chemically Functionalized
  Phosphorene: Two-Dimensional Multiferroics with Vertical Polarization and
  Mobile Magnetism}}}.
\newblock {\emph{\JournalTitle{Journal of the American Chemical Society}}}
  \textbf{\bibinfo{volume}{139}}, \bibinfo{pages}{11506--11512},
  \doiprefix\url{10.1021/jacs.7b04422} (\bibinfo{year}{2017}).

\bibitem{Xu_PRL2020}
\bibinfo{author}{Xu, M.} \emph{et~al.}
\newblock \bibinfo{journal}{\bibinfo{title}{{Electrical Control of Magnetic
  Phase Transition in a Type-I Multiferroic Double-Metal Trihalide
  Monolayer}}}.
\newblock {\emph{\JournalTitle{Physical Review Letters}}}
  \textbf{\bibinfo{volume}{124}}, \bibinfo{pages}{67602},
  \doiprefix\url{10.1103/PhysRevLett.124.067602} (\bibinfo{year}{2020}).

\bibitem{Li_SciAdv2020}
\bibinfo{author}{Li, J.} \emph{et~al.}
\newblock \bibinfo{journal}{\bibinfo{title}{{Intrinsic magnetic topological
  insulators in van der Waals layered MnBi$_2$Te$_4$ -family materials}}}.
\newblock {\emph{\JournalTitle{Science Advances}}}
  \textbf{\bibinfo{volume}{5}}, \bibinfo{pages}{eaaw5685},
  \doiprefix\url{10.1126/sciadv.aaw5685} (\bibinfo{year}{2019}).

\bibitem{Wu_SciAdv2020}
\bibinfo{author}{Wu, J.} \emph{et~al.}
\newblock \bibinfo{journal}{\bibinfo{title}{{Natural van der Waals
  heterostructural single crystals with both magnetic and topological
  properties}}}.
\newblock {\emph{\JournalTitle{Science Advances}}}
  \textbf{\bibinfo{volume}{5}}, \doiprefix\url{10.1126/sciadv.aax9989}
  (\bibinfo{year}{2019}).

\bibitem{Hu_NComm2020}
\bibinfo{author}{Hu, C.} \emph{et~al.}
\newblock \bibinfo{journal}{\bibinfo{title}{{A van der Waals antiferromagnetic
  topological insulator with weak interlayer magnetic coupling}}}.
\newblock {\emph{\JournalTitle{Nature Communications}}}
  \textbf{\bibinfo{volume}{11}}, \bibinfo{pages}{97},
  \doiprefix\url{10.1038/s41467-019-13814-x} (\bibinfo{year}{2020}).

\bibitem{Deng2020}
\bibinfo{author}{Deng, Y.} \emph{et~al.}
\newblock \bibinfo{journal}{\bibinfo{title}{{Quantum anomalous Hall effect in
  intrinsic magnetic topological insulator MnBi$_2$Te$_4$}}}.
\newblock {\emph{\JournalTitle{Science}}} \textbf{\bibinfo{volume}{367}},
  \bibinfo{pages}{895--900}, \doiprefix\url{10.1126/science.aax8156}
  (\bibinfo{year}{2020}).
\newblock \eprint{1904.11468}.

\bibitem{Mellnik_Nature2014}
\bibinfo{author}{Mellnik, A.~R.} \emph{et~al.}
\newblock \bibinfo{journal}{\bibinfo{title}{{Spin-transfer torque generated by
  a topological insulator}}}.
\newblock {\emph{\JournalTitle{Nature}}} \textbf{\bibinfo{volume}{511}},
  \bibinfo{pages}{449--451}, \doiprefix\url{10.1038/nature13534}
  (\bibinfo{year}{2014}).

\bibitem{Yang_SciAdv2020}
\bibinfo{author}{Yang, M.} \emph{et~al.}
\newblock \bibinfo{journal}{\bibinfo{title}{{Creation of skyrmions in van der
  Waals ferromagnet Fe$_3$GeTe$_2$ on (Co/Pd)n superlattice}}}.
\newblock {\emph{\JournalTitle{Science Advances}}}
  \textbf{\bibinfo{volume}{6}}, \doiprefix\url{10.1126/sciadv.abb5157}
  (\bibinfo{year}{2020}).

\bibitem{Wu_NComm2020}
\bibinfo{author}{Wu, Y.} \emph{et~al.}
\newblock \bibinfo{journal}{\bibinfo{title}{{N{\'{e}}el-type skyrmion in
  WTe$_2$/Fe$_3$GeTe$_2$ van der Waals heterostructure}}}.
\newblock {\emph{\JournalTitle{Nature Communications}}}
  \textbf{\bibinfo{volume}{11}}, \bibinfo{pages}{3860},
  \doiprefix\url{10.1038/s41467-020-17566-x} (\bibinfo{year}{2020}).
\newblock \eprint{1907.11349}.

\bibitem{Park_PRB2021}
\bibinfo{author}{Park, T.-E.} \emph{et~al.}
\newblock \bibinfo{journal}{\bibinfo{title}{{N{\'{e}}el-type skyrmions and
  their current-induced motion in van der Waals ferromagnet-based
  heterostructures}}}.
\newblock {\emph{\JournalTitle{Physical Review B}}}
  \textbf{\bibinfo{volume}{103}}, \bibinfo{pages}{104410},
  \doiprefix\url{10.1103/PhysRevB.103.104410} (\bibinfo{year}{2021}).
\newblock \eprint{1907.01425}.

\bibitem{Zhang_NMater2020}
\bibinfo{author}{Zhang, X.-X.} \emph{et~al.}
\newblock \bibinfo{journal}{\bibinfo{title}{{Gate-tunable spin waves in
  antiferromagnetic atomic bilayers}}}.
\newblock {\emph{\JournalTitle{Nature Materials}}}
  \textbf{\bibinfo{volume}{19}}, \bibinfo{pages}{838--842},
  \doiprefix\url{10.1038/s41563-020-0713-9} (\bibinfo{year}{2020}).

\bibitem{McCreary_NComm2020}
\bibinfo{author}{McCreary, A.} \emph{et~al.}
\newblock \bibinfo{journal}{\bibinfo{title}{{Distinct magneto-Raman signatures
  of spin-flip phase transitions in CrI$_3$}}}.
\newblock {\emph{\JournalTitle{Nature Communications}}}
  \textbf{\bibinfo{volume}{11}}, \bibinfo{pages}{3879},
  \doiprefix\url{10.1038/s41467-020-17320-3} (\bibinfo{year}{2020}).
\newblock \eprint{1910.01237}.

\bibitem{Cenker_NPhys2021}
\bibinfo{author}{Cenker, J.} \emph{et~al.}
\newblock \bibinfo{journal}{\bibinfo{title}{{Direct observation of
  two-dimensional magnons in atomically thin CrI$_3$}}}.
\newblock {\emph{\JournalTitle{Nature Physics}}} \textbf{\bibinfo{volume}{17}},
  \bibinfo{pages}{20--25}, \doiprefix\url{10.1038/s41567-020-0999-1}
  (\bibinfo{year}{2021}).

\bibitem{Huang2020}
\bibinfo{author}{Huang, B.} \emph{et~al.}
\newblock \bibinfo{journal}{\bibinfo{title}{{Emergent phenomena and proximity
  effects in two-dimensional magnets and heterostructures}}}.
\newblock {\emph{\JournalTitle{Nature Materials}}}
  \textbf{\bibinfo{volume}{19}}, \bibinfo{pages}{1276--1289},
  \doiprefix\url{10.1038/s41563-020-0791-8} (\bibinfo{year}{2020}).

\end{thebibliography}

\section*{Acknowledgements}
H. K. and S. K. acknowledge supports from EPSRC via EP/T006749/1 and also help by Oscar Lee for producing graphical images. S. R. and J. H. G. acknowledge funding from the European Union Seventh Framework Programme under Grant No. 881603 (Graphene Flagship). ICN2 is funded by the CERCA Programme/Generalitat de Catalunya and supported by the Severo Ochoa programme (MINECO Grant. No. SEV-2017-0706).

\section*{Author contributions}
The authors contributed to all aspects of the article. 

\section*{Competing interests}
The authors declare no competing interests.

\section*{Publisher’s note}
Springer Nature remains neutral with regard to jurisdictional claims in published maps and institutional affiliations.

\end{document}